\documentclass[aps,eqsecnum,nofootinbib,superscriptaddress,showpacs,prd]{revtex4}
\usepackage{graphicx,amssymb,amsbsy,bm,amsmath}
\renewcommand{\Im}{\ensuremath{\mathrm{Im}}}
\renewcommand{\Re}{\ensuremath{\mathrm{Re}}}
\newcommand{\bnabla}{\ensuremath{\boldsymbol{\nabla}}}
\newcommand{\Tr}{\ensuremath{\mathrm{Tr}}}
\newcommand{\PV}{\ensuremath{\mathrm{PV}}}

\newcommand{\intp}{\ensuremath{\int\frac{d^3 p}{(2\pi)^3}}}
\newcommand{\intq}{\ensuremath{\int\frac{d^3 q}{(2\pi)^3}}}

\newcommand{\intpo}{\ensuremath{\int^{+\infty}_{-\infty} dp_0}}
\newcommand{\intqo}{\ensuremath{\int^{+\infty}_{-\infty} dq_0}}
\newcommand{\intso}{\ensuremath{\int^{+\infty}_{-\infty} ds_0}}
\newcommand{\intto}{\ensuremath{\int^{+\infty}_{-\infty} dt_0}}
\newcommand{\intinfty}{\ensuremath{\int^{+\infty}_{-\infty}}}

\newcommand{\bx}[1][]{\ensuremath{\mathbf{x}{#1}}}
\newcommand{\bk}{\ensuremath{\mathbf{k}}}
\newcommand{\bp}{\ensuremath{\mathbf{p}}}
\newcommand{\bq}{\ensuremath{\mathbf{q}}}

\newcommand{\bhp}{\ensuremath{\widehat{\mathbf{p}}}}
\newcommand{\bhq}{\ensuremath{\widehat{\mathbf{q}}}}
\newcommand{\xt}[1][]{\ensuremath{\mathbf{x}{#1},t{#1}}}

\newcommand{\bgamma}{\ensuremath{\boldsymbol{\gamma}}}
\newcommand{\nn}{\nonumber}
\newcommand{\be}{\begin{equation}}
\newcommand{\ee}{\end{equation}}
\newcommand{\bea}{\begin{eqnarray}}
\newcommand{\eea}{\end{eqnarray}}
\begin{document}
\title{Dynamical renormalization group approach to transport in
ultrarelativistic plasmas:\\the electrical conductivity in high
temperature QED}
\author{Daniel Boyanovsky}
\email{boyan@pitt.edu} \affiliation{Department of Physics and
Astronomy, University of Pittsburgh, Pittsburgh, Pennsylvania
15260, USA} \affiliation{LPTHE, Universit\'e Pierre et Marie Curie
(Paris VI) et Denis Diderot (Paris VII), UMR 7589 CNRS, Tour 16,
1er.\ \'etage, 4, Place Jussieu, 75252 Paris, Cedex 05, France}
\author{Hector J. de Vega}
\email{devega@lpthe.jussieu.fr} \affiliation{LPTHE, Universit\'e
Pierre et Marie Curie (Paris VI) et Denis Diderot (Paris VII), UMR
7589 CNRS, Tour 16, 1er.\ \'etage, 4, Place Jussieu, 75252 Paris,
Cedex 05, France} \affiliation{Department of Physics and
Astronomy, University of Pittsburgh, Pittsburgh, Pennsylvania
15260, USA}
\author{Shang-Yung Wang}
\email{swang@lanl.gov} \affiliation{Theoretical Division, MS B285,
Los Alamos National Laboratory, Los Alamos, NM 87545, USA}
\date{\today}

\begin{abstract}
The DC electrical conductivity of an ultrarelativistic QED plasma
is studied in real time by implementing the dynamical
renormalization group. The conductivity is obtained from the
real-time dependence of a dissipative kernel closely related to
the retarded photon polarization. Pinch singularities in the
imaginary part of the polarization are manifest as secular terms
that grow in time in the perturbative expansion of this kernel.
The leading secular terms are studied explicitly and it is shown
that they are insensitive to the anomalous damping of hard
fermions as a result of a cancellation between self-energy and
vertex corrections. The resummation of the secular terms via the
dynamical renormalization group leads directly to a
renormalization group equation in real time, which \emph{is} the
Boltzmann equation for the (gauge invariant) fermion distribution
function. A direct correspondence between the perturbative
expansion and the linearized Boltzmann equation is established,
allowing a direct identification of the self-energy and vertex
contributions to the collision term. We obtain a Fokker-Planck
equation in momentum space that describes the dynamics of the
departure from equilibrium to leading logarithmic order in the
coupling. This equation determines that the transport time scale
is given by $t_\mathrm{tr}=\frac{24\,\pi}{e^4 T \ln(1/e)}$. The
solution of the Fokker-Planck equation approaches asymptotically
the steady-state solution as $ \sim e^{-t/(4.038\ldots \,
t_\mathrm{tr})}$. The steady-state solution leads to the
conductivity $\sigma = \frac{15.698\,T}{e^2\ln(1/e)}$ to leading
logarithmic order. We discuss the contributions beyond leading
logarithms as well as beyond the Boltzmann equation. The dynamical
renormalization group provides a link between linear response in
quantum field theory and kinetic theory.
\end{abstract}

\preprint{LA-UR-02-7780} 

\maketitle
\tableofcontents
\section{Introduction}
Transport phenomena play a fundamental role in the dynamics of the
formation and evolution of an ultrarelativistic quark-gluon plasma
as well as in electromagnetic plasmas in the early universe. The
viscosity of a quark-gluon plasma enters in a hydrodynamic
description~\cite{hydrobooks,hydro} and energy losses depend in
general on transport coefficients of the plasma~\cite{loss}. In
early universe cosmology the electric conductivity plays a
fundamental role in the formation and decay (diffusion) of
primordial magnetic fields~\cite{magfields}.

A reliable calculation of transport coefficients in QED and QCD
requires a deeper understanding of screening by the medium in
order to treat the infrared properties of gauge theories
systematically. A study of transport coefficients for hot and
dense QED and QCD plasmas in kinetic theory including screening
corrections has been presented in Refs.~\cite{baym,heiselberg}. In
these references transport coefficients are computed by solving
the Boltzmann kinetic equation for the single particle
distribution functions with a collision term that includes
screening corrections. In an ultrarelativistic plasma the
propagators for soft quasiparticles have to include
nonperturbative resummations in terms of hard thermal loops
(HTL)~\cite{brapis,book:lebellac}. This resummation includes
consistently Debye screening for the longitudinal component of the
gauge field (Coulomb interaction) and dynamical screening by
Landau damping for the transverse
component~\cite{brapis,book:lebellac}. Screening via the HTL
resummation of the propagators for  gauge fields renders the
transport cross sections infrared finite~\cite{baym,heiselberg}.

The program of calculating transport coefficients from a kinetic
Boltzmann equation with HTL corrections to the collisional cross
sections has been pursued further, and a number of transport
coefficients have been calculated to leading logarithmic order in
the coupling constant or in the limit of the large number of
flavors~\cite{arnold}. An alternative approach to transport from a
more microscopic point of view is based on Kubo's linear response
formulation~\cite{kubo}. In this formulation, transport
coefficients are related to two-point correlation functions of
composite operators in the long-wavelength, low-frequency
limit~\cite{hosoya,jeon}. This approach provides a direct link to
transport phenomena from the underlying microscopic quantum field
theory, and formally a correspondence between Kubo's linear
response and the Boltzmann equation has been established in the
literature for some scalar field theories~\cite{jeon,kuboltz}.

The calculation of transport coefficients from a microscopic
quantum field theory at finite temperature or density involves
non-perturbative resummations~\cite{jeon} of a select type of
Feynman diagrams. As discussed in Ref.~\cite{jeon} for scalar
theories this resummation is necessary because in the limit of
vanishing external momentum and frequency, there emerges
\emph{pinch singularities} corresponding to the propagation of
on-shell intermediate states over arbitrarily long times and
distances. This propagation will be damped by collisional
processes which result in a perturbative width in the propagators.
Each order in the perturbative expansion which would be naively
suppressed by another power of the coupling, introduces a small
denominator that is controlled by the width, thus cancelling the
powers of the coupling in the numerators. Therefore, all terms in
the perturbative expansion from a select group of ladder-type
diagrams contribute at the same order~\cite{jeon}. The sum of
ladder diagrams leads to an integral equation which is similar to
a Boltzmann equation for the single-particle distribution function
in scalar theories~\cite{jeon,kuboltz,jakovac}, a result that was
anticipated in Ref.~\cite{holstein} within the context of
transport in electron-phonon systems.

Whereas the Boltzmann equation is a very efficient method to
extract transport coefficients without complicated resummations,
few attempts had been made to understand transport phenomena and
to extract transport coefficients directly from the underlying
microscopic quantum field theory in \emph{gauge theories}.
Recently, Kubo's formulation and the sum of ladder diagrams with a
width for the fermion propagator was carried out for the case of
the color conductivity~\cite{basagoiti1}  confirming the results
obtained via the kinetic approach~\cite{seli}. This program was
extended to study the  electric conductivity in an
ultrarelativistic QED plasma~\cite{basagoiti2,aarts}. In
Ref.~\cite{basagoiti2} it was shown that the sum of ladder
diagrams is equivalent to the steady-state form (i.e., no time
derivatives) of the linearized Boltzmann equation obtained in
Refs.~\cite{baym,arnold} in the leading logarithmic approximation.
This result was confirmed in Ref.~\cite{aarts}, where it was also
shown that new diagrams must be included to fulfill the Ward
identities, but that these diagrams do not contribute to leading
logarithmic order. While the important contributions of
Refs.~\cite{basagoiti1,basagoiti2,aarts} established the
\emph{equivalence} between the ladder resummation and the
steady-state form of the Boltzmann equation, a direct relationship
between linear response, the quantum field theoretical approach to
resummation of pinch singularities in \emph{real time} and the
time dependent Boltzmann equation in gauge field theories was not
yet available.

There is a fundamental interest in transport phenomena in
ultrarelativistic plasmas which warrants an understanding of the
main physical phenomena from different points of view. There are
systematic resummation schemes in quantum field theory which could
provide an alternative to the kinetic description or lead to a
systematic calculation of higher order effects that are not
captured by the kinetic approach. Furthermore, a microscopic
approach should allow a clear understanding of when a kinetic
description is valid, and also provide a framework to study
strongly out of equilibrium phenomena outside of the realm of
validity of kinetic theory. A resummation program that begins from
the equations of motion for correlation function is the
Schwinger-Dyson approach which leads to a hierarchy of equations
for higher order correlation functions. Suitable truncations of
this hierarchy justified by the particular physical case would
lead to a systematic calculation of transport coefficients. Such
program was initiated in Ref.~\cite{mottola} for the case of the
DC electrical conductivity in a high temperature QED plasma.

An alternative program to study relaxation and transport is based
on a real-time implementation of the renormalization
group~\cite{boyanqed,scalar}. Just as in the usual renormalization
group, this approach is based on a wide separation of scales. In
real-time these scales are the microscopic and the transport time
scales, which are widely separated in weakly coupled theories. The
dynamical renormalization group equations determine the evolution
of correlation functions and expectation values on the \emph{long}
time scales. In particular, the Boltzmann kinetic equation can be
interpreted as a \emph{renormalization group equation} for the
single-particle distribution function where the renormalization
group parameter is \emph{real time}~\cite{scalar,boyanqed}. In
Refs.~\cite{boyanqed,scalar} it was pointed out that the pinch
singularities that are ubiquitous in finite temperature field
theory and that require the resummation of the perturbative
expansion are manifest in real time as \emph{secular terms},
namely terms that grow in time in the perturbative expansion of
real-time kernels. The resummation of these pinch singularities in
the Fourier transform of the correlation functions in the limit of
small momentum and frequency leads to integral
equations~\cite{jeon,jakovac}, while directly in real time this
resummation is achieved by the dynamical renormalization
group~\cite{boyanqed,scalar}.

The main point of the dynamical renormalization group (DRG)
approach is that time acts as an infrared cutoff, the correlation
functions do not have singularities at any finite time since
singularities only arise in the infinite time limit. In transport
phenomena, as discussed above, pinch singularities arise from the
propagation of on-shell intermediate states over arbitrarily long
time (or distances). In the same manner as the usual
renormalization group resums the infrared behavior of correlation
functions in critical phenomena, the dynamical renormalization
group resums the long-time behavior of correlators or expectation
values~\cite{scalar}. In the case of the single-particle
distribution functions, the DRG equation has been shown to be the
Boltzmann equation~\cite{scalar,boyanqed}.

In this article we begin a program to study transport phenomena
directly from the underlying quantum field theory in \emph{real
time}. The main goal of the program is to provide an understanding
of transport and relaxation implementing concepts and ideas from
the renormalization group description (either in real time or
frequency and momentum space). Such program will lead to a
description of transport phenomena much in the same manner as in
critical phenomena and in deep-inelastic scattering, where physics
at widely different scales is studied via the renormalization
group. An example of this interpretation has been recently
provided in the identification of the Altarelli-Parisi-Lipatov
equations, which describe the evolution of parton distribution
functions in invariant momentum transfer $Q^2$, as DRG (or
Boltzmann) equations~\cite{boydglap}. Thus the dynamical
renormalization group links transport phenomena with critical
phenomena. Furthermore, a description of transport phenomena
directly from the underlying quantum field theory in real time
lead to further understanding of the validity of the kinetic
description as well as to deal with strongly out of equilibrium
phenomena where a kinetic description may not be suitable or
reliable.

\textbf{Goals and main results of this article}. In this article
we focus on studying the DC electrical conductivity in an
ultrarelativistic QED plasma. We study transport phenomena
directly from quantum field theory in \emph{real time} by
implementing a resummation of the perturbative expansion via the
dynamical renormalization group. This program allows to establish
a direct relationship between linear response, the resummation of
pinch singularities and the time-dependent Boltzmann equation. The
solution of the linearized dynamical renormalization group
equation leads to the transport coefficients, in this case the DC
electric conductivity. Along the way this program also establishes
several important aspects: (i) a direct relationship between pinch
singularities in the perturbative expansion in linear response and
their resummation via the Boltzmann equation in \emph{real time},
(ii) a direct identification of the single-particle distribution
function with the nonequilibrium expectation value of a
\emph{gauge invariant} bilinear operator, (iii) the relevant
approximations that determine the validity of the kinetic
description, (iv) a direct identification of the self-energy and
vertex corrections in quantum field theory and the different terms
in the linearized Boltzmann equation, highlighting that the
transport time scale is a consequence of the cancellation of the
anomalous damping rate between the self-energy and vertex
corrections, and (v) a pathway to include corrections to the
leading logarithmic approximation and to the Boltzmann equation,
with a firm ground on quantum field theory.

The main results of this article are summarized as follows.
\begin{itemize}
\item[(i)]{We begin by studying the real-time dynamics of an
initially prepared magnetic field fluctuation as an initial value
problem in linear response. We relate the DC conductivity to the
time integral of a kernel closely related to the retarded photon
polarization. A perturbative evaluation of this kernel in
\emph{real time} reveals secular terms, namely terms that grow in
time and invalidate the perturbative expansion. These secular
terms are a manifestation of pinch singularities in the Fourier
transform of this kernel. We highlight how these secular terms are
manifest in the perturbative solution of the Boltzmann equation in
real time. This is an important and revealing aspect of the study
of quantum field theory in real time that establishes a direct
link between perturbation theory in the quantum field theoretical
approach, secular terms (pinch singularities) and perturbation
theory at the level of the Boltzmann equation. This study also
illuminates the resummation of secular terms performed by the
Boltzmann equation.}

\item[(ii)]{In early studies of the DC conductivity in QED plasmas
it was recognized that a subtle cancellation between self-energy
and vertex corrections makes the conductivity insensitive to the
anomalous fermion damping rate~\cite{lebedev}. This cancellation
for a QED plasma, originally noted in Ref.~\cite{lebedev} has
since been found in many other
contexts~\cite{carrington,kraemmer,aurenche,gale} and was
explicitly shown to occur in the ladder resummation approach to
extract the conductivity~\cite{basagoiti2,aarts}. Here we study
the cancellation between the self-energy and vertex corrections
with HTL photon propagators for the DC conductivity and transport
phenomena directly in real time. An important consequence of the
perturbative approach in real time is that the secular terms
directly indicate at which time scale perturbation theory breaks
down. This time scale is associated with the transport time
scale~\cite{scalar,boyanqed} beyond which a resummation program
like the DRG must be used. Our analysis in real time clearly
indicates that the cancellation of the contribution from
\emph{ultrasoft photon exchange} between the self-energy and
vertex corrections makes the transport time scale insensitive to
the anomalous damping rate of hard fermions~\cite{lebedev}. Within
the framework of transport phenomena this cancellation is at the
heart of the distinction between the quasiparticle relaxation time
scale and the transport time scale, which makes the calculation of
the DC \emph{electrical conductivity} much more subtle and
complicated than that for the \emph{color conductivity} in QCD.
For color transport there is \emph{no} such cancellation and the
leading contribution to the color conductivity arises from the
anomalous damping of the hard quarks~\cite{basagoiti1}.}

\item[(iii)]{From the study in linear response, we identify the
quantum field theoretical equivalent of the gauge invariant
single-particle distribution function whose equation of motion is
related to the dissipative kernel that leads to the conductivity.
We obtain the equation of motion for this distribution function in
perturbation theory and show that its solution features secular
terms at large times. We then implement the dynamical
renormalization group to resume the leading secular terms. The
dynamical renormalization group equation for the single-particle
distribution function is recognized as the time-dependent
Boltzmann equation. Furthermore, we show that the anomalous
damping of hard fermions is manifest as \emph{anomalous} secular
terms in the perturbative solution of the Boltzmann equation in
the \emph{relaxation time approximation}. The perturbative
solution of the \emph{linearized} time-dependent Boltzmann
equation is shown to feature exactly the same \emph{linear}
secular terms as those found in the perturbative evaluation of the
dissipative kernel. This study clarifies the cancellation of the
anomalous damping between the different contributions in the
linearized Boltzmann equation, and confirms the identification of
the various terms in the linearized collision term with the
self-energy and vertex corrections to the dissipative kernel in
perturbation theory. This is yet another link between the
perturbative framework in quantum field theory and that at the
level of the Boltzmann equation, establishing directly the
resummation implied by the Boltzmann equation. Our study clarifies
directly in real time how the Boltzmann equation resums the pinch
singularities found in perturbation theory. This detailed and
clear link between quantum field theory and the Boltzmann equation
cannot be extracted from the simplified equivalence between ladder
diagrams and the steady-state Boltzmann equation and requires the
time-dependent Boltzmann equation.}

\item[(iv)]{After recognizing that the leading logarithmic
contribution to the collision term of the linearized Boltzmann
equation is dominated by the kinematic region of momentum exchange
$eT\lesssim q \lesssim T$ between particles with typical momenta
$p \gtrsim T$, we expand the collision kernel in powers of $q/p$
to obtain a Fokker-Planck equation which describes the \emph{time}
and \emph{momentum} evolution of the departure from equilibrium of
the distribution function in the linearized approximation. The
time dependent Fokker-Planck equation is solved by expanding in
the eigenfuntions of a positive definite Hamiltonian. For late
times, its solution approaches asymptotically the steady-state
solution as $\sim e^{- t/(4.038\ldots \, t_\mathrm{tr})}$, where
$t_\mathrm{tr}=\frac{24 \,\pi}{e^4 T \ln(1/e)}$. We solved
analytically the steady-state Fokker-Planck equation for small and
large momenta which describes the small- and large-momentum
behavior of the departure from equilibrium in the steady state. We
use these analytic asymptotic solutions to calculate numerically
the DC conductivity. We find the leading logarithmic expression
for the DC conductivity which agrees to within less than $0.1\,\%$
with the results of Ref.~\cite{arnold}. Furthermore, the
Fokker-Planck equation allows to establish contact with the
variational formulation used in Ref.~\cite{arnold}.}

\item[(v)]{We discuss the diagrams that are necessary to be
included in the collision term to next to leading logarithmic
order in the coupling, and the range of validity of the Boltzmann
equation. It is pointed out that in order to go beyond the simple
Boltzmann equation, terms that describe spin precession must be
included in the set of kinetic equations.}
\end{itemize}

The article is organized as follows. In Sec.~\ref{sec:relax} we
introduce the real-time approach to extract the DC conductivity
from the hydrodynamic relaxation of long-wavelength magnetic
fields. The DC conductivity is determined by the time integral of
a dissipative kernel directly related to the photon polarization.
This kernel will provide the link between quantum field theory and
kinetic theory. In Sec.~\ref{sec:perttheory} we analyze the kernel
that defines the conductivity to lowest order in the
hard-thermal-loop approximation and describe the strategy to resum
the perturbative series by extracting the leading secular terms in
time. Then we study in detail the resummed one-loop self-energy
and vertex in the hydrodynamic limit. In this section we make a
connection with the results of
Refs.~\cite{blaizot,boyanqed,scalar} for the real-time behavior of
the fermion propagator and its anomalous damping in order to
identify the secular terms in the dissipative kernel associated
with the anomalous fermion damping. The Ward identity between the
self-energy and vertex is shown to be fulfilled in this
approximation. In Sec.~\ref{sec:twopoop} we compute the imaginary
part of the resummed two-loop transverse photon polarization using
the resummed self-energy and vertex obtained above. We extract the
hydrodynamic poles (or pinch denominators) which in real time are
manifest as secular terms that grow in time. Then we analyze the
cancellation alluded to above and extract the leading logarithmic
behavior in the gauge coupling of the leading secular terms. We
discuss in detail which diagrams and which region of the exchanged
momentum contribute to the leading secular term to leading
logarithmic accuracy.

In Sec.~\ref{section:boltz} we establish contact with the
Boltzmann equation approach by identifying the single-particle
distribution functions and obtain their equations of motion in
perturbation theory. The perturbative solutions to these equations
feature secular terms. The dynamical renormalization group is
introduced to resum the perturbative equation of motion, and the
DRG equation is identified with the Boltzmann equation. We
establish direct contact between the linearized Boltzmann equation
and the results obtained from the perturbative quantum field
theory of the previous section. We obtain a Fokker-Planck
equation, we solve it in an eigenfunction expansion and find its
asymptotic late time behavior yielding a  steady-state solution.
>From it we extract the leading logarithmic contribution to the DC
conductivity. In this section we also discuss the regime of
validity of the Boltzmann approach and the contributions that must
be included to go beyond leading logarithmic order as well as
beyond the Boltzmann equation. Finally, we present our conclusions
and a discussion of future avenues in Sec.~\ref{conclusions}. Two
appendices are devoted to summarizing the imaginary-time and
real-time propagators used in the main text.

\section{Linear response}\label{sec:relax}

In this section we relate the conductivity to the relaxation of
the gauge mean field as well as to the current induced by an
external electric field using linear response. The reason for
delving on linear response is to recognize the main
\emph{real-time quantity} that leads to the conductivity and is
the link between Kubo's linear response, the dynamical
renormalization group and the Boltzmann approach.

The Lagrangian density of QED, the theory under consideration, is
given by
\begin{equation}
\mathcal{L}= \bar{\psi}(i\!\not\!\partial -e \not\!\!A)\psi
-\frac{1}{4}F_{\mu \nu}F^{\mu \nu} \; ,\label{lagra}
\end{equation}
where the zero-temperature mass of the fermion $m$ has been
neglected in the high temperature limit $T\gg m$. We begin by
casting our study directly in a manifestly gauge invariant form
(see also Ref.~\cite{scalar}). In the Abelian case it is
straightforward to reduce the Hilbert space to the gauge invariant
states and to define gauge invariant fields. This is best achieved
within the canonical Hamiltonian formulation in terms of primary
and secondary class constraints. In the Abelian case there are two
first class constraints:
\begin{equation}
\pi_0=0 \; ,\quad
\bnabla\cdot\boldsymbol{\pi}=-e\,\psi^\dagger\psi \; ,
\label{firstclass}
\end{equation}
where $\pi_0$ and $\boldsymbol{\pi}=-\mathbf{E}$ are the canonical
momenta conjugate to $A^0$ and $\mathbf{A}$, respectively.
Physical states are those which are simultaneously annihilated by
the first class constraints and physical operators \emph{commute}
with the first class constraints. Writing the gauge field in terms
of transverse and longitudinal components as
$\mathbf{A}=\mathbf{A}_L+\mathbf{A}_T$ with
$\bnabla\times\mathbf{A}_L=\bnabla\cdot\mathbf{A}_T=0$ and
defining
\begin{equation}
\Psi(x)=\psi(x) \; e^{ie\int d^3y
\bnabla_\mathbf{x}G(\mathbf{x}-\mathbf{y})\cdot\mathbf{A}_L(y)} \;
,
\end{equation}
where $G(\bx-\mathbf{y})$ the Coulomb Green's function satisfying
$\bnabla_{\bx}^2 G(\bx-\mathbf{y})=\delta^3(\bx-\mathbf{y})$,
after some algebra using the canonical commutation relations one
finds that $\mathbf{A}_T(x)$ and $\Psi(x)$ are \emph{gauge
invariant} field operators.

The Hamiltonian can now be written solely in terms of these gauge
invariant operators and when acting on gauge invariant states the
resulting Hamiltonian is equivalent to that obtained in Coulomb
gauge. However we emphasize that we have \emph{not} fixed any
gauge, this treatment, originally introduced by Dirac is
manifestly gauge invariant. The instantaneous Coulomb interaction
can be traded for a gauge invariant Lagrange multiplier field
which we call $A^0$, leading to the following Lagrangian
density\cite{boyanqed}
\begin{equation}
\mathcal{L} = \bar{\Psi}(i\!\not\!\partial -e\gamma^0 A^0 +
e\boldsymbol{\gamma}\cdot \mathbf{A}_T)\Psi+ \frac{1}{2}
\left[(\partial_{\mu} \mathbf{A}_T)^2+ (\bnabla
A^0)^2\right]\label{lagrainv} \; .
\end{equation}
We emphasize that $A^0$ should \emph{not} be confused with the
temporal gauge field component.

The main reason to introduce the gauge invariant formulation is
that we will establish contact with the Boltzmann equation for the
single particle distribution function which must be defined in a
gauge invariant manner.

\subsection{Relaxation of gauge mean field}

The strategy is first to prepare the  system at equilibrium in the
remote past and then to introduce an adiabatic external source
$\mathbf{J}_\mathrm{ext}$ coupled to the transverse  gauge field
$\mathbf{A}_T$. The external source will induce an expectation
value for the gauge field, representing a small departure from
equilibrium. The external field is switched-off at $ t = 0 $ and
the induced expectation value relaxes then towards equilibrium.
The real-time dynamics of relaxation is studied as an
\emph{initial value problem} and the conductivity is extracted
from the relaxation rate of long-wavelength perturbations. This
approach has already been used in Refs.~\cite{boyanqed,scalar},
where more details can be found.

Introducing an external source $\mathbf{J}_\mathrm{ext}$ with the
following time dependence
\begin{equation}
J^i_\mathrm{ext}(\xt)=J^i_\mathrm{ext}(\bx) \; e^{\epsilon t} \;
\Theta(-t) \; ,\label{source}
\end{equation}
we find the equation of motion for the expectation value
$\mathbf{a}_T(\xt)=\langle \mathbf{A}_T(\xt)
\rangle_{J_\mathrm{ext}}$ to be given by~\cite{boyanqed}
\begin{equation}
\ddot{a}^i_T(\xt)-\bnabla^2 { a}^i_T(\xt)+\int d^4x' \;
\Pi^{ij}_\mathrm{ret}(\bx-\bx', t-t') \; {a}^j_T(\bx',t')=
J^i_\mathrm{ext}(\bx) \; e^{\epsilon t} \; \Theta(-t) \;
,\label{eqnofmotion}
\end{equation}
where $\Pi^{ij}_\mathrm{ret}(\bx-\bx',t-t')$ is the retarded
polarization~\cite{boyanqed}
\begin{eqnarray}
\Pi^{ij}_\mathrm{ret}(\bx-\bx',t-t')&=& -i\,\langle\left[J^i(\xt),
J^j(\bx',t')\right]\rangle\,\Theta(t-t')\nn\\
&=& i \intinfty \frac{d\omega}{\pi} \;
\mathrm{Im}\Pi^{ij}(\omega,k)
 \; e^{-i[\omega(t-t')-\bk\cdot(\bx-\bx')]} \; \Theta(t-t') \; ,
\label{polarization}
\end{eqnarray}
where $\mathbf{J}=e\bar{\Psi}\boldsymbol{\gamma}\Psi$ is the
electromagnetic current and $k=|\bk|$. Using the Fourier
representation of $\Theta(t-t')$, we find that the spatial and
temporal Fourier transform of the retarded polarization can be
written in the form
\begin{equation}
{\Pi}^{ij}(\omega,k) = \intinfty\frac{dk_0}{\pi}
\frac{\mathrm{Im}{\Pi}^{ij}(k_0,k)}{k_0-\omega-i0} \; .
\label{retpolarization}
\end{equation}
Taking the spatial Fourier transform of the equation of motion
(\ref{eqnofmotion}) and denoting the spatial Fourier transforms of
the expectation value of the gauge field and the external source
term as $\mathbf{a}_T(\bk,t)$ and
$\mathbf{J}_\mathrm{ext}(\bk,t)$, respectively, we find that for
$t<0$ the solution of the equation of motion (\ref{eqnofmotion})
with the source eq.~(\ref{source}) is given by
\begin{equation}
\mathbf{a}_T(\bk,t<0) = \mathbf{a}_T(\bk,0) \; e^{\epsilon t} \; ,
\label{solTlesszero}
\end{equation}
where $\mathbf{a}_T(\bk,0)$ and $\mathbf{J}_\mathrm{ext}(\bk)$ are
related by eq.~(\ref{eqnofmotion}) for $t<0$. In the limit
$\epsilon \to 0$ this solution entails that
$\dot{\mathbf{a}}_T(\bk,t\leqslant 0)=0$. Introducing the
function,
\begin{equation}
\mathcal{G}^{ij}(k,t) \equiv
\frac{1}{\pi}\intinfty\frac{dk_0}{k_0}
 \; \mathrm{Im}{\Pi}^{ij}(k_0,k) \; e^{-ik_0 t} \; , \label{Goft}
\end{equation}
carrying out an integration by parts in the last term of the
equation of motion (\ref{eqnofmotion}) and changing variables
$t-t'\to t'$ we find that for $t>0$ the equation of motion
(\ref{eqnofmotion}) becomes
\begin{equation}
\ddot{a}^i_T(\bk,t)+
\left[k^2\delta^{ij}+\mathcal{G}^{ij}(k,0)\right]
a^j_T(\bk,t)-\int^{t}_{0}dt' \; \mathcal{G}^{ij}(k,t') \;
\dot{a}^j_T(\bk,t-t')= 0 \; ,\label{motion}
\end{equation}
where we have used $\dot{\mathbf{a}}_T(\bk,t\leq 0)=0$ in the
limit $\epsilon \rightarrow 0$. Eq.~(\ref{motion}) manifestly
describes the dynamics of the induced expectation value as an
\emph{initial value problem} in real time, which can be solved via
Laplace transform once the kernel $\mathcal{G}^{ij}(k,t)$ is
determined. An important aspect of this equation is that it
illuminates the connection with relaxation and dissipation.

\emph{If} the kernel $\mathcal{G}^{ij}(k,t)$ is localized in time
in the region $0<t<t_\mathrm{mem}(k)$, where $t_\mathrm{mem}(k)$
determines the memory of the kernel, then for $t \gg
t_\mathrm{mem}(k)$ we can expand in derivatives
$\dot{\mathbf{a}}_T(\bk,t-t')= \dot{\mathbf{a}}_T(\bk,t)-
\ddot{\mathbf{a}}_T(\bk,t) \; t'+\cdots$ and the equation of
motion (\ref{eqnofmotion}) becomes an infinite series of higher
time derivatives,
\begin{equation}
\ddot{a}^i_T(\bk,t)+ \left[k^2
\delta^{ij}+\mathcal{G}^{ij}(k,0)\right]
a^j_T(\bk,t)+\sigma^{ij}(k)\,\dot{a}^j_T(\bk,t)+
\sigma_1^{ij}(k)\, \ddot{a}^j_T(\bk,t) +\cdots =0 \; .
\label{localeqn}
\end{equation}
In the above expression we have introduced,
\begin{equation}
\sigma^{ij}(k) =-\int_0^{\infty}dt \; \mathcal{G}^{ij}(k,t) \;,
\quad \sigma_1^{ij}(k)=\int_0^{\infty}dt' \;  t' \;
\mathcal{G}^{ij}(k,t') \; , \label{condofk}
\end{equation}
where the upper limit in the time integrals is taken to infinity
since by assumption $t\gg t_\mathrm{mem}(k)$.

We will now focus on the relaxation of gauge mean field in the
long-wavelength limit. For $k\to 0$ we expect that
$\mathcal{G}^{ij}(0,t) =\mathcal{G}(t) \; \delta^{ij}$, with
$\mathcal{G}(t) =\frac{1}{3}\sum_{i=1}^3\mathcal{G}^{ii}(0,t)$.
Hence, we \emph{define} the DC electrical conductivity as
\begin{equation}
\sigma=\frac{1}{3}\lim_{k\to 0}\sum_{i=1}^3 \sigma^{ii}(k)=
-\int_0^{\infty}dt'  \; \mathcal{G}(t') \; .\label{conductivity}
\end{equation}
The function $\mathcal{G}^{ij}(k,0)$ has a simple interpretation
$\mathcal{G}^{ij}(k,0)= \mathrm{Re}\Pi^{ij}(0,k)$,\footnote{Since
$\mathrm{Im}\Pi_T(k_0,k) $ is an odd function of $k_0$, it
vanishes as $-\sigma_k k_0$ for $k_0\to 0$ and there is no need to
append a principal part prescription.} which can be understood
from the spectral representation of the Fourier transform of the
polarization eq.~(\ref{retpolarization}). In the Matsubara
representation, the inverse transverse photon propagator in the
static limit ($\omega_n =0$) is given by $k^2+\Pi_T(k,0)$. Thus
the fact that there is \emph{no} magnetic mass in thermal
QED~\cite{book:lebellac} entails that $\lim_{k \to
0}\mathrm{Re}\Pi_T(0,k)=0$. As a result we can neglect
$\mathcal{G}^{ij}(k,0)$ in eq.~(\ref{localeqn}) since only its
transverse component enters the equation of motion for
$\mathbf{a}_T$. Therefore, eq.~(\ref{localeqn}) becomes
\begin{equation} \label{apro} \ddot{\mathbf{a}}_T(\bk,t) + \sigma
\; \dot{\mathbf{a}}_T(\bk,t) + k^2 \; \mathbf{a}_T(\bk,t) = 0 \; ,
\end{equation} with the solution for the long-time relaxation of small
departures from equilibrium for transverse gauge fields of long
wavelengths $k \ll \sigma$,
\begin{equation} \mathbf{a}_T(\bk,t) \sim
e^{-k^2t/\sigma} \; .\label{relaxation}
\end{equation}
This purely diffusive relaxation is the hallmark of slow decay of
magnetic fields in the hydrodynamic limit, an important aspect of
the generation and evolution of cosmic magnetic
fields~\cite{magfields}.

The validity of the derivative expansion leading to the local
equation of motion (\ref{apro}) can now be assessed. Since
$\ddot{\mathbf{a}}_T =-(k^2/\sigma)^2 \; \mathbf{a}_T$, the term
$\ddot{\mathbf{a}}_T(\bk,t)$ in eq.~(\ref{apro}) is subleading for
$k\ll \sigma$. Furthermore, the  coefficient of the second
derivative term in the derivative expansion eq.~(\ref{localeqn})
is given by
\begin{equation}
\int^{\infty}_0 dt \; t \; \mathcal{G}^{ij}(k,t) \sim
\sigma^{ij}(k) \; t_\mathrm{mem}(k) \; . \label{coeff2ndder}
\end{equation}
Hence, the second derivative term in the derivative expansion will
be much smaller than the first derivative term displayed in
 eq.~(\ref{localeqn}) when
\begin{equation}
\frac{k^2  \; t_\mathrm{mem}(k)}{\sigma}\ll 1 \; .\label{validity}
\end{equation}
Anticipating that after a resummation program and up to logarithms
of the electromagnetic coupling $\sigma \propto T/\alpha$ and that
$t_\mathrm{mem}$ is the transport relaxation time (to be confirmed
later) with $t_\mathrm{mem} \propto 1/\alpha^2 T$,  the validity
of the derivative expansion is warranted for long wavelength $k
\ll e T$. In this regime we can also drop the usual kinetic
(second derivative) term since $k\ll \sigma$ and the long-time
limit is completely determined by the hydrodynamic form
eq.~(\ref{relaxation}). The conductivity $\sigma$ is also
identified with the inverse of the magnetic diffusion coefficient,
which in turn determines the Reynolds number in the equations of
magnetohydrodynamics and the decay of magnetic fields in
cosmology.

Introducing a convergence factor $\epsilon \to 0^+$ in the
definition of $\sigma_k$ in eq.~(\ref{condofk}), we find
\begin{equation} \sigma^{ij}(k) = -\int^{\infty}_0 dt \;
\mathcal{G}^{ij}(k,t) \; e^{-\epsilon t} =
i\intinfty\frac{dk_0}{\pi} \;
\frac{\mathrm{Im}{\Pi}^{ij}(\bk,k_0)}{k_0(k_0-i\epsilon)} \;
.\label{condrep} \end{equation} Using the dispersive
representation eq.~(\ref{retpolarization}) for the retarded
polarization, we find that
\begin{equation}
\sigma^{ij}(k) = \left.\frac{i}{\omega}
\left[{\Pi}^{ij}(\omega,k)-{\Pi}^{ij}(\bk,0) \right]
\right|_{\omega=0}. \label{identity}
\end{equation}
Since $\mathrm{Im}{\Pi}(\omega,k)$ and
$\mathrm{Re}{\Pi}(\omega,k)$ are odd and even in $\omega$,
respectively, we find,
\begin{equation}
\sigma^{ij}(k)=-\left.\frac{\mathrm{Im}{\Pi}^{ij}(\omega,k)}{\omega}
\right|_{\omega=0},\quad \sigma= \frac{1}{3}\lim_{k\to 0}
\sum_{i=1}^3 \sigma^{ii}(k) \; . \label{relationimpol}
\end{equation}
Eq.~(\ref{relationimpol}) is the well known result of Kubo's
linear response, and the main point of revisiting it here is the
relationship between the usual result and the time integral of the
kernel $\mathcal{G}$ which will be shown to have a simple
correspondence with the Boltzmann approach.

\subsection{Induced current}\label{indcurr}

Consider introducing an external electric field
$\boldsymbol{\mathcal{E}}=-\dot{\boldsymbol{\mathcal{A}}_T}$ with
$\boldsymbol{\mathcal{A}}_T$ switched-on adiabatically
\begin{equation}
\mathcal{A}^i(\xt)=\mathcal{A}^i(\bx)\;e^{\epsilon t} \;
\Theta(-t)\quad\mathrm{for}\; t<0 \; ,\label{Asource}
\end{equation}
in the Lagrangian density this corresponds to the shift
$\mathbf{A}_T \to \mathbf{A}_T+ \boldsymbol{\mathcal{A}}_T$. In
linear response, the induced current is given by
\begin{equation}
\langle J^i(\xt)\rangle= i\int d^4x'  \; \langle \left[J^i(\xt),
J^j(\bx',t')\right]\rangle \; \mathbf{\mathcal{A}}^j_T(\bx',t') \;
\Theta(t-t') \; .
\end{equation}
In terms of the polarization eq.~(\ref{polarization}), integrating
by parts in time and taking the spatial Fourier transform on both
sides we find
\begin{equation}
\langle J^i(k,t)\rangle = -\int_{-\infty}^t dt' \;
\mathcal{G}^{ij}(k,t-t') \; \mathcal{E}^j(k,t') \; .
\label{linresp}
\end{equation}
Since the background gauge field $\boldsymbol{\mathcal{A}}_T$ is
switched-on adiabatically, the external electric field
$\boldsymbol{\mathcal{E}}$ vanishes for $t<0$. Furthermore
assuming that the external electric field is constant in time for
$t>0$, we find
\begin{equation}
\langle J^i(k,t)\rangle = -\int_{0}^t dt' \;
\mathcal{G}^{ij}(k,t')\;\mathcal{E}^j(k) \label{linres2} \; ,
\end{equation}
where we have relabelled $t'\to t-t'$. At this stage it is
convenient to define
\begin{equation}
J^i( t) \equiv \langle J^i(k=0,t)\rangle \; ,\quad
\mathcal{G}^{ij}(t)\equiv\mathcal{G}^{ij}(k=0,t) \; ,
\quad\mathcal{E}^i\equiv\mathcal{E}^i(k=0) \; ,\label{defs}
\end{equation}
hence for a spatially constant external electric field the induced
drift current at asymptotically long times is given by
\begin{equation}
J^i( t\to \infty)= \sigma\;\mathcal{E}^i\; , \label{drift}
\end{equation}
where we have used eq.~(\ref{conductivity}). The important aspect
of eq.~(\ref{linres2}) is that
\begin{equation}\frac{d}{dt} J^i(t)=-\mathcal{G}^{ij}(t)\;\mathcal{E}^j\; ,
\label{timederi}
\end{equation}
a relation that will allow us to establish contact with the
Boltzmann equation (see Sec.~\ref{section:boltz}). In the limit
$k\to 0$, it must be that
\begin{equation}
\mathcal{G}^{ij}(t)= \delta^{ij} \; \mathcal{G}(t)\; ,
\quad\mathcal{G}(t)= \frac{1}{3} \; \mathcal{G}^{ii}(t) \; ,
\label{calGe}
\end{equation}
in terms of which one obtains
\begin{equation}
\label{sigfin} \sigma = -\int^{\infty}_0 \mathcal{G}(t) \; dt \; .
\end{equation}

The linear response analysis both for the relaxation of a gauge
mean field as well as for the induced current clearly suggests
that the important quantity to study in \emph{real time} is the
long time behavior of the kernel $\mathcal{G}^{ij}(k,t)$ which in
fact will be the link with the kinetic approach to be studied
later.

\subsection{Strategy to extract DC conductivity}

Eq.~(\ref{relationimpol}) is the usual definition of the
conductivity via Kubo's linear response~\cite{kubo,baym}. The
computation of the conductivity extracted from the long-wavelength
and low-frequency limit of the imaginary part of the polarization
features infrared divergences in perturbation theory. These
divergences are manifest as pinch singularities and are a
consequence of the propagation of intermediate states nearly
on-shell for arbitrarily long time and distance~\cite{jeon}.
Including a width for the particles via a partial resummation of
the perturbative series regulates these pinch singularities but
each power of the coupling constant is compensated by powers of
the (perturbative width) in the denominator implying the
non-perturbative nature of the transport coefficient and requiring
a resummation scheme~\cite{jeon,basagoiti1,basagoiti2}.

Instead of following this route, we propose here a different
approach. The discussion above highlights that in the real-time
framework to extract the relaxation of long-wavelength fields the
important quantity is the kernel $\mathcal{G}(k,t)$ defined by
eq.~(\ref{Goft}). In particular the DC conductivity
eq.~(\ref{conductivity}) will be finite if the asymptotic long
time behavior of $\mathcal{G}(k,t)$ is such that the total time
integral is finite for $k \to 0$. This requires that the kernel
$\mathcal{G}(k,t)$ be localized in time, namely it has short
memory in the long-wavelength limit. The important aspect of
focusing on the kernel $\mathcal{G}(k,t)$ is that it is
\emph{finite} for any \emph{finite} $k$ and $t$. Infrared
divergences can only appear in the \emph{infinite time limit}
which will result in an infinite integral in time. The finite time
argument plays the role of a regulator, in fact for finite time,
the intermediate states can only propagate during this finite
duration of time and the pinch singularities are therefore
regulated~\cite{scalar,boyanqed}. These singularities will emerge
in the long-time limit in the form of secular terms, i.e., terms
in the perturbative expansion that grow in
time~\cite{scalar,boyanqed}. This situation is similar to that in
critical phenomena and in field theory, the perturbative series
for the $n$-point functions is finite for a finite momentum
cutoff, only when the cutoff is taken to be very large does the
perturbative expansion diverges.

The dynamical renormalization group introduced in
Refs.~\cite{scalar,boyanqed} provides a resummation of the secular
terms and  leads to an improved asymptotic long-time behavior of
real time quantities (see Refs.~\cite{scalar,boyanqed} for details
and examples). Thus the strategy that we propose is the following:
(i) We first obtain the perturbative expansion of
$\mathrm{Im}\Pi^{ij}(\omega,k)/\omega$ for $\omega,k \to 0$. This
perturbative expansion will feature singular denominators of the
form $(\bk\cdot\bhp-\omega)^{-n}$, where $\bp$ is some loop
momentum to be integrated out and $n\geq 1$. The Fourier transform
in time to obtain $\mathcal{G}^{ij}(k,t)$ of these denominators
will feature terms of the form $t^{n-1}$ which are \emph{secular}
for $n=2,3, \dots$, signalling the breakdown of the perturbative
expansion at long times. Thus the kernel $\mathcal{G}(k,t)$ is
finite for any finite time and the pinch singularities are
manifest in the infinite time limit. (ii) After extracting the
leading secular terms (terms that grow the fastest in time at a
given order in perturbation theory), we will  use the dynamical
renormalization group resummation introduced in
Refs.~\cite{scalar,boyanqed} to resum these secular terms and
improve the asymptotic long-time behavior. After this real time
renormalization group resummation the kernel $\mathcal{G}(k,t)$
will have an improved and bound long-time behavior and the
conductivity can now be extracted via the time integral of the
kernel in the zero-momentum limit. This is akin to the resummation
and improvement of the perturbative series provided by the usual
renormalization group in field theory and critical phenomena.

As we will see in detail below, the resummation of the secular
terms via the dynamical renormalization group provides the bridge
between linear response and the Boltzmann equation in real time.
We now carry out this program, first by extracting the leading
secular terms in perturbation theory and then invoking the
dynamical renormalization group resummation, which will lead to
the Boltzmann equation.

\section{Perturbation theory}\label{sec:perttheory}

This section is devoted to analyzing the HTL contribution to the
photon polarization and the one-loop hard fermion self-energy,
hard fermion-soft photon vertex with HTL-resummed soft internal
propagator as well as the corresponding Ward identity in the
hydrodynamic limit.

\subsection{Polarization in the hard thermal loop
approximation}\label{oneHTL}

While there are several alternative methods to obtain the
imaginary part of the polarization, we will carry out our
perturbative study in the imaginary-time (Matsubara) formulation
of finite temperature field theory~\cite{book:lebellac}. The most
transparent manner to compute higher order corrections is to
introduce dispersive representations for propagators and
self-energies. To one-loop order, the leading order in the high
temperature limit (hard thermal loop) contribution is obtained by
using the free field fermion propagators in the one-loop
polarization. The polarization is found to be given by
\begin{equation}
\Pi^{ij}(i\nu_k,k)= e^2 T \sum_{\omega_m} \Tr \intp \; \gamma^i \;
S(P) \; \gamma^j \;  S(P+K) \; ,\label{onelup}
\end{equation}
where $K=(i\nu_k,\bk)$ and $P=(i\omega_m,\bp)$ with $\nu_k$ and
$\omega_m$ being the bosonic and fermionic Matsubara frequencies,
respectively (see Appendix~\ref{app:itf} for notations) .
Performing the Matsubara frequency sums (see
Appendix~\ref{app:itf}) and then taking the analytic continuation
$i\nu_k \to \omega+i0$, we find in the HTL limit $k,\omega\ll T$
\begin{equation}
\mathrm{Im}\Pi^{\mathrm{HTL},ij}(\omega,k)= 2 \pi e^2\omega\intp
\; \frac{dn_F(p)}{dp} \;  \hat{p}^i \;  \hat{p}^j \; [\delta(\bk
\cdot\hat{\bp}-\omega)+\delta(\bk\cdot\hat{\bp}+\omega)] \; ,
\label{imagHTL}
\end{equation}
where in the product of fermionic spectral functions only the
terms of the form $\rho_+ \; \rho_+$ and $\rho_- \; \rho_-$  [see
eq.~(\ref{ferprop})] contribute in the limit $k,\omega\to 0$.
Hence, we find in the HTL approximation
\begin{equation}
\mathcal{G}^{\mathrm{HTL},ij}(t) = 4 \;  e^2\int
\frac{d^3p}{(2\pi)^3} \frac{d n_F(p)}{dp} \;  \hat{p}^i \;
\hat{p}^j = - \delta_{ij} \; \frac{e^2 \, T^2}{9} \; .
\label{ghtl}
\end{equation}
>From the linear response relation eq.~(\ref{timederi}), we find in
the hard thermal loop approximation
\begin{equation}
\frac{d}{dt} J^i(t) = -4 e^2\int \frac{d^3p}{(2\pi)^3} \;
\hat{p}^i \; \boldsymbol{\mathcal{E}} \cdot\!\bnabla_\bp n_F(p)=
\frac{e^2 \, T^2}{9} \; \boldsymbol{\mathcal{E}}^i  \;
.\label{timederihtl}
\end{equation}
It is clear that to this order the induced drift current will grow
linearly in time. This is a result of the fact that this lowest
order approximation does not include collisions. As we will see
later, this low order result will emerge also in the Boltzmann
equation (see Sec.~\ref{section:boltz}).

\begin{figure}[t]
\includegraphics[width=2.5in,keepaspectratio=true]{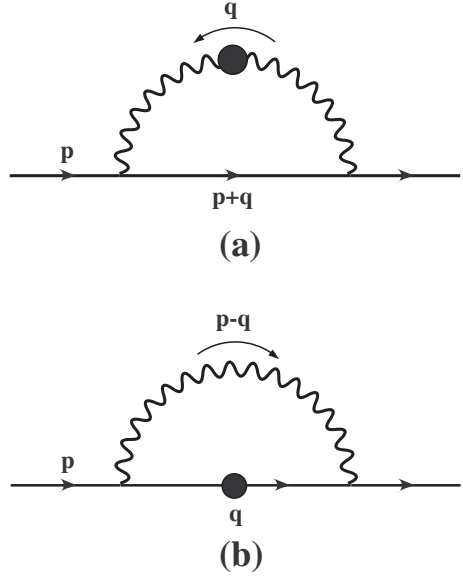}
\caption{Resummed one-loop hard fermion self-energy with (a) soft
internal photon and (b) soft internal fermion. The line with a dot
denotes the HTL-resummed propagator.}\label{fig:se}
\end{figure}

\subsection{Resummed one-loop self-energy of hard fermions}
\label{subsec:selfenergy}

For massless fermions, it proves convenient to decompose the
self-energy into positive and negative helicity components as
\begin{equation}
\Sigma(p_0,p)=\gamma_-(\bhp) \; \Sigma_+(p_0,p) +\gamma_+(\bhp) \;
\Sigma_-(p_0,p) \quad  ,\quad \Sigma_\pm(p_0,p)=
\frac{1}{2}\Tr[\gamma_\pm(\bhp) \; \Sigma(p_0,p)] \;
,\label{sigmaI}
\end{equation}
where
$\gamma_\pm(\bhp)=(\gamma^0\mp\boldsymbol{\gamma}\cdot\bhp)/2$ are
the positive and negative helicity projectors, respectively. The
resummed one-loop self-energy $\Sigma(P)$ of a hard fermion with
momentum $p\sim T$, as depicted in Fig.~\ref{fig:se},  it receives
contributions arising from soft internal photon and soft internal
fermion lines. Let us write
\begin{equation}
\Sigma(P)=\Sigma^\mathrm{sp}(P)+\Sigma^\mathrm{sf}(P) \; ,
\end{equation}
where the superscripts `sp' and `sf' denote the soft-photon and
soft-fermion contributions, respectively. Using the imaginary time
propagators given in Appendix~\ref{app:itf}, we find
\begin{eqnarray}
\Sigma^\mathrm{sp}(P)&=&e^2 \;  T \sum_{\nu_n}\intq [\gamma^0 \;
S(P+Q) \gamma^0 \;  {}^\ast\!D_L(Q)+\gamma^i \;  S(P+Q) \;
\gamma^j \;
{}^\ast\!D_T(Q) \; \mathcal{P}_T^{ij}(\bhq)] \; ,\label{Sigma}\\
\Sigma^\mathrm{sf}(P)&=&e^2 \;  T \sum_{\omega_n}\intq [\gamma^0
\; {}^\ast\!S(Q') \; \gamma^0 \;  D_L(P-Q')+\gamma^i \;
{}^\ast\!S(Q') \; \gamma^j  \; D_T(P-Q') \;
\mathcal{P}_T^{ij}(\widehat{\bp-\bq})] \; ,\label{Sigmasf}
\end{eqnarray}
where $P=(i\omega_m,\bp)$ with $p\sim T$, $Q=(i\nu_n,\bq)$,
$Q'=(i\omega_n,\bq)$ with $q\ll T$, and
$\mathcal{P}_T^{ij}(\bhq)=\delta^{ij}-\hat{q}^i\hat{q}^j$ is the
transverse projector.  For later convenience, we have assigned the
loop momentum $Q$ to the soft internal propagators. Furthermore,
we will neglect the instantaneous Coulomb interaction here and
henceforth since it does not contribute to the imaginary part.
After performing the Matsubara frequency sums, we can rewrite the
self-energy as a spectral representation
\begin{equation}
\Sigma(i\omega_m,p)=\intinfty\frac{dp_0}{\pi} \;
\frac{\Im\Sigma(p_0,p)}{p_0-i\omega_m} \; , \label{Sigmasr}
\end{equation}
 This spectral representation will be useful to carry
out the Matsubara sums when the self-energy is inserted in the
polarization. Decomposing $\Im\Sigma(p_0,p)$ onto
$\gamma_\pm(\bp)$ analogous to eq.~(\ref{sigmaI}) and using the
properties (see Appendix A for the spectral densities)
${}^\ast\!\rho_{L,T}(-q_0,q)=-{}^\ast\!\rho_{L,T}(q_0,q)$,
${}^\ast\!\rho_\pm(-q_0,q)={}^\ast\!\rho_\mp(q_0,q)$,
$n_B(-q_0)=-[1+n_B(q_0)]$ and $n_F(-q_0)=1-n_F(q_0)$, we obtain
\begin{equation}
\Im\Sigma_\pm(p_0,p)=\Im\Sigma^\mathrm{sp}_\pm(p_0,p)+
\Im\Sigma^\mathrm{sf}_\pm(p_0,p) \; ,\label{ImSigma}
\end{equation}
where
\begin{eqnarray}
\Im\Sigma^\mathrm{sp}_\pm(p_0,p) &=&
\pi e^2 \intq\intqo \; [n_B(q_0)+n_F(|\bp+\bq|)]\nn \\
&&\times\bigl\{ \big[K^{\pm}_1(\bp,\bq) \; {}^\ast\!\rho_L(q_0,q)+
K^{\pm}_2(\bp,\bq) \; {}^\ast\!\rho_T(q_0,q)\big] \;
\delta(p_0+q_0-|\bp+\bq|)\nn \\
&&+\big[K^{\mp}_1(\bp,\bq) \; {}^\ast\!\rho_L(q_0,q)+
K^{\mp}_2(\bp,\bq) \; {}^\ast\!\rho_T(q_0,q)\big] \;
\delta(p_0-q_0+|\bp+\bq|)\bigr\} \; ,
\label{SEplus2}\\
\Im\Sigma^\mathrm{sf}_\pm(p_0,p) &=&\pi e^2
\intq\intqo\frac{1+n_B(|\bp-\bq|)-n_F(q_0)}{2|\bp-\bq|}\nn\\
&&\times\big[K^{\pm}_3(\bp,\bq) \; {}^\ast\!\rho_+(q_0,q)+
K^{\mp}_3(\bp,\bq) \; {}^\ast\!\rho_-(q_0,q)\big] \;
\delta(p_0-q_0-|\bp+\bq|)\nn\\
&&+\big[K^{\mp}_3(\bp,\bq) \; {}^\ast\!\rho_+(q_0,q)+
K^{\pm}_3(\bp,\bq) \; {}^\ast\!\rho_-(q_0,q)\big] \;
\delta(p_0+q_0+|\bp+\bq|)\big\} \; , \label{SEplus2sf}
\end{eqnarray}
with
\begin{eqnarray}
K^{\pm}_1(\bp,\bq)&=&\frac{1}{2} \;
[1\pm\bhp\cdot\widehat{\bp+\bq}] \; , \; K^{\pm}_2(\bp,\bq)=
1\mp(\bhp\cdot\bhq)(\widehat{\bp+\bq}\cdot\bhq) \; ,\;
K^{\pm}_3(\bp,\bq)=1\mp(\bhp\cdot\widehat{\bp-\bq})
(\bhq\cdot\widehat{\bp-\bq}) \; .\label{Ks}
\end{eqnarray}
>From the above expressions, one finds that
$\Im\Sigma_\pm(-p_0,p)=\Im\Sigma_\mp(p_0,p)$. If
$\Im\Sigma(p_0,p)$ is finite on the particle and antiparticle mass
shells $p_0=\pm p$, respectively, then
$\Gamma(p)=\Im\Sigma_\pm(\pm p,p)$ is the damping rate for the
hard (anti)fermion~\cite{book:lebellac}.

\subsection{Anomalous damping of fermions}\label{subsec:anomadamping}

The leading contribution to the imaginary part of the hard fermion
self-energy arises from the \emph{ultrasoft} region of the
transverse photon exchange for which $q\ll e
T$~\cite{blaizot,boyanqed,scalar}. This is so because in a high
temperature QED plasma the fermionic excitation receives an
effective thermal mass $m_f=eT/\sqrt{8}$ and the instantaneous
Coulomb interaction is Debye screened by the electric (or Debye)
mass $\omega_D=eT/\sqrt{3}$, whereas the magnetic interaction is
only dynamically screened by Landau damping. Hence, we will focus
only on the soft-photon contribution $\Im\Sigma^\mathrm{sp}_\pm$
in this subsection.

In the ultrasoft region of the loop momentum $q$, $q_0\ll eT $ in
eq.~(\ref{SEplus2}), we can approximate  the distribution
$n_B(q_0)+n_F(|\bp+\bq|)\simeq T/q_0$ and the HTL spectral
function for the transverse photon~\cite{blaizot,boyanqed,scalar}
\begin{equation}
\frac{{}^\ast\!\rho_T(q_0,q)}{q_0}\simeq \frac{\delta(q_0)}{q^2}
\; ,
\end{equation}
which as pointed out in Ref.~\cite{blaizot} can be interpreted as
the exchange of a \emph{magnetostatic} transverse photon.
Furthermore, in this ultrasoft limit and for hard fermion momentum
$p \gtrsim T$ we can replace $\widehat{\bp+\bq}\simeq \bhp$. After
some algebra, we find near the particle and antiparticle mass
shells $p_0\sim \pm p$~\cite{blaizot,boyanqed}
\begin{equation}
\Im\Sigma^\mathrm{sp}_{\pm}(p_0\approx \pm p,p)= \alpha T
\int^{\omega_D}_{0}dq\int_{-1}^{1}dx \; (1-x^2) \; \delta(p_0\mp
p-qx) \; , \label{anodamp}
\end{equation}
where $x=\bhp\cdot\bhq$. The integration is straightforward and
yields the following leading order result
\begin{equation}
\Im\Sigma^\mathrm{sp}_{\pm}(p_0\approx\pm p,p)= \alpha T
\ln\left|\frac{\omega_D}{p_0 \mp p}\right|+ {\cal O}\left( |{p_0
\mp p}|^0 \right) \; .\label{logadamp}
\end{equation}
While the fermion damping rate is ill-defined, several resummation
schemes, either based on the thermal eikonal (Bloch-Nordsieck)
approximation that sums rainbow diagrams~\cite{blaizot} or via the
dynamical renormalization group~\cite{boyanqed,scalar}, lead to
the conclusion that the fermion propagator at large times decays
in real time as
\begin{equation}
\langle\Psi(\xt)\bar{\Psi}(\mathbf{0},0)\rangle \propto e^{-\alpha
\;  T \; t \; \ln(\omega_Dt)}\label{resuprop} \; ,
\end{equation}
which corresponds to an inverse relaxation time scale for the hard
fermion
\begin{equation}
\Gamma = \alpha \;  T \;  \ln(1/e) \; .\label{hardgamma}
\end{equation}
This is known as the \emph{anomalous fermion
damping}~\cite{lebedev} and is determined by the kinematic region
of ultrasoft transverse photon exchange with momentum $q\ll eT$.
The region $q\gtrsim eT$ of transverse photon exchange leads to a
subleading contribution to the damping of hard fermions of the
order $\alpha^2 T$~\cite{blaizot,basagoiti2,aarts}.

The  reasons to repeat this analysis here are twofold: (i)  to
clarify in \emph{real time} that the anomalous damping rate
eq.~(\ref{hardgamma}) \emph{does not} contribute to the
conductivity, as a consequence of a cancellation between
self-energy and vertex corrections as anticipated in
Ref.~\cite{lebedev}, and (ii) we will study the solution of the
Boltzmann equation in the \emph{relaxation time approximation}
which will also feature the anomalous time dependence.

As it will be discussed in detail below in Sec.~\ref{sec:pertsec},
there is a precise cancellation of the contribution from ultrasoft
photon momentum exchange between the self-energy and the vertex
corrections to the  polarization in perturbation theory. This
cancellation will also be manifest in the linearized Boltzmann
equation, and the analysis presented above will lead us to a
direct identification of self-energy and vertex contributions in
the linearized Boltzmann equation.

\begin{figure}[t]
\includegraphics[width=2.0in,keepaspectratio=true]{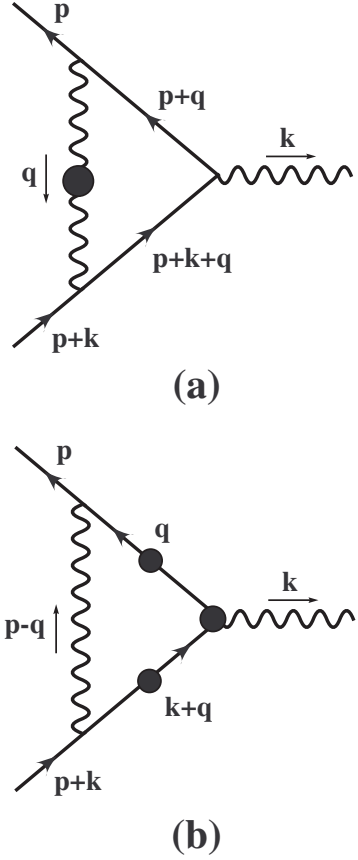}
\caption{Resummed one-loop hard fermion-soft photon vertex with
(a) soft internal photon and (b) soft internal fermion. The line
(vertex) with a dot denotes the HTL-resummed propagator
(vertex).}\label{fig:v}
\end{figure}

\subsection{Resummed one-loop fermion-photon vertex}

We now evaluate the resummed one-loop hard fermion-soft photon
vertex, which consistently with the Ward identity\cite{aarts} also
receives contributions from soft internal photon and soft internal
fermion (see Fig.~\ref{fig:v}). Writing,
\begin{equation}
\Gamma^\mu(P+K,P)=\Gamma^{\mathrm{sp},\mu}(P+K,P)+
\Gamma^{\mathrm{sf},\mu}(P+K,P) \; ,
\end{equation}
we find in the imaginary time formalism,
\begin{eqnarray}
e \; \Gamma^{\mathrm{sp},\mu}(P+K,P)&=&e^3 \;
T\sum_{\nu_n}\intq\,\big[\gamma^0 \;
S(P+Q) \; \gamma^\mu \;  S(P+K+Q) \; \gamma^0\;{}^\ast\!D_L(Q) +\nn\\
&&+\,\gamma^i \;  S(P+Q) \; \gamma^\mu \;  S(P+K+Q) \;  \gamma^j
\;
{}^\ast\!D_T(Q) \; \mathcal{P}_T^{ij}(\bhq)\big] \; ,\label{eGammamu}\\
e \; \Gamma^{\mathrm{sf},\mu}(P+K,P)&=&e^3 \;
T\sum_{\omega_n}\intq\,\big[\gamma^0 \; {}^\ast\!S(Q') \;
{}^\ast\Gamma^\mu(Q'+K,Q') \; {}^\ast\!S(Q'+K) \;
\gamma^0 \;  D_L(P-Q')+\nn\\
&&+\,\gamma^i \; {}^\ast\!S(Q') \; {}^\ast\Gamma^\mu(Q'+K,Q') \;
{}^\ast\!S(Q'+K) \; \gamma^j \; D_T(P-Q) \;
\mathcal{P}_T^{ij}(\widehat{\bp-\bq})\big] \; ,\label{eGammamusf}
\end{eqnarray}
where ${}^\ast\Gamma^\mu$ is the HTL-resummed fermion-photon
vertex, $K=(i\nu_k,\bk)$ with $k\ll T$, and the rest of the
four-momenta are the same as those defined for the self-energy.

Indeed, using the Ward identities satisfied by the respective free
and HTL-resummed quantities~\cite{book:lebellac}
\begin{eqnarray}
&&K_\mu \gamma^\mu = S^{-1}(Q)-S^{-1}(K+Q) \; , \quad K_\mu \;
{}^\ast\Gamma^\mu(Q'+K,Q')
={}^\ast\!S^{-1}(Q')-{}^\ast\!S^{-1}(Q'+K) \; ,\label{WI1}
\end{eqnarray}
and the expressions for the vertices given by
eqs.~(\ref{eGammamu}) and (\ref{eGammamusf}), one can easily show
that the resummed one-loop self-energy and vertex satisfy the Ward
identities
\begin{eqnarray}
&&K_\mu \;  \Gamma^{\mathrm{sp},\mu}(P+K,P) =
\Sigma^\mathrm{sp}(P+K)-\Sigma^\mathrm{sp}(P) \; , \quad K_\mu \;
\Gamma^{\mathrm{sf},\mu}(P+K,P) =
\Sigma^\mathrm{sf}(P+K)-\Sigma^\mathrm{sf}(P) \; ,\label{WI2}
\end{eqnarray}
 hence
\begin{equation}
K_\mu \;  \Gamma^\mu(P+K,P) =\Sigma(P+K)-\Sigma(P).\label{WI3}
\end{equation}
We note that these Ward identities for the resummed one-loop
self-energy and vertex  are exact in the kinematic region under
consideration $p\sim T$ and $k\ll T$.

Let us first concentrate on the soft-photon contribution
$\Gamma^{\mathrm{sp},\mu}$. The sum over the Matsubara frequencies
can be done straightforwardly. Using the identity
\begin{equation}
\frac{1}{ab}=\mathrm{PV}\left(\frac{1}{b-a}\right)
\left(\frac{1}{a}-\frac{1}{b}\right), \label{oneoverAB}
\end{equation}
where the principal value (PV) prescription is necessary  to
define the limit $b-a\to 0$, we find that,
\begin{eqnarray}
\Gamma^{\mathrm{sp},\mu}(P+K,P)&=&e^2\intq\intqo\intpo\intso
\big[\gamma^i \; \rho_F(p_0,\bp+\bq) \;
\gamma^\mu \; \rho_F(s_0,\bp+\bk+\bq) \; \gamma^j \; \nn\\
&&\times\,{}^\ast\!\rho_T(q_0,q) \; \mathcal{P}_T^{ij}(\bhq)+
\gamma^0 \; \rho_F(p_0,\bp+\bq) \; \gamma^\mu
\;\rho_F(s_0,\bp+\bk+\bq)\gamma^0 \; {}^\ast\!\rho_L(q_0,q)\big]\nn\\
&&\times\,\frac{1}{s_0-p_0-i\nu_k}
\left[\frac{n_B(q_0)+n_F(p_0)}{p_0-q_0-i\omega_m}-
\frac{n_B(q_0)+n_F(s_0)}{s_0-q_0-i\omega_m-i\nu_k}\right] \;
,\label{Gammanu}
\end{eqnarray}
where the denominator $1/(s_0-p_0-i\nu_k)$ in the above expression
should be understood with the principal value prescription as in
eq.~(\ref{oneoverAB}).

The kinematic region of interest for the polarization and the
conductivity corresponds to hard external fermion $p \sim T$ and
soft external photon $k \ll T$. Furthermore after the analytic
continuation of the external photon frequency $i\nu_k\to\omega+i0$
we need $\omega \ll T$, and eventually we will take the
long-wavelength, low-frequency limit $\omega,k\to 0$.

The form of the free fermion spectral function $\rho_F$ suggests
that the above result can be written as a sum of the products
$\rho_+ \; \rho_+$, $\rho_-  \; \rho_-$, $\rho_+  \; \rho_-$ and
$\rho_-
 \; \rho_+$, where $\rho_\pm$ are the respective spectral functions
for free fermion and antifermion [see eq.~(\ref{ferprop})]. The
integrals over the dispersive variables $p_0$ and $s_0$ can be
done trivially using the spectral functions for free fermions. For
hard external fermion $p\sim T$ and soft external photon $k\ll T$
we can make the following kinematic approximations
\begin{eqnarray}
|\bp+\bk+\bq|-|\bp+\bq|&\simeq&\widehat{\bp+\bq}\cdot\bk \sim k \;
, \quad |\bp+\bk+\bq|+|\bp+\bq|\simeq2|\bp+\bq| \sim T \; .
\end{eqnarray}
After the analytic continuation of the external photon frequency
$i\nu_k\to\omega+i0$ with $\omega\sim k\ll T$, we find the
products $\rho_\pm \rho_\pm$ lead to denominators of the form
\begin{equation}
\frac{1}{\widehat{\bp+\bq}\cdot\bk\mp\omega} \; ,\label{hydroploe}
\end{equation}
whereas the products $\rho_\pm \rho_\mp$ lead to denominators of
the form
\begin{equation}
\frac{1}{2|\bp+\bq|\mp\omega}\sim \frac{1}{T} \; .
\end{equation}
Hence in the long-wavelength, low-frequency (or hydrodynamic)
limit of the external photon $k$, $\omega\to 0$, the products
$\rho_\pm \rho_\pm$ lead to singular denominators (pinch
singularities) which furnish the leading contributions, whereas
products of the form $\rho_\pm \rho_\mp$ lead to contributions
that are suppressed by inverse powers of the temperature. We note
that in our notation the pole-like singularities of the form given
by eq.~(\ref{hydroploe}) must be understood with a principal value
prescription as discussed above. Since these poles emerge in the
hydrodynamic limit and play an important role in our discussion
below, hereafter we will refer to them as  hydrodynamic poles.

Keeping only terms with the products $\rho_\pm \rho_\pm$ that
eventually lead to hydrodynamic poles, we find after some algebra
that $\Gamma^{\mathrm{sp},\mu}(P+K,P)$  in the hydrodynamic limit
can be written in the spectral representation as
\begin{equation}
\Gamma^{\mathrm{sp},\mu}(P+K,P)=\intso\left[
\frac{\digamma^\mu(s_0,\bp;i\nu_k,\bk)}{s_0-i\omega_m}
-\frac{\digamma^\mu(s_0,\bp+\bk;i\nu_k,\bk)}{s_0-i\omega_m-i\nu_k}\right]\;
, \label{GammamuF}
\end{equation}
where
\begin{equation}
\digamma^\mu(s_0,\bp;k_0,\bk)=\gamma_-(\bhp) \;
\digamma_+^\mu(s_0,\bp;k_0,\bk) +\gamma_+(\bhp) \;
\digamma_-^\mu(s_0,\bp;k_0,\bk) \; .\label{F}
\end{equation}
The components $\digamma_\pm^\mu$ are given by
\begin{eqnarray}
\digamma_\pm^\mu(s_0,\bp;i\nu_k,\bk)&=&e^2
\intq\int^{+\infty}_{-\infty}dq_0\,
\bigg\{\frac{L^\mu_+(\widehat{\bp+\bq})}{\widehat{\bp+\bq}\cdot\bk-i\nu_k}
\big[K^{\pm}_1(\bp,\bq) \; {}^\ast\!\rho_L(q_0,q)+
K^{\pm}_2(\bp,\bq) \; {}^\ast\!\rho_T(q_0,q)\big]\nn\\
&&\times\,\delta(|\bp+\bq|-q_0-s_0)
-\frac{L^\mu_-(\widehat{\bp+\bq})}{\widehat{\bp+\bq}\cdot\bk+i\nu_k}
\big[K^{\mp}_1(\bp,\bq) \; {}^\ast\!\rho_L(q_0,q)+
K^{\mp}_2(\bp,\bq)  \;   {}^\ast\!\rho_T(q_0,q)\big]\nn\\
&&\times\,\delta(|\bp+\bq|-q_0+s_0)\bigg\}\;
[n_B(q_0)+n_F(|\bp+\bq|)] \; , \label{Fpm}
\end{eqnarray}
with $L_\pm(\bhp)=(1,\pm\bhp)$ being lightlike four-vectors.

The soft-fermion contribution to the vertex has been studied in
detail in Ref.~\cite{aarts}. The conclusion of the detailed study
of the vertex is that whereas the soft-fermion contribution is
needed to fulfill the Ward identity, its contribution to the
polarization is subleading~\cite{aarts}. Anticipating that in
agreement with this conclusion that the soft-fermion contribution
to the vertex will not contribute at leading logarithmic order, we
will neglect this contribution for the moment and postpone a
detailed discussion until we study the vertex corrections to the
resummed two-loop photon polarization (see
Sec.~\ref{vcontribution}).

\subsection{Resummed one-loop Ward identity in the hydrodynamic
limit}\label{sec:WardId}

In the program that we advocate here, namely the implementation of
a resummation of the secular terms via the dynamical
renormalization group in real time, we must confirm that the main
ingredients in such a resummation fulfill the Ward identity, which
guarantees that the result of the resummation will be gauge
invariant.

While eqs.~(\ref{WI1})-(\ref{WI2}) and (\ref{WI3}) assert the
fulfillment of the Ward identities, our focus below will be to
extract the secular terms that are associated with hydrodynamic
poles of the form $(\bk\cdot\bhp-\omega)^{-n}$ in the limit $\bk ;
\omega \rightarrow 0$. These poles are the manifestation of the
pinch singularities and in real time they lead to secular terms in
the perturbative expansion as discussed above. To lowest order
with HTL resummed propagators these poles correspond to  $ n =1,2
$. As it will become clear in the detailed discussion below, the
dynamical renormalization group will provide a resummation of the
leading order secular terms. Thus the building blocks of the
resummation program are precisely these singular terms arising
from the insertions of one loop self-energy and vertex with HTL
resummed propagators.

Thus we must confirm that the Ward identity between the resummed
one-loop self-energy and vertex is fulfilled for these singular
contributions. Once the Ward identity of these building blocks is
confirmed, the dynamical renormalization group program that leads
to a resummation of these leading order singular contribution is
gauge invariant.

As per the discussion above we will neglect the soft-fermion
contribution to the vertex in the analysis that follows since its
contribution to the polarization is subleading~\cite{aarts} (see
below). The spectral representation of $\Gamma^{\mathrm{sp},\mu}$
given by eqs.~(\ref{GammamuF})--(\ref{Fpm}) is particularly useful
to establish the Ward identity between the resummed one-loop
self-energy and vertex in the hydrodynamic limit. With the
spectral representation for the self-energy eq.~(\ref{Sigmasr}) in
terms of $\Im\Sigma^\mathrm{sp}$ given by eq.~(\ref{SEplus2}) and
the above spectral representation for the vertex
$\Gamma^{\mathrm{sp},\mu}$, it is straightforward to check that
the resummed one-loop Ward identity \begin{equation}
K_\mu\Gamma^{\mathrm{sp},\mu}(P+K,P)=\intinfty
\frac{ds_0}{\pi}\bigg[
\frac{\Im\Sigma^\mathrm{sp}(s_0,\bp+\bk)}{s_0-i\omega_m-i\nu_k}-
\frac{\Im\Sigma^\mathrm{sp}(s_0,\bp)}{s_0-i\omega_m}\bigg]
=\Sigma^\mathrm{sp}(P+K)-\Sigma^\mathrm{sp}(P) \end{equation} is
fulfilled in the hydrodynamic limit. The next step in the program
is to insert the resummed one-loop self-energy and vertex in the
transverse photon polarization, the fulfillment of the Ward
identity between the self-energy and the vertex will ensure
current conservation and gauge invariance.

\section{Resummed two-loop transverse photon
polarization}\label{sec:twopoop}

\begin{figure}[t]
\includegraphics[width=2.5in,keepaspectratio=true]{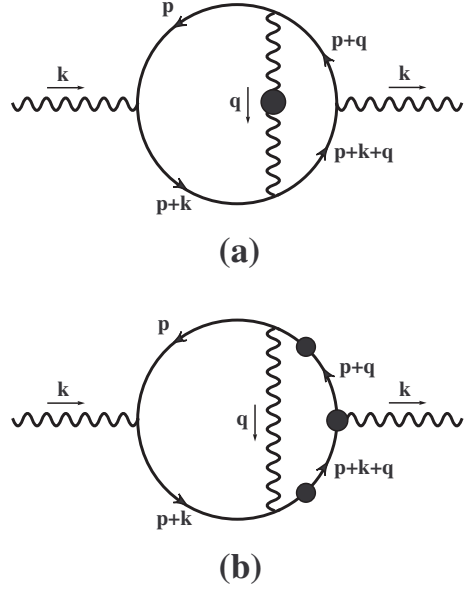}
\caption{Resummed two-loop photon polarization with the
fermion-photon vertex correction.}\label{fig:pv}
\end{figure}

At two-loop order the calculation of the photon self-energy
involves two kinds of topologically different diagrams, namely the
self-energy and vertex correction diagrams as depicted in
Figs.~\ref{fig:pse} and \ref{fig:pv}, respectively. Similar
diagrams have been studied for a hot \emph{QCD plasma} within the
context of photon or dilepton production~\cite{diagrams} and more
recently in Ref.~\cite{gale} but in a very different kinematic
region. The color algebra, which introduces important differences,
and the very different kinematic region of interest for dilepton
production,  which is not the relevant for transport phenomena,
make these previous calculations not useful for the purpose of
studying transport coefficients in the hot QED plasma.

\subsection{The self-energy contribution}

\begin{figure}[t]
\includegraphics[width=2.25in,keepaspectratio=true]{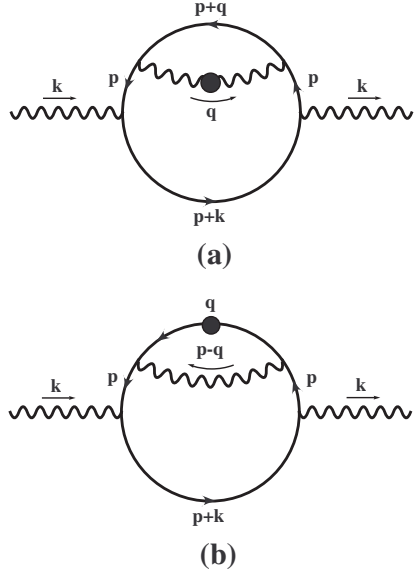}
\caption{Resummed two-loop photon polarization with the fermion
self-energy correction.}\label{fig:pse}
\end{figure}

The two-loop diagrams for the fermion self-energy contribution to
the transverse polarization are depicted in Fig.~\ref{fig:pse}. In
the imaginary-time formalism the self-energy correction to the
polarization is given by \begin{equation}
\Pi^{\mathrm{SE},ij}(K)=e^2T\sum_{\omega_m} \Tr\intp \; [\gamma^i
\; S(P) \; \Sigma(P) \;  S(P) \; \gamma^j \;  S(P+K) +\gamma^i \;
S(P) \; \gamma^j  \; S(P+K) \;  \Sigma(P+K)  \;  S(P+K)] \; ,
\end{equation} where $\Sigma(P)$ is one-loop resummed fermion
self-energy in the spectral representation given by
eq.~(\ref{Sigmasr}) and for later convenience we have written down
explicitly the self-energy insertion for each fermion line. The
Matsubara sum over the internal fermion frequency can be performed
easily following the method illustrated in the previous section.
After using the identity eq.~(\ref{oneoverAB}) we obtain
\begin{eqnarray}
\Pi^{\mathrm{SE},ij}(K)&=&\frac{e^2}{\pi}\,\Tr \intp\intpo\intqo
\intso\intto\bigg\{\gamma^i \; \rho_F(p_0,\bp) \;
\Im\Sigma(q_0,p) \; \rho_F(s_0,\bp) \; \gamma^j\nn\\
&&\times\,\rho_F(t_0,\bp+\bk)
\bigg[\frac{n_F(t_0)-n_F(p_0)}{(p_0-s_0)
(p_0-q_0)(t_0-p_0-i\nu_k)}+\frac{n_F(t_0)-n_F(s_0)}{(s_0-p_0)
(s_0-q_0)(t_0-s_0-i\nu_k)}+ \nn\\
&&+\,\frac{n_F(t_0)-n_F(q_0)}{(q_0-p_0)
(q_0-s_0)(t_0-q_0-i\nu_k)}\bigg]+\gamma^i \; \rho_F(t_0,\bp) \;
\gamma^j \; \rho_F(p_0,\bp+\bk) \; \Im\Sigma(q_0,p+k)\nn\\
&&\times\,\rho_F(s_0,\bp+\bk)\bigg[\frac{n_F(t_0)-n_F(p_0)}{(p_0-s_0)
(p_0-q_0)(t_0-p_0+i\nu_k)}+\frac{n_F(t_0)-n_F(s_0)}{(s_0-p_0)
(s_0-q_0)(t_0-s_0+i\nu_k)} + \nn\\
&&+\,\frac{n_F(t_0)-n_F(q_0)}{(q_0-p_0)
(q_0-s_0)(t_0-q_0+i\nu_k)}\bigg]\bigg\} \; ,
\end{eqnarray}
where $\Im\Sigma(q_0,p)$ is given by eq.~(\ref{ImSigma}) and all
the frequency denominators are understood with the principal value
prescription.

The imaginary part $\Im\Pi^{\mathrm{SE},ij}(\omega,k)$ is obtained
through the analytical continuation $i\nu_k\to\omega+i0$ and can
be interpreted as cutting the photon self-energy diagrams in
Fig.~\ref{fig:pse} in all possible manners. However, just as in
the discussion of the vertex above, terms with products of the
form $\rho_\pm \rho_\pm \rho_\pm$ give the leading contributions
that feature hydrodynamic poles (or pinch denominators) in the
hydrodynamic limit, while terms that involve mixed products of
$\rho_\pm$ will be subleading by powers of $1/T$. Therefore, only
the products with the same particle or antiparticle components in
\emph{all} fermionic lines will yield the leading contribution.

Focusing only on the leading contributions arising from terms
where the fermion spectral functions are either all particle
($\rho_+ $) or all antiparticle ($\rho_-$) [see
eq.~(\ref{ferprop})], we find
\begin{eqnarray}
\Im\Pi^{\mathrm{SE},ij}(\omega,k)&=& \pi e^2\,\mathrm{PV}
\intp\intpo\intqo\intso\intto \;
\frac{n_F(t_0)-n_F(q_0)}{(q_0-p_0)
(q_0-s_0)}\nn\\
&&\times\,\Tr\left[\gamma^i \; \rho_F(p_0,\bp) \; \Im\Sigma(q_0,p)
\; \rho_F(s_0,\bp) \; \gamma^j \;
\rho_F(t_0,\bp+\bk) \; \delta(t_0-q_0-\omega)\right.\nn\\
&&\left. -\,\gamma^i \; \rho_F(t_0,\bp) \; \gamma^j \;
\rho_F(p_0,\bp+\bk) \; \Im\Sigma(q_0,p+k) \; \rho_F(s_0,\bp+\bk)
\; \delta(t_0-q_0+\omega)\right] \; ,\label{ImPiSEij}
\end{eqnarray}
where the fermion spectral functions are either all particle or
antiparticle contributions. Since the dispersive variable $q_0$ is
constrained by the delta functions $\delta(t_0-q_0\mp\omega)$, the
leading terms that feature hydrodynamic poles are obtained from
cutting the polarization diagrams in Fig.~\ref{fig:pse} through
the fermion self-energy and hence the internal soft photon line.
The other possible cuts do not lead to hydrodynamic poles, hence
no secular terms arise from the other cuts.

\begin{figure}[t]
\includegraphics[width=2.25in,keepaspectratio=true]{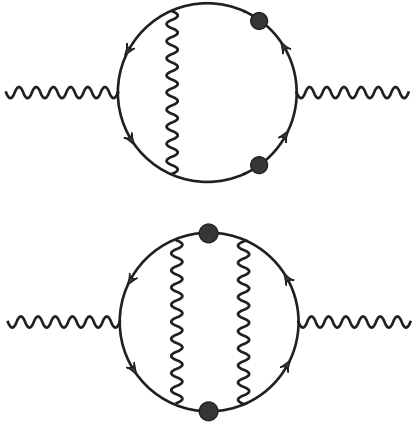}
\caption{Diagrams corresponding to the resummed two-loop photon
polarization in Fig.~\ref{fig:pv}(b).}\label{fig:pvsf}
\end{figure}

Performing the integrals over $p_0$, $q_0$, $s_0$, and $t_0$ by
using the delta functions, we find the leading terms in $k$,
$\omega\to 0$ limit to be given by
\begin{equation}
\Im\Pi^{\mathrm{SE},ij}(\omega,k)=2\pi \;  e^2 \; \omega \; \PV
\intp\frac{dn_F(p)}{dp} \; \hat{p}^i \;  \hat{p}^j  \;
\bigg[\frac{\Im\Sigma_+(|\bp+\bk|-\omega,p)+
\Im\Sigma_+(p+\omega,|\bp+\bk|)}{(\bk\cdot\bhp-\omega)^2}+(\omega\to
-\omega)\bigg] \; , \label{ImPiSET1}
\end{equation}
where we have used the property
$\Im\Sigma_-(-\omega,p)=\Im\Sigma_+(\omega,p)$. The above
expression shows clearly that in $k$, $\omega\to 0$ limit the
leading contributions to $\Im\Pi^{\mathrm{SE},ij}(\omega,k)$ arise
from terms that feature hydrodynamic poles of the form
\begin{equation}
\frac{1}{(\bk\cdot\bhp\mp\omega)^2}.
\end{equation}
\noindent which will lead to linear secular terms in time.  For
$k,\omega \ll p \sim T$ we see that the imaginary parts of the
fermion self-energy in eq.~(\ref{ImPiSET1}) above are nearly on
the mass shell of fermions or antifermions, the off-shellness
being of order $|\omega-\bk\cdot\bhp|$. Upon inserting
$\Im\Sigma_+$ given by  eq.~\eqref{SEplus2} into
eq.~\eqref{ImPiSET1} and noticing that for $q \ll T$ the second
delta functions in eq.~\eqref{SEplus2} cannot be satisfied near
the fermion mass shell, we obtain
\begin{eqnarray}
&&\Im\Pi^{\mathrm{SE},ij}(\omega,k)= 2\pi^2 e^4 \omega\,\PV
\intp\frac{dn_F(p)}{dp} \; \hat{p}^i \;
\hat{p}^j \;  \intq \intqo \;  \bigg\{[n_B(q_0)+n_F(|\bp+\bq|)]\nn\\
&&\times\big[K^+_1(\bp,\bq) \; {}^\ast\!\rho_L(q_0,q)+
K^+_2(\bp,\bq) \; {}^\ast\!\rho_T(q_0,q)\big]+\frac{1}{2|\bp-\bq|}
[1+n_B(|\bp-\bq|)-n_F(q_0)] \\
&&\times\,\big[K^{+}_3(\bp,\bq) \; {}^\ast\!\rho_+(q_0,q)+
K^{-}_3(\bp,\bq) \; {}^\ast\!\rho_-(q_0,q)\big]\bigg\}
\bigg[\frac{\delta(\bhp\cdot\bq-q_0-\bk\cdot\bhp+\omega)+
\delta(\bhp\cdot\bq-q_0+\bk\cdot\bhp-\omega)}{(\bk\cdot\bhp-\omega)^2}
+(\omega\to -\omega)\bigg] \; .\label{ImPiSET}\nn
\end{eqnarray}
As will be shown below, the hydrodynamic poles in
$\Im\Pi^{\mathrm{SE},ij}(\omega,k)$ correspond to secular terms in
time in the function $\mathcal{G}^{ij}(t)$ and signal the need for
resummation in real time.

\subsection{Anomalous secular term from the self-energy}\label{sec:secterms}

At this stage we can begin establishing a connection with the
program of resummation in real time by analyzing the leading
secular terms arising in the perturbative expansion. As analyzed
in detail in Sec.~\ref{subsec:anomadamping}, the \emph{leading}
contribution to the imaginary part of the self-energy near the
fermion (or antifermion) mass shell is determined by the exchange
of an ultrasoft transverse photon with momentum $q\ll eT$. Keeping
only this leading contribution for the moment, we find that for $k
\ll p$ [see eq.~(\ref{logadamp})]
\begin{equation}
\Im\Sigma^\mathrm{sp}_+(|\bp+\bk|-\omega,p)=
\Im\Sigma^\mathrm{sp}_+(p+\omega,|\bp+\bk|)=-\alpha \; T \;
\ln\left|\frac{\omega-\bk\cdot\bhp}{\omega_D}\right| \; .
\label{leadingsigI}
\end{equation}
Introducing this leading estimate for the imaginary part of the
fermion (antifermion) self-energy into eq.~(\ref{ImPiSET1}), we
obtain
\begin{equation}
\Im\Pi^{\mathrm{SE},ij}(\omega,k) = -4e^2 \alpha T
\omega\,\mathrm{PV}\intp\frac{dn_F(p)}{dp}
\left[\frac{\hat{p}^i\hat{p}^j}{(\omega-\bk \cdot
\bhp)^2}\ln\left|\frac{\omega-\bk \cdot \bhp}{\omega_D}
\right|+(\omega \to -\omega)\right] \; .\label{imagSElead}
\end{equation}
We now compute the function $\mathcal{G}^{ij}(k,t)$ by performing
the Fourier transform in $\omega$ as per eq.~(\ref{Goft}). Using
the properties of the principal value, we find
\begin{equation}
\mathcal{G}^{\mathrm{SE},ij}(t) = -2 \alpha \,
\mathcal{G}^{\mathrm{HTL},ij}
\int^{+\infty}_{-\infty}\frac{1-\cos(yt)}{\pi y^2} \;
\ln\left|\frac{y}{\omega_D} \right|\; ,\label{GSE}
\end{equation}
where $\mathcal{G}^{\mathrm{HTL},ij}(t)$ is given by
eq.~(\ref{ghtl}). With the help of the results established in
Refs.~\cite{boyanqed,scalar}, we find the asymptotic long time
limit to be given by
\begin{equation}
\mathcal{G}^{\mathrm{SE},ij}(t) = \mathcal{G}^{\mathrm{HTL},ij} \;
\left[-2 \alpha  T t
  \ln(\omega_D t) + \mathrm{non~secular~terms}\right] \; . \label{asyGSE}
\end{equation}
Therefore combining the one-loop (HTL) result with the leading
contribution from ultrasoft transverse photon exchange in the
resummed self-energy correction to the two-loop polarization, we
find
\begin{equation}
\mathcal{G}^{ij}(t)=\mathcal{G}^{\mathrm{HTL},ij} \;
\left[1-2\alpha T t \ln(\omega_D t) +
\mathrm{non~secular~terms}\right] \; . \label{sum}
\end{equation}
Let us \emph{assume} for a moment that the exponentiation of the
leading secular term in the form
\begin{equation}
\mathcal{G}^{ij}(t) = \mathcal{G}^{\mathrm{HTL},ij}(t) \;
e^{-2\alpha T t \ln(\omega_D t)} = - \delta_{ij} \; \frac{e^2 \,
T^2}{9} \; e^{-2 \alpha T  t \ln(\omega_D t)} \; ,\label{Goftres}
\end{equation}
and explore the consequences of this result. [We used here
eq.~(\ref{ghtl})]. Such exponentiation \emph{would} be justified
considering only the self-energy insertion  on both fermionic
lines in the loop and on the basis of the resummed form of the
propagator eq.~(\ref{resuprop})~\cite{blaizot,boyanqed} since the
function $\mathcal{G}(k,t)$ behaves as the square of the
propagator. Furthermore, as it has become customary in the
literature, let us \emph{approximate}
\begin{equation}
e^{-2 \alpha T  t \ln(\omega_D \; t)}\approx e^{-2 \Gamma t} \; ,
\label{appxexpo}
\end{equation}
with the \emph{anomalous damping rate} given by
eq.~(\ref{hardgamma}). The time integral of $\mathcal{G}^{ij}(t)$
[see eq.~(\ref{sigfin})] to obtain the conductivity can now be
done in closed form and we find
\begin{equation}
\sigma \approx \frac{e^2  T^2}{18 \, \Gamma} = \frac{2\pi}{9}
\frac{T}{\ln(1/e)} +\mathrm{subleading} \; .\label{condSE}
\end{equation}
This discussion highlights several important points:
\begin{itemize}
\item[(i)]{In the limit $k$, $\omega \to 0$ the hydrodynamic poles
of the form $1/(\omega \mp\bk\cdot\bhp)^2$, which are a
consequence of the pinch terms~\cite{jeon}, result in secular
perturbations for $\mathcal{G}(k, t)$, i.e., terms that grow in
time. The relation between pinch singularities and secular terms
in perturbative theory and their resummation via the dynamical
renormalization group has been previously discussed in
Refs.~\cite{boyanqed,scalar}.

>From eq.~(\ref{GSE}) and the results obtained in
Refs.~\cite{boyanqed,scalar} we see that the hydrodynamic pole
$1/(\omega -\bk\cdot \bhp)^2$ leads to a secular term linear in
$t$, while the threshold singularity $\ln|\omega -\bk\cdot \bhp|$
leads to the $\ln(\omega_D \,t)$ enhancement of the secular term.
Thus, the extra logarithmic contribution to the secular term
originates in the ultrasoft momentum region $q\ll eT$ of the
exchanged transverse photon, which is only dynamically screened by
Landau damping. Without this infrared, logarithmic enhancement the
hydrodynamic pole would lead to a secular term linear in time.
This observation will be important when we combine the
contributions from the self-energy and vertex.}

\item[(ii)]{\emph{If} the imaginary part of the fermion
self-energy on the mass shell were a finite constant $\Gamma\ll T$
a Dyson resummation of self-energy insertions would lead to a
Breit-Wigner form for  the fermion and antifermion spectral
functions near the mass shell, namely
\begin{equation}
\rho_\pm(p_0,p) = \frac{1}{\pi} \frac{\Gamma}{(p_0 \mp
p)^2+\Gamma^2} \; .
\end{equation}
Using these spectral functions to compute the imaginary part of
the polarization to \emph{one-loop} order, keeping only the
products $\rho_\pm \; \rho_\pm$ and taking the limits $k$, $\omega
\to 0$, it is straightforward to find that the conductivity is
independent of these limits and given by
\begin{equation}
\sigma = \frac{e^2 T^2}{18 \, \Gamma} \; .\label{con}
\end{equation}

This simple exercise shows that the resummation of the secular
terms in eq.~(\ref{Goftres}) by taking the coefficient of $t$ to
be a constant $2\Gamma$ gives the correct answer for the
conductivity in the case where the damping rate is a finite
constant and the spectral function near the quasiparticle poles is
of the Breit-Wigner form, providing a reassuring check in a
simpler case.} \item[(iii)] {In order to avoid confusion, we want
to stress that the eqs.~(\ref{condSE})-(\ref{con}) are a result of
the self-energy correction \emph{only}. As will be discussed in
detail below, the vertex correction cancels the ultrasoft photon
exchange contribution from the self-energy and as a result the
eqs.~(\ref{condSE})-(\ref{con}) \emph{do not apply} to the hot QED
plasma. The correct expression is given by eq.~(\ref{sigmaval}).
The main point of the derivation of
eqs.~(\ref{condSE})-(\ref{con}) is the following: (a) to
illustrate what would be the result if only the self-energy
correction is taken into account but without the vertex
correction, (b) as we will show below that this approximation is
equivalent to the relaxation time approximation in the Boltzmann
equation.}
\end{itemize}

As it will be seen in detail later when we implement the dynamical
renormalization group the resummed form of $\mathcal{G}(t)$ with
only self-energy correction is much more subtle than the simple
exponentiation assumed above in eq.~(\ref{Goftres}). While the
result obtained above is correct for a constant damping rate, the
anomalous logarithmic time dependence will prevent the existence
of a drift current at long times leading to a vanishing
conductivity from \emph{only} self-energy corrections from
ultrasoft photon exchange. We will discuss these issues in more
detail in Sec.~\ref{sec:reltimeappx}, where we introduce the
dynamical renormalization group equation in the relaxation time
approximation.

While this discussion has highlighted these important points,
keeping only the self-energy corrections is not consistent with
the Ward identities and the vertex correction must be included. As
advanced in Ref.~\cite{lebedev} and discussed explicitly below in
Sec.~\ref{sec:pertsec}, the contribution from the anomalous
damping to the conductivity is in fact \emph{cancelled} by the
vertex correction. This cancellation, which will also be made
manifest in the Boltzmann equation below is the reason that the
conductivity is determined not by the quasiparticle relaxation
time scale but by the transport time scale.

\subsection{The vertex contribution}\label{vcontribution}
The vertex corrections to the photon polarization are displayed in
Fig.~\ref{fig:pv}. We focus first on the vertex correction  with
hard thermal loop resummed photon exchange,  displayed in
Fig.~\ref{fig:pv}(a).

In the imaginary-time formulation the contribution to the
polarization from this diagram is given by
\begin{equation}
\Pi_a^{\mathrm{V},ij}(K)=e^2T\sum_{\omega_m}\Tr\intp\; \gamma^i \;
S(P) \;  \Gamma^j(P+K,P) \;  S(P+K) \; ,
\end{equation}
where $\Gamma^j(P+K,P)$ is one-loop resummed fermion-photon vertex
in the spectral representation given by eq.~\eqref{GammamuF}. The
Matsubara sum over the internal fermion frequency is facilitated
by the dispersive representation of the vertex $\Gamma^j(P+K,P)$
and of the fermion propagators. After using the summation formula
eq.~(\ref{matsums}) and the identity eq.~(\ref{oneoverAB})
repeatedly to combine factors, we obtain
\begin{eqnarray}\label{polvert}
\Pi_a^{\mathrm{V},ij}(K)&=&-e^2\,\Tr\intp\intpo\intso\intto \;
\gamma^i \, \rho_F(p_0,\bp)\nn\\
&&\times\,\bigg\{\bigg[\frac{n_F(t_0)-n_F(s_0)}{t_0-s_0-i\nu_k}+
\frac{n_F(p_0)-n_F(t_0)}{t_0-p_0-i\nu_k}\bigg]
\frac{\digamma^j(s_0,\bp;i\nu_k,\bk)}{s_0-p_0}+\\
&&+\,\bigg[\frac{n_F(t_0)-n_F(p_0)}{t_0-p_0-i\nu_k}+
\frac{n_F(p_0)-n_F(s_0)}{s_0-p_0-i\nu_k}\bigg]
\frac{\digamma^j(s_0,\bp+\bk;i\nu_k,\bk)}{s_0-t_0}\bigg\}\;
\rho_F(t_0,\bp+\bk) \; , \nonumber
\end{eqnarray}
where $\digamma^j(s_0,\bp;i\nu_k,\bk)$ is given by eqs.~(\ref{F})
and (\ref{Fpm}) and all the frequency denominators \emph{without}
$i\nu_k$  should be understood with the principal value
prescription.

Upon the analytic continuation $i\nu_k\to\omega+i0$, the imaginary
parts of the polarization arise from the following contributions:
(i) The imaginary part of the denominator (delta functions)
multiplies the real part of the  function $\digamma^j$. (ii) The
real part of the denominator multiplies the imaginary part of the
function $\digamma^j$. We begin by analyzing the first
contribution, namely those arising from the imaginary part of the
denominators in the square brackets in eq.~(\ref{polvert}):
\begin{itemize}
\item[(i)]{Terms proportional to $\delta(t_0-s_0-\omega)$ and
$\delta(s_0-p_0-\omega)$. These terms arise from cutting the
polarization diagram in Fig.~\ref{fig:pv}(a) through the internal
soft photon line because the dispersive variable $s_0$ is
constrained. For $\omega \ll T$ the delta functions will have
support only for the products $\rho_\pm \; \rho_\pm$. For such
products of fermion spectral functions $\rho_\pm \rho_\pm$ the
denominator $s_0-p_0=t_0-p_0-\omega$ and $s_0-t_0=p_0-t_0+\omega$
lead to the hydrodynamic pole of the form
$1/(\omega\mp\bk\cdot\bhp)$ since $p_0=\pm p$ and $t_0=\pm
|\bp+\bk|$ for these products of fermion (or antifermion) spectral
functions. Thus these discontinuities will lead to secular terms.}

\item[(ii)]{Terms proportional to $\delta(t_0-p_0-\omega)$. These
term arise from cutting the polarization diagram in
Fig.~\ref{fig:pv}(a) \emph{only} through the internal fermion
lines because the dispersive variable $s_0$ is not constrained.
Again the delta function will have support only for the products
$\rho_\pm \; \rho_\pm$ in which case the contribution of this cut
is of the form
\begin{equation}
\omega \frac{dn_F(p)}{dp}\bigg[
\frac{\mathrm{Re}\digamma^j(s_0,\bp;\omega,\bk)}{s_0 \mp p}-
\frac{\mathrm{Re}\digamma^j(s_0,\bp+\bk;\omega,\bk)}{s_0
\mp|\bp+\bk|}\bigg] \; , \label{cut2}
\end{equation}
where we have integrated over dispersive variables $p_0$ and
$t_0$. Since the only hydrodynamic pole appears in the function
$F$ and is a single pole, these discontinuities will not lead to
secular terms. Furthermore, upon relabelling $\bp \to \bp+\bk$ in
the first term in the momentum integral, we find that these
contributions cancel each other.}
\end{itemize}
We now analyze the second contribution, namely those arising from
of the imaginary part of the function $\digamma^j$. The imaginary
part of the function $\digamma^j$ is obtained by replacing
$\widehat{\bp+\bq}\cdot\bk \mp k_0 \to \pm i\pi
\delta(\widehat{\bp+\bq}\cdot\bk \mp \omega)$. The terms that can
give rise to hydrodynamic poles of the form
$1/(\bk\cdot\bhp\mp\omega)$ are the second and third terms in
eq.~(\ref{polvert}) when the product of fermion spectral functions
is either $\rho_+ \; \rho_+$ or $\rho_- \;  \rho_-$. Combining
these two terms we obtain contributions of the form
\begin{equation}\frac{d n_F(p)}{d p}
\frac{\bk\cdot\bhp}{\omega\mp\bk\cdot\bhp}\left[
\frac{\mathrm{Im}\digamma^j(s_0,\bp;\omega,\bk)}{s_0-p}
-\frac{\mathrm{Im}\digamma^j(s_0,\bp+\bk;k_0,\bk)}{s_0-|\bp+\bk|}
\right] \; .\label{contri}
\end{equation}
There is a single hydrodynamic denominator in this expression
since the hydrodynamic denominator in $\digamma^j$ became a delta
function. Hence there is no secular term arising from this
discontinuity. Furthermore, upon relabelling $\bp \to \bp+\bk$
with $k \ll p$ in the first term in the momentum integral, we find
that these contributions vanish and hence they do not contribute
in the limit $k\to 0$.

After this analysis, we conclude that the leading contributions
from the vertex correction in Fig.~\ref{fig:pv}(a) that feature
hydrodynamic poles  leading to secular terms and therefore
contribute to the conductivity are given by
\begin{eqnarray}
&&\Im\Pi_a^{\mathrm{V},ij}(\omega,k)=-\pi e^2\,
\Tr\intp\intpo\intso\intto \; \gamma^i \; \rho_F(p_0,\bp) \;
\rho_F(t_0,\bp+\bk) \\
&&\times \bigg[\frac{n_F(t_0)-n_F(s_0)}{s_0-p_0}\,
\Re\digamma^j(s_0,\bp;\omega,\bk)\;\delta(t_0-s_0-\omega)
+\,\frac{n_F(p_0)-n_F(s_0)}{s_0-t_0}\,
\Re\digamma^j(s_0,\bp+\bk;\omega,\bk)\,\delta(s_0-p_0-\omega)\bigg]
\; , \nn
\end{eqnarray}
where $\Re\digamma^j(s_0,\bp;\omega,\bk)$ can be extracted from
eqs.~\eqref{F} and \eqref{Fpm} through the analytical continuation
$i\nu_k\to\omega+i0$ and all frequency denominators [including
those in $\Re\digamma^j(s_0,\bp;\omega,\bk)$] should be understood
as the principal values. Furthermore, the fermionic spectral
functions in the above expression are both $\rho_+$ or both
$\rho_-$ since only these products lead to the hydrodynamic poles.

Using the delta functions to perform the integrals over the
dispersive variables $s_0$, $p_0$, and $t_0$, we find the leading
terms in $k$, $\omega\to 0$ limit to be given by
\begin{eqnarray}
\Im\Pi_a^{\mathrm{V},ij}(\omega,k)&=&-2\pi e^2\omega\,\PV \intp \;
\frac{dn_F(p)}{dp}\bigg\{\frac{\hat{p}^i}{\bk\cdot\bhp-\omega}\,
\big[\Re\digamma_+^j(|\bp+\bk|-\omega,\bp;\omega,\bk)\nn\\
&&+\,\Re\digamma_+^j(p+\omega,\bp+\bk;\omega,\bk)\big] +(\omega\to
-\omega)\bigg\} \; ,
\end{eqnarray}
where we have used the property
$\Re\digamma_-(-s_0,\bp;\omega,\bk)=\Re\digamma_+(s_0,\bp;-\omega,\bk)$.
Upon inserting $\Re\digamma_+$ obtained from eqs.~\eqref{F} and
\eqref{Fpm} and keeping in $\Re\digamma_+$ only the delta
functions that can be satisfied for $\omega,k\to 0$, we find
\begin{eqnarray}
\Im\Pi_a^{\mathrm{V},ij}(\omega,k)&=& -2\pi e^4\omega\,\PV
\intp\frac{dn_F(p)}{dp}\intq\int^{+\infty}_{-\infty}
{dq_0}\left[n_B(q_0)+n_F(|\bp+\bq|)\right]\nn\\
&&\times\,\hat{p}^i \; \widehat{p+q}^j \big[K^+_1(\bp,\bq) \;
{}^\ast\!\rho_L(q_0,q)+
K^+_2(\bp,\bq) \; {}^\ast\!\rho_T(q_0,q)\big]\nn\\
&&\times\bigg[\frac{\delta(\bhp\cdot\bq-q_0-\bk\cdot\bhp+\omega)
+\delta(\bhp\cdot\bq-q_0+\bk\cdot\bhp-\omega)}
{(\bk\cdot\bhp-\omega)(\bk\cdot\widehat{\bp+\bq}-\omega)}
+(\omega\to -\omega)\bigg] \; .\label{ImPiVT}
\end{eqnarray}

The contribution from the diagram in Fig.~\ref{fig:pv}(b) to the
resummed two-loop photon polarization corresponds to those from
the two diagrams displayed in Fig.~\ref{fig:pvsf}, which can be
now calculated handily by writing each \emph{one-loop} vertex in
terms of the spectral representation similar to that given by
eq.~(\ref{GammamuF}) but with the free photon spectral functions
replacing those in eq.~(\ref{Fpm}). In order to simplify the
computation we will set the external photon momentum $k=0$. The
sum over the Matsubara frequency in the soft fermion loop can be
easily done. The resulting expression is rather lengthy but the
imaginary part can be computed straightforwardly following the
analysis presented above. In the limit $\omega\rightarrow 0$, we
find
\begin{eqnarray}
&&\Im\Pi_b^{\mathrm{V},ij}(\omega,k=0) = -e^2 \omega\int
\frac{d^3p}{(2\pi)^3}\intpo\intso\intto \;
\mathrm{Re}\widetilde{\digamma}^i_+(s_0,\bp) \;
{}^\ast\!\rho_+(p_0,p) \; {}^\ast\!\rho_+(t_0,p)\nn\\
&&\times\bigg\{\frac{\hat{p}^j}{\omega}\bigg(\frac{1}{t_0-p_0-\omega}
-\frac{1}{t_0-p_0+\omega}\bigg) \; \frac{dn_F(s_0)}{ds_0} \;
[\delta(p_0-s_0)+\delta(t_0-s_0)]
+\,2\intinfty du_0 \; \mathrm{Re}\widetilde{\digamma}^j_+(u_0,\bp)\nn\\
&&\times\bigg[\frac{2}{\omega^2}
\frac{\delta(u_0-s_0)}{(t_0-s_0)(p_0-s_0)}\;
\frac{dn_F(s_0)}{ds_0}-\frac{1}{\omega(u_0-s_0)^2}\;
\bigg(\frac{1}{t_0-p_0-\omega}-\frac{1}{t_0-p_0+\omega}\bigg)\nn\\
&&\times\bigg(\frac{dn_F(p_0)}{dp_0}
\;[\delta(p_0-s_0)+\delta(p_0-u_0)] +\,\frac{dn_F(t_0)}{dt_0}
[\delta(t_0-s_0) +\delta(t_0-u_0)] \bigg) \bigg] \bigg\} \;
,\label{Pib}
\end{eqnarray}
where $\widetilde{\digamma}^j_\pm(s_0,\bp)\equiv
\omega\,\digamma^j_\pm(s_0,\bp;\omega,\mathbf{0})$ and
$\digamma^j_\pm(s_0,\bp;\omega,\mathbf{0})$ is given by
eq.~(\ref{Fpm}) but with the replacement
${}^\ast\!\rho_T\to\rho_B$ and ${}^\ast\!\rho_L\to 0$. In
obtaining the above expression, we have used the property
$\widetilde{\digamma}^j_-(-s_0,\bp)=-\widetilde{\digamma}^j_+(s_0,\bp)$,
${}^\ast\!\rho_-(-p_0,p)={}^\ast\!\rho_+(p_0,p)$ and safely set
$\omega=0$ inside the arguments of the delta functions and in the
denominators which do not vanish in this limit.

Let us first focus on terms proportional to $1/(t_0-p_0\mp\omega)$
in eq.~(\ref{Pib}). Whereas there are pole-pole, pole-cut and
cut-cut contributions in the product ${}^\ast\!\rho_+(t_0,p) \;
{}^\ast\!\rho_+(p_0,p)$, only the pole-pole contribution will lead
to pinch denominators. However, upon inspecting the support of the
delta functions in $\widetilde{\digamma}$ [see eq.~(\ref{Fpm}) and
recall the above-mentioned replacement], it becomes clear that the
pole-pole contribution vanishes because the quasiparticle pole
$\omega_+(p)$ lies above the light cone but the delta functions in
$\widetilde{\digamma}$ only have support below the light cone.
Thus only the terms proportional to $1/\omega^2$ feature pinch
denominator and hence lead to \emph{linear} secular terms in
$\mathcal{G}(t)$. However, will argue below that these linear
secular terms are subleading in the leading logarithmic
approximation. Therefore, we conclude that while the contribution
from the diagram in Fig.~\ref{fig:pv}(b) is necessary to fulfill
the Ward identity, it actually does not contribute to leading
logarithmic order in agreement with the conclusion of
Ref.~\cite{aarts}.

\subsection{The cancellation between vertex and self-energy
leading contributions and the extraction of leading secular
terms}\label{sec:pertsec}

Gathering the above results for the self-energy and vertex
contributions [see eqs.~(\ref{ImPiSET}) and (\ref{ImPiVT})], we
find that the resummed two-loop photon polarization
$\Im\Pi^{ij}(\omega,k)$ can be written in a compact form as
\begin{eqnarray}
\Im\Pi^{\text{2-loop},ij}(\omega,k)&=&
\Im\Pi^{\mathrm{SE},ij}(\omega,k)+\Im\Pi^{\mathrm{V},ij}(\omega,k)\nn\\
&=&-2\pi
e^4\omega\,\PV\intp\frac{dn_F(p)}{dp}\intq\intqo\bigg\{\bigg[
\big[\mathcal{F}^{ij}(\omega,\bk,\bp,\bq)-
\mathcal{F}^{ij}(\omega,\bk,\bp,\mathbf{0})\big]\nn\\
&&\times\left[n_B(q_0)+n_F(|\bp+\bq|)\right] \big[K^+_1(\bp,\bq)
\; {}^\ast\!\rho_L(q_0,q)+ K^+_2(\bp,\bq) \;
{}^\ast\!\rho_T(q_0,q)\big]-
\frac{\mathcal{F}^{ij}(\omega,\bk,\bp,\mathbf{0})}{2|\bp-\bq|}\nn\\
&&\times\,[1+n_B(|\bp-\bq|)-n_F(q_0)] \big[K^+_3(\bp,\bq) \;
{}^\ast\!\rho_+(q_0,q)+
K^-_3(\bp,\bq) \; {}^\ast\!\rho_-(q_0,q)\big]\bigg]\nn\\
&&\times[\delta(\bhp\cdot\bq-q_0-\bk\cdot\bhp+\omega)
+\delta(\bhp\cdot\bq-q_0+\bk\cdot\bhp-\omega)]+ (\omega\to
-\omega)\bigg\} \; ,\label{ImPiT1}
\end{eqnarray}
where
\begin{equation}
\mathcal{F}^{ij}(\omega,\bk,\bp,\bq)= \frac{\hat{p}^i \;
\widehat{p+q}^j}{(\bk\cdot\bhp-\omega)
(\bk\cdot\widehat{\bp+\bq}-\omega)} \; .\label{Fij}
\end{equation}
This expression makes clear that the region of \emph{ultrasoft
photon momentum} $q\ll eT\ll p\sim T$ which leads to the anomalous
damping eq.~(\ref{hardgamma}) is \emph{cancelled} between the
contributions of the self-energy and those of the vertex to the
polarization. This is the cancellation that was found in
Ref.~\cite{lebedev}, which our analysis makes manifest in the
perturbative framework. Whereas the region of ultrasoft photon
momentum is suppressed in the electric conductivity as a result of
the explicit cancellation between the self-energy and vertex
corrections, there will be contributions to the conductivity from
the region of exchanged momentum $eT \lesssim q \lesssim T$, which
within our real-time framework will arise as \emph{linear} secular
terms in the perturbative expansion.

The analysis of the secular terms in the polarization studied in
the case of the self-energy correction in Sec.~\ref{sec:secterms}
revealed that the hydrodynamic poles (or pinch denominators) of
the form $\ln|\omega -\bk\cdot \bhp|/(\omega -\bk\cdot \bhp)^2$
give rise to a secular term of the form $t\ln t$ in
$\mathcal{G}(t)$. The logarithmic enhancement of such secular term
is due to the logarithmic, infrared divergence in the self-energy
that arises from the region of ultrasoft photon exchange $q \ll
eT$. However, since the soft-photon contribution in
eq.~(\ref{ImPiT1}) involves the difference of the function
$\mathcal{F}$ which vanishes for $q=0$, for $q\ll p$ we can expand
the integrand in powers of $q/p$. Rotational invariance dictates
that only even powers will survive the angular integration. Thus,
the lowest order term in the $q$-integral will have an extra power
of $q^2$ in the numerator which will render finite any potential
logarithmic, infrared divergence that is responsible for the
anomalous damping. This extra power of $q^2$ in the numerator is a
manifestation of the fact that the electrical conductivity is
determined by the transport time scale~\cite{baym}.

We can now proceed to calculate the corresponding
$\mathcal{G}(k,t)$ function in the limit $k\to 0$. Since
$\Im\Pi^{\text{2-loop},ij}(\omega,k)$ is free of logarithmic,
infrared singularity responsible for the anomalous damping, the
Fourier transform in eq.~(\ref{Goft}) can be performed easily by
using contour integration. In particular, we need
\begin{eqnarray}
\mathrm{PV}\int^{+\infty}_{-\infty} d\omega\, e^{-i\omega t} \;
\frac{\delta(\bhp\cdot\bq-q_0\mp\bk\cdot\bhp\pm\omega)}
{(\bk\cdot\bhp-\omega)(\bk\cdot\widehat{\bp+\bq}-\omega)} &\simeq&
\frac{i\pi \; \delta(\bhp\cdot\bq-q_0)}
{\bk\cdot(\widehat{\bp+\bq}-\bhp)} \;
\big[e^{-i\,\bk\cdot\bhp\,t}-
e^{-i\,\bk\cdot\widehat{\bp+\bq\,}t}\big]\nn\\
&\buildrel{q\ll p }\over\simeq& -\pi\; t \;
e^{-i\,\bk\cdot\bhp\,t}\; \delta(\bhp\cdot\bq-q_0) \; ,\label{FT}
\end{eqnarray}
where in the second approximation we have extracted the leading
secular term in the small $k$, $q$ limit. We find to two-loop
order with hard thermal loop propagators and in $k\to 0$ limit
\begin{eqnarray}
&&\mathcal{G}^{\text{2-loop},ij}(t) = -8\pi \;  e^4 \; t \intp
\frac{dn_F(p)}{dp}\intq\intqo \; \bigg\{
(\hat{p}^i\hat{p}^j-\hat{p}^i\widehat{p+q}^j) \;
[n_B(q_0)+n_F(|\bp+\bq|)]\nn\\
&&\big[K^+_1(\bp,\bq) \; {}^\ast\!\rho_L(q_0,q)+ K^+_2(\bp,\bq) \;
{}^\ast\!\rho_T(q_0,q)\big] +\,\frac{\hat{p}^i \; \hat{p}^j}{2 \;
|\bp-\bq|}\big[K^+_3(\bp,\bq) \;
{}^\ast\!\rho_+(q_0,q)+K^-_3(\bp,\bq) \; {}^\ast\!\rho_-(q_0,q)\big]\nn\\
&&\times\,[1+n_B(|\bp-\bq|)-n_F(q_0)]\bigg\}\;
\delta(q_0-\bhp\cdot\bq) \; . \label{G2lups}
\end{eqnarray}
For soft momentum $q\ll p$ we can expand the integrand in powers
of $q/p$. The cancellation between the self-energy and vertex for
ultrasoft photon momentum transfer $q\to 0$ thus entails an extra
power of $q^2$ in the soft-photon contribution to the integrand in
eq.~(\ref{G2lups}), which as remarked above is the origin of the
transport time scale.

To compute $\sigma$ from eq.~(\ref{sigfin}) we ultimately need
$\mathcal{G}(t)=\sum_i\mathcal{G}^{ii}(t)/3$. We can obtain a
reliable estimate for the leading logarithmic contribution by
utilizing the following approximations. Since
${}^\ast\!\rho_{L,T}(q_0,q)$ are odd functions in $q_0$, we
approximate
\begin{eqnarray}
n_B(q_0)+n_F(|\bp+\bq|)\approx T/q_0 \quad ,\quad
1+n_B(|\bp-\bq|)-n_F(q_0) \approx  1/2 +n_B(p) \; .
\end{eqnarray}
Furthermore, following Baym et al.~\cite{baym}, we neglect the HTL
contributions in the \emph{denominators} of the respective HTL
spectral functions
\begin{eqnarray}
{}^\ast\!\rho_L(q_0,q) & \simeq &
-\frac{1}{\pi}\frac{\mathrm{Im}\Pi^\mathrm{HTL}_L(q,q_0)}{(q^2)^2}=
\frac{e^2 T^2}{6 q^4} \;
\frac{q_0}{q} \; \Theta(q^2-q^2_0) \; ,\nn \\
{}^\ast\!\rho_T(q_0,q) & \simeq & -\frac{1}{\pi}\frac{\mathrm{Im}
\Pi^\mathrm{HTL}_T(q,q_0)}{(q^2_0-q^2)^2}= \frac{e^2 T^2}{12 q^4}
\; \frac{q_0}{q} \;
\frac{1}{1-\frac{q^2_0}{q^2}} \; \Theta(q^2-q^2_0) \; ,\nn\\
{}^\ast\!\rho_\pm(q_0,q) & \simeq &
-\frac{1}{\pi}\frac{\mathrm{Im}
\Sigma^\mathrm{HTL}_\pm(q,q_0)}{(q_0\mp q)^2}=\frac{e^2 T^2}{16
q^3} \; \frac{1}{1\mp \frac{q_0}{q}} \; \Theta(q^2-q^2_0) \; .
\label{leadlogapx}
\end{eqnarray}
Performing the trivial $q_0$ integral in eq.~(\ref{G2lups}) and
keeping the leading terms in $q/p$ in the integrand, we find that
the leading logarithms arise from the momentum integral of the
form $\int_{q_\mathrm{min}}^{q_\mathrm{max}}dq/q=
\ln(q_\mathrm{max}/q_\mathrm{min})$, with the upper and lower
momentum cutoffs $q_\mathrm{max}\sim T$ and $q_\mathrm{min}\sim
eT$, respectively, which limit the validity of the approximations
made. We thus see that the leading logarithmic terms emerge from
the region of exchanged momentum $q_\mathrm{min}\lesssim q
\lesssim q_\mathrm{max}$ and appears in the form
$\ln(q_\mathrm{max}/q_\mathrm{min}) \sim \ln(1/e)$. In particular,
this entails that in order to extract the leading logarithms we
can simply use the \emph{perturbative} spectral functions.

Finally, upon collecting the results we find
\begin{equation}
\mathcal{G}^\text{2-loop}(t) =
-\frac{32+3\pi^2}{128\pi^3}\;\mathcal{G}^\mathrm{HTL}\; e^4\; T \;
\ln\left(\frac{1}{e}\right)\, t  \; , \label{G2lups2}
\end{equation}
where [see eqs.~(\ref{ghtl}) and (\ref{calGe})]
\begin{equation}
\mathcal{G}^\mathrm{HTL}=-\frac{e^2T^2}{18}\,.
\end{equation}
Therefore, up to three-loop order we obtain that
\begin{equation}
\mathcal{G}(t)= \mathcal{G}^\mathrm{HTL} \left[1-
\frac{32+3\pi^2}{128\pi^3}\; e^4 \; T \;
\ln\left(\frac{1}{e}\right)\,t +\cdots \right] \; .\label{fullG}
\end{equation}
This expression clearly displays the \emph{linear} secular term
arising from denominators of the form $1/(\omega-\bk\cdot \bhp)^2$
\emph{without} the logarithm of time featured by the anomalous
damping precisely because of the cancellation between self-energy
and vertex for ultrasoft photon momentum transfer. It also
highlights that the \emph{transport time scale} is given by
$t_\mathrm{tr}\sim[e^4 T \ln(1/e)]^{-1}$, which is much longer
than the fermionic quasiparticle relaxation time scale
$t_\mathrm{qp} \sim [e^2 T\ln(1/e)]^{-1}$. The origin of the
logarithms in these expressions is also different: in the
transport time scale the logarithm originates from momenta $eT
\lesssim q \lesssim T$, while in the quasiparticle relaxation time
scale it is momenta $e^2 T \lesssim q \lesssim eT$ that lead to
the logarithm. Our analysis reveals that this difference is a
consequence of the cancellation between self-energy and vertex
corrections for ultrasoft photon momentum transfer.

\begin{figure}[t]
\includegraphics[width=2.25in,keepaspectratio=true]{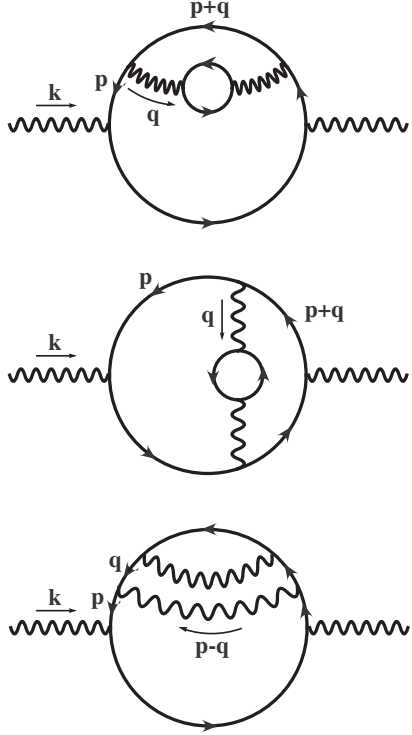}
\caption{Three-loop diagrams for the photon polarization in the
leading logarithmic approximation.}\label{fig:3loop1}
\end{figure}

\subsubsection*{Origin of the leading logarithms}

\begin{figure}[t]
\includegraphics[width=2.25in,keepaspectratio=true]{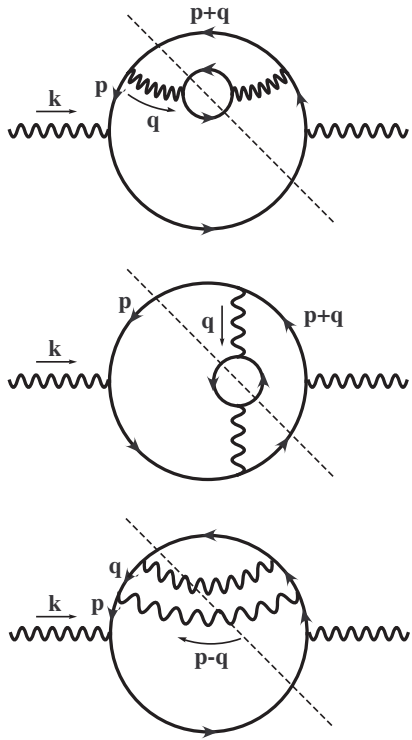}
\caption{Cuts in the three-loop diagrams for the photon
polarization in the leading logarithmic approximation that lead to
the secular term. The dashed lines determine the
cuts.}\label{fig:3lupcuts}
\end{figure}

An important aspect gleaned from the above result is that the
leading logarithmic correction can be extracted by using the
\emph{perturbative} spectral functions given by
eq.~(\ref{leadlogapx}) and by cutting off the integral over the
transfer momentum $q$ with upper and lower cutoffs
$q_\mathrm{max}\sim T$ and $q_\mathrm{min}\sim e T$, respectively.
This simplification corresponds to computing the \emph{three-loop}
diagrams depicted in Fig.~\ref{fig:3loop1} with cutoff momentum
integrals. The imaginary parts of the polarization that lead to
the secular term and leading logarithmic contribution correspond
to the cuts displayed in Fig.~\ref{fig:3lupcuts}. The cut-diagrams
with the photon self-energy correspond to M{\o}ller scattering
$e^\pm\,e^\pm\rightarrow e^\pm\,e^\pm$, since the singular
denominators that lead to secular terms arise from the products
$\rho^\pm_F  \; \rho^\pm_F$ and the photon spectral function below
the light cone in the HTL approximation arises also from these
type of products in the photon polarization bubble. The internal
longitudinal (transverse) photon propagators with momentum $q$
contributes the factor $(q^2)^{-2}$ [$(q^2_0-q^2)^{-2}$] that
leads to the logarithms. The cut-diagram with the fermion
self-energy corresponds to Compton scattering $\gamma\,e^\pm
\rightarrow \gamma\,e^\pm$ as well as pair annihilation and
creation processes $e^+\,e^-\leftrightarrow \gamma + \gamma$,
where the electrons, positrons and photons are all hard. The
internal fermion propagators with momentum $q$ give rise the
factor $(q_0\mp q)^{-2}$ that leads to the logarithms.

The diagrams that yield the leading logarithms in the coupling are
\emph{formally} corrections of $\mathcal{O}(e^6)$ to the photon
polarization. The logarithms arise from the \emph{free}
intermediate state propagators with momentum restricted by the
upper and lower cutoffs $q_\mathrm{max}\sim T$ and
$q_\mathrm{\min} \sim e~T$, respectively. Thus, only the
polarization diagrams depicted in Fig.~\ref{fig:3lupcuts} with the
simplified spectral functions eq.~(\ref{leadlogapx}) and the upper
and lower momentum cutoffs $q_{max} \sim T$, $q_{min} \sim e~T$
(or, alternatively, only the cut diagrams shown in
Fig.~\ref{fig:3lupcuts} with restricted momentum transfer
$eT\lesssim q \lesssim T$) contribute linear secular terms to
leading logarithmic order. Such simplification in the leading
logarithmic approximation has already been noticed in
Ref.~\cite{baym}. In turn, this implies that both Debye screening
for the longitudinal photon and dynamical screening by Landau
damping for the transverse photon propagators are
\emph{irrelevant} to extract the leading logarithmic contribution
and we can simply extract the leading secular terms to leading
logarithmic accuracy by computing the \emph{three-loop}
corrections to the polarization with \emph{free vacuum photon
propagators} and upper and lower momentum cutoffs of order $T$ and
$eT$, respectively. This observation will be important below where
we establish the correspondence with the Boltzmann equation.

\begin{figure}[t]
\includegraphics[width=2.25in,keepaspectratio=true]{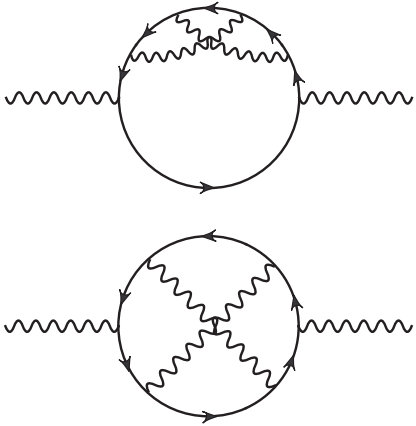}
\caption{Three-loop photon polarization diagrams of
$\mathcal{O}(e^6)$ that feature a linear secular term but no
logarithms in the coupling. }\label{fig:nolead3loops}
\end{figure}

This analysis also allows to identify the diagrams in the
perturbative expansion that yield the leading logarithmic
contribution and the linear secular terms. To leading logarithmic
order there are additional corrections to the polarization
formally of $\mathcal{O}(e^6)$, as depicted in
Fig.~\ref{fig:nolead3loops}. They correspond to a vertex
correction to the fermion self-energy and a \emph{crossed} ladder
diagram. The calculation of these diagrams is very complicated and
beyond the scope of this article. However, from the discussion
above it is clear that these diagrams will feature a linear
secular term but no logarithms of the coupling.

This above discussion highlights the origin of the leading
logarithmic contributions and we can now argue firmly that the
contribution from the diagram in Fig.~\ref{fig:pv}(b) (or,
equivalently, those depicted in Fig.~\ref{fig:pvsf}) is subleading
at leading logarithmic order. As argued after eq.~(\ref{Pib}), it
is the term proportional to $1/\omega^2$ that can lead to secular
terms. To leading order we can replace the HTL fermion spectral
function ${}^\ast\!\rho_+(p_0,p)$ in eq.~(\ref{Pib}) by that given
in eq.~(\ref{leadlogapx}). As a result, the contributions from the
diagrams in Fig.~\ref{fig:pvsf} are of $\mathcal{O}(e^8)$ and
$\mathcal{O}(e^{10})$, respectively. Therefore, the soft-fermion
contribution to the vertex is subleading in agreement with the
conclusion of Ref.~\cite{aarts}.

\section{Dynamical renormalization group and Boltzmann equation}
\label{section:boltz}

In the previous sections we have extracted the corrections to the
photon polarization that feature linear secular terms in time to
leading logarithmic approximation. The next task is to sum the
secular terms in the perturbative expansion by using the dynamical
renormalization group approach introduced in
Ref.~\cite{boyanqed,scalar}. The dynamical renormalization group
applies to the equation of motion of expectation values, either of
the Heisenberg fields or of composite
operators~\cite{boyanqed,scalar} which lead to an initial value
problem. Thus the first step is to obtain the corresponding
equations of motion for $\mathcal{G}(t)$ or an alternative
quantity from which it can be derived. In order to understand what
are the equations of motion (and corresponding initial value
problem) for $\mathcal{G}(t)$ (or a related quantity), we begin
with equation eq.~(\ref{timederi}) which basically identifies
$\mathcal{G}(t)$ with the time derivative of the expectation value
of the current in the zero-momentum limit. Thus we now study the
expectation value of the current to identify the proper equation
of motion in order to implement the DRG resummation.

We begin the analysis by writing
\begin{equation}
\Psi(\xt) =\sum_{\lambda}\intp \left[ b_{\bp,\lambda}(t) \;
u_{\bp,\lambda} \; e^{-i(p t-\bp\cdot\bx)}
+d^{\dagger}_{\bp,\lambda}(t) \; v_{\bp,\lambda} \; e^{i(p
  t-\bp\cdot\bx)}\right] \; , \label{Psi}
\end{equation}
where $u$ and $v$ are the free massless Dirac spinors. The time
evolution of the operators $b_{\bp,\lambda}(t)$ and
$d_{\bp,\lambda}(t)$ is determined by the Heisenberg equations of
motion. Here we have factored out the rapidly varying phases
$e^{\mp ipt}$, therefore for momentum $p \sim T \gg m$ these
operators vary \emph{slowly} only due to the interaction. An
important aspect of the gauge invariant formulation introduced
earlier is that the operators $b$ and $d$ are \emph{manifestly}
gauge invariant, since the field operator $\Psi$ has been
constructed to be manifestly gauge invariant.

Using the properties of the free massless Dirac spinors $u$ and
$v$, we find that the zero momentum limit of the expectation value
of the electromagnetic current is given by
\begin{equation} \label{jota}
J^i(t)=J^i_c(t)+J^i_s(t) \; ,
\end{equation}
with
\begin{eqnarray}
J^i_c(t)&=&e\sum_{\lambda}\intp \; \hat{p}^i \; \Big[\langle
b^\dagger_{\bp,\lambda}(t) \; b_{\bp,\lambda}(t)\rangle -\langle
d_{-\bp,\lambda}(t) \; d^{\dagger}_{-\bp,\lambda}(t) \rangle
\Big] \; ,\label{conveccurr} \\
J^i_s(t)&=&-ie\sum_{\lambda,\lambda'}\intp \Big[\langle
b^\dagger_{\bp,\lambda}(t) \; d^{\dagger}_{\bp,\lambda'}(t)\rangle
 \; u^\dagger_{\bp,\lambda} \; (\bp\times\!\boldsymbol{\Sigma})^i \;
v_{-\bp,\lambda'} \; e^{2ipt}\Big]+\mathrm{c.c.}
\;,\label{spinccurr}
\end{eqnarray}
where $\Sigma^i$ are the spin matrices. The term $\mathbf{J}_c(t)$
is the \emph{convection current} and $\mathbf{J}_s(t)$ is the
\emph{spin current}. The convection current term is time
independent in the free theory and is therefore varying slowly in
the interacting theory, the time variation is solely due to the
interactions. On the other hand, the spin current term mixes
particles and antiparticles and oscillates very fast (even in free
field theory), on time scales $p^{-1} \sim 1/T$. This term is
responsible for the phenomenon of \emph{zitterbewegung} associated
with the mixture of positive and negative energy solutions in
relativistic fermionic wave packets. Thus under the assumption
that processes associated with photon exchange will not mix
particles and antiparticles, namely that the exchanged photon
momenta are $\ll T$, we can neglect the contribution from the spin
current term since it oscillates very fast and will average out on
time scales $\sim 1/T$.

Introducing the spatial Fourier transforms of the fermion fields
\begin{equation}
\Psi(\xt)=\intp \; \Psi(\bp,t) \; e^{i\bp\cdot\bx} \;, \quad
\Psi^{\dagger}(\xt)= \intp \; \Psi^\dagger(\bp,t) \;
e^{-i\bp\cdot\bx} \; ,\label{FTDirac} \end{equation} we find the
following expressions for the expectation values that enter in the
convection current
\begin{equation}
\sum_{\lambda}\langle b^\dagger_{\bp,\lambda}(t) \;
b_{\bp,\lambda}(t)\rangle= \langle\bar{\Psi}(\bp,t) \;
\gamma_-(\bhp) \; \Psi(\bp,t)\rangle\; , \quad
\sum_{\lambda}\langle d_{-\bp,\lambda}(t) \;
d^\dagger_{-\bp,\lambda}(t)\rangle= \langle\bar{\Psi}(\bp,t) \;
\gamma_+(\bhp) \; \Psi(\bp,t)\rangle \; . \label{expec}
\end{equation} These are identified as the
contributions from positive and negative energy components to the
convection current. Under the assumption that all the processes
involved do not lead to a mixture of positive and negative energy
components, the two contributions will not mix in any order of
perturbation theory. This is manifest in the perturbative
calculation of the previous section in that only products of
spectral functions for particles or antiparticles enter in the
secular terms. Thus the \emph{slowly} varying part of the current
at zero momentum is determined by the convection term and given by
the expression
\begin{equation} J^i_c(t) = e\intp \; \hat{p}^i \; \left[\langle
\bar{\Psi}(\bp,t) \; \gamma_-(\bhp) \; \Psi(\bp,t)\rangle -\langle
\bar{\Psi}(\bp,t) \; \gamma_+(\bhp) \; \Psi(\bp,t)\rangle \right]
= e\intp \; \hat{p}^i\;\langle\bar{\Psi}(\bp,t) \;
\boldsymbol{\gamma}\cdot\bhp
 \; \Psi(\bp,t)\rangle \; . \label{conveccurr2}
\end{equation}
It proves convenient to introduce the spin-averaged expectation
values of the number of particles and antiparticles as
\begin{equation}
f(\bp,t) = \frac{1}{2} \sum_{\lambda}\langle
b^\dagger_{\bp,\lambda}(t) \; b_{\bp,\lambda}(t) \rangle \; ,
\quad \overline{f}(\bp,t) =  \frac{1}{2} \sum_{\lambda}\langle
d^\dagger_{\bp,\lambda}(t) \; d_{\bp,\lambda}(t)\rangle \;
,\label{distri}
\end{equation}
which as emphasized above are gauge invariant. In terms of these,
the convection part of the current is given by
\begin{equation}
J^i_c(t) = 2e\int \frac{d^3p}{(2\pi)^3}\,
\hat{p}^i\left[f(\bp,t)+\overline{f}(-\bp,t)\right]\;,
\label{kinetcur}
\end{equation}
which is the same form as that in kinetic theory, hence we call
this the \emph{kinetic current}. This analysis clearly highlights
two important aspects: (i) The kinetic current is identified with
the expectation value of the convection current in Dirac theory,
this is the component that varies slowly on the time scale $1/T$,
the contribution from the spin current averages out on time scales
larger than $1/T$ provided there are no processes with transferred
momenta $\sim 2 T$  that can mix particles with antiparticles.
(ii) The distribution functions $f$ and $\overline{f}$ are
identified as the expectation values of the number operators for
particles and antiparticles, respectively. Our gauge invariant
formulation guarantees that these are indeed gauge invariant
physical observables.

>From now on we will refer to the current simply as the convection
part and the discussion above clearly indicates that this is the
kinetic current. Thus we can establish the following
correspondence between the quantum field theory and the kinetic
approach [see eq.~(\ref{jota})]:
\begin{eqnarray}
&&J^i(t)=-ie\int \frac{d^3p}{(2\pi)^3}\; \hat{p}^i\;
\mathrm{Tr}\left[\boldsymbol{\gamma}\cdot\bhp\;S^<(\bp;t,t)
\right] \; ,\nn\\
&&\; f(\bp,t) = -\frac{i}{2} \mathrm{Tr}\left[\gamma_-(\bhp) \;
S^<(\bp;t,t) \right]\; ,\nn\\
&&\overline{f}(-\bp,t) = 1+\frac{i}{2}
\mathrm{Tr}\left[\gamma_+(\bhp)  \; S^<(\bp;t,t)\right] \;
.\label{currcor}
\end{eqnarray}
The correspondence given by eq.~(\ref{currcor}) makes manifest
that the kinetic current is determined by the equal-time limit of
the spatial Fourier transform of the \emph{full} fermion
propagator. The full fermion propagator is given by a Dyson sum
\begin{equation}
S=S_0+S_0 \;  \Sigma \;  S_0+S_0 \;  \Sigma \;  S_0 \; \Sigma
S_0+\cdots.
\end{equation}
If an external electric field
$\boldsymbol{\mathcal{E}}=-\dot{\boldsymbol{\mathcal{A}}_T}$ is
present, then $S_0$ is the free propagator in the presence of the
background gauge field $\boldsymbol{\mathcal{A}}_T$ but
\emph{without} corrections from interaction due to the fluctuating
gauge fields, which are contained in the self-energy $\Sigma$.
This observation for the kinetic current makes explicit that the
collision term that enters the Boltzmann equation is determined by
the fermion self-energy (see below).

Furthermore, at this stage we can make contact with the function
$\mathcal{G}^{ij}(t)$ defined by the linear response relation
eq.~(\ref{timederi}). For this purpose we define to linear order
in the external electric field
\begin{equation}
f(\bp,t)=n_F(p)+\boldsymbol{\mathcal{E}}\cdot\bhp
 \; \Lambda(p,t) \; , \quad
\overline{f}(-\bp,t)=n_F(p)+\boldsymbol{\mathcal{E}}\cdot\bhp
 \; \overline{\Lambda}(p,t) \; .\label{deflin}
\end{equation}
We therefore identify,
\begin{equation}\label{idG} \mathcal{G}^{ij}(t) = -2e \int
\frac{d^3p}{(2\pi)^3}\; \hat{p}^i \;  \hat{p}^j \;
\left[\dot{\Lambda}(p,t)+\dot{\overline{\Lambda}}(p,t)\right],
\end{equation}
where the dot stands for derivative with respect to time.
eq.(\ref{sigfin}) then leads to the following expression for the
conductivity
\begin{equation}\label{condulam}
\sigma = \frac{2e}{3}\int \frac{d^3p}{(2\pi)^3}\,
\left[{\Lambda}(p,t=\infty)+{\overline{\Lambda}}(p,t=\infty)\right],
\end{equation}
\noindent where we have used the initial condition
${\Lambda}(p,t=0)={\overline{\Lambda}}(p,t=0)=0$.

We now derive the Boltzmann equation for the distribution
functions $f$ and $\overline{f}$ in the presence of the background
gauge field by implementing the dynamical renormalization group.
We first calculate the time derivative of these distribution
functions to linear order in the background gauge field as in
Sec.~\ref{indcurr} and extract the positive and negative energy
components of the coefficient of $\hat{p}^i$ from the HTL photon
polarization. As discussed in Sec.~\ref{oneHTL} the polarization
for small external frequency and momentum only receives
contributions from the products of fermion spectral functions of
the form $\rho_+  \; \rho_+$ and $\rho_- \; \rho_-$, corresponding
to the positive and energy fermion mass shells respectively. This
explicitly shows that there is \emph{no} mixing between positive
and negative energy components in the small transferred momentum
limit, thus justifying the neglect of the spin current
contribution.

We assume that at a given time $t_0$ the initial density matrix is
diagonal in the number basis and that the number operators for
fermion and antifermion have respective expectation values
$f(\bp,t_0)$ and $\overline{f}(\bp,t_0)$ as given by
eq.~(\ref{distri}). This choice of the initial density matrix
entails that the fermion propagators have the free field form as
given in Appendix~\ref{app:rtf} but in terms of the nonequilibrium
distributions $f(\bp,t_0)$ and $\overline{f}(\bp,t_0)$ at time
$t_0$ (see also Ref.~\cite{boyanqed}). From the results of
Sec.~\ref{oneHTL} we find to lowest order in the background field
and leaving the calculation of the collision terms (arising from
the interaction with the photon fields, namely the self-energy
corrections) for the moment,  that the time derivative of the
distribution functions are determined by the HTL photon
polarization. Using the nonequilibrium fermion propagators given
in Appendix~\ref{app:rtf}, we find
\begin{equation} \frac{\partial
}{\partial t}f(\bp,t) = -
e\,\boldsymbol{\mathcal{E}}\cdot\!\bnabla_\bp f(\bp,t_0) +
\mbox{collision~terms}\; , \quad \frac{\partial }{\partial
t}\overline{f}(-\bp,t)=
-e\,\boldsymbol{\mathcal{E}}\cdot\!\bnabla_\bp\overline{f}(-\bp,t_0)
+ \mbox{collision~terms}\; . \label{vlasovpart}
\end{equation}
The next step is to obtain the contribution to the time derivative
of the distribution functions from the interaction, namely the
\emph{collision term}. From the definition of the distribution
functions eq.~(\ref{distri}), it follows that their time
derivatives are completely determined by the Heisenberg equations
for the fermion fields, namely,
\begin{equation}
\frac{\partial }{\partial
t}\Psi(\bp,t)=-i\,\boldsymbol{\alpha}\cdot\bp \;
\Psi(\bp,t)+ie\intq\left[\boldsymbol{\alpha}\cdot
\mathbf{A}_T(\bq,t)-A^0(\bq,t)\right] \Psi(\bp-\bq,t) \;
.\label{Heiseq}
\end{equation}
Taking the time derivative of the distribution functions and using
eq.~(\ref{Heiseq}), we find that the collision term for $f(\bp,t)$
is given by~\cite{boyanqed}
\begin{eqnarray}
\frac{\partial f(\bp,t)}{\partial t}\bigg|_\mathrm{coll}&=&
\frac{ie}{2}\intq
\big[\big\langle\bar{\Psi}^-(\bp+\bq,t)A^{0,-}(\bq,t)\gamma_-(\bhp)
\Psi^+(\bp,t)\big\rangle\nn\\
&&-\big\langle\bar{\Psi}^-(\bp+\bq,t)
\boldsymbol{\gamma}\cdot\mathbf{A}^-_T(\bq,t)
\gamma^0\gamma_-(\bhp)\Psi^+(\bp,t)\big\rangle \big]
+\mathrm{c.c.} \; ,\label{fdot}
\end{eqnarray}
where the superscripts `$\pm$' refer to fields in the forward and
backward branches in closed-time-path formalism, respectively (see
~\cite{boyanqed} for details) . The collision term for
$\overline{f}(-\bp,t)$ has a similar expression but with
$\gamma_-(\bhp) \to -\gamma_+(\bhp)$. The expectation values in
these expressions are in an initial density matrix and can be
systematically computed in perturbation theory. In order to
compute the time derivative of the distribution function in
eq.~(\ref{fdot}) we now follow the procedure outlined in Sec.~IV B
in Ref.~\cite{boyanqed}.

\begin{figure}[t]
\includegraphics[width=1.75in,keepaspectratio=true]{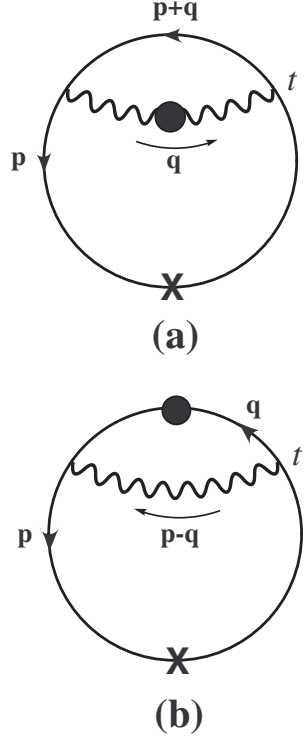}
\caption{Diagrams corresponding to the collision term $\partial
f(\bp,t)/\partial t|_\mathrm{coll}$ up to one-loop order: (a) the
soft-photon contribution and (b) the soft-fermion contribution.
The solid line denotes the free fermion propagator in the presence
of the background gauge field. The heavy cross in the fermion line
represents the insertion of $\gamma^0\gamma_-$ at the time
$t$.}\label{fig:kinecurr}
\end{figure}

Up to one-loop order, as depicted in Fig.~\ref{fig:kinecurr}, the
collision term receives contributions from soft internal photon
and soft internal fermion
\begin{eqnarray}
\frac{\partial f(\bp,t)}{\partial t}\bigg|_\mathrm{coll}&=&
C[f(.,t_0),\overline{f}(.,t_0),t-t_0] \equiv
C^\mathrm{sp}[f(.,t_0),\overline{f}(.,t_0),t-t_0]+
C^\mathrm{sf}[f(.,t_0),\overline{f}(.,t_0),t-t_0] \;
.\label{collparts}
\end{eqnarray}
Following the steps detailed in Ref.~\cite{scalar,boyanqed} with
the propagators given in Appendix~\ref{app:rtf}, we find that the
collision term can be written in an illuminating form which
displays the familiar gain minus loss form as
\begin{equation}
C[f(.,t_0),\overline{f}(.,t_0);t-t_0] = 2\intpo\;
\frac{\sin\left[(p-p_0)(t-t_0)\right]}{\pi(p-p_0)}\left\{
\left[1-f(\bp,t_0)\right] \;
\Sigma_>(p_0,p)-f(\bp,t_0)\Sigma_<(p_0,p)\right\} \;  ,
\label{collgainloss}
\end{equation}
where
\begin{equation}
\Sigma_>(p_0,p)=\Sigma^\mathrm{sp}_>(p_0,p)+
\Sigma^\mathrm{sf}_>(p_0,p) \; , \quad
\Sigma_<(p_0,p)=\Sigma^\mathrm{sp}_<(p_0,p)+
\Sigma^\mathrm{sf}_<(p_0,p) \; ,
\end{equation}
with the following self-energy components
\begin{eqnarray}
\Sigma^\mathrm{sp}_>(p_0,p) &=& e^2\pi \intq \intqo
\big\{\big[K^+_1(\bp,\bq) \; {}^\ast\!\rho_L(q_0,q)+
K^+_2(\bp,\bq) \; {}^\ast\!\rho_T(q_0,q)\big]\nn\\
&&\times\,f(\bp+\bq,t_0) \; n_B(q_0) \;\delta(p_0-q_0-|\bp+\bq|)
+\,\big[K^-_1(\bp,\bq) \; {}^\ast\!\rho_L(q_0,q)+
K^-_2(\bp,\bq) \; {}^\ast\!\rho_T(q_0,q)\big]\nn\\
&&\times\,[1-\overline{f}(-\bp-\bq,t_0)] \;
n_B(q_0) \; \delta(p_0-q_0+|\bp+\bq|)\big\} \; ,\nn\\
\Sigma^\mathrm{sp}_<(p_0,p) &=& e^2\pi \intq\intqo \;
\big\{\big[K^+_1(\bp,\bq) \; {}^\ast\!\rho_L(q_0,q)+
K^+_2(\bp,\bq) \; {}^\ast\!\rho_T(q_0,q)\big]\nn\\
&&\times\,[1-f(\bp+\bq,t_0)][1+n_B(q_0)]
\;\delta(p_0-q_0-|\bp+\bq|)
+ \, \big[K^-_1(\bp,\bq) \; {}^\ast\!\rho_L(q_0,q)+
K^-_2(\bp,\bq) \; {}^\ast\!\rho_T(q_0,q)\big]\nn\\
&&\times\,\overline{f}(-\bp-\bq,t_0) \; [1+n_B(q_0)] \;
\delta(p_0-q_0+|\bp+\bq|)\big\} \; ,
\end{eqnarray}
and
\begin{eqnarray}
\Sigma^\mathrm{sf}_>(p_0,p) &=& \pi e^2
\intq\frac{1}{2|\bp-\bq|}\intqo \; \big\{ \big[K^+_3(\bp,\bq) \;
{}^\ast\!\rho_+(q_0,q)+
K^-_3(\bp,\bq) \; {}^\ast\!\rho_-(q_0,q)\big]\nn\\
&&\times\,n_F(q_0) \; n_B(|\bp-\bq|) \; \delta(p_0-q_0-|\bp-\bq|)
+\,\big[K^-_3(\bp,\bq) \; {}^\ast\!\rho_+(q_0,q)+
K^+_3(\bp,\bq) \; {}^\ast\!\rho_-(q_0,q)\big]\nn\\
&&\times\,[1-n_F(q_0)][1+n_B(|\bp-\bq|)] \;
\delta(p_0+q_0+|\bp-\bq|)\big\} \; ,\nn\\
\Sigma^\mathrm{sf}_<(p_0,p) &=& \pi e^2
\intq\frac{1}{2|\bp-\bq|}\intqo \; \big\{ \big[K^+_3(\bp,\bq) \;
{}^\ast\!\rho_+(q_0,q)+
K^-_3(\bp,\bq) \; {}^\ast\!\rho_-(q_0,q)\big]\nn\\
&&\times\,[1-n_F(q_0)][1+n_B(|\bp-\bq|)]\;
\delta(p_0-q_0-|\bp-\bq|)\nn\\
&&+\,\big[K^-_3(\bp,\bq) \; {}^\ast\!\rho_+(q_0,q)+
K^+_3(\bp,\bq){}^\ast\!\rho_-(q_0,q)\big]
\; n_F(q_0) \; n_B(|\bp-\bq|) \; \delta(p_0+q_0+|\bp-\bq|)\big\}
\; . \label{sigless}
\end{eqnarray}
For antiparticles we find that
\begin{equation}
\frac{\partial \overline{f}(\bp,t)}{\partial t}
\bigg|_\mathrm{coll}=
\overline{C}[f(.,t_0),\overline{f}(.,t_0);t-t_0] \;
,\label{collantiparts}
\end{equation}
where
\begin{equation}\label{collkernanti}
\overline{C}[f(\bp,t_0),\overline{f}(-\bq,t_0);t-t_0] =
C[\overline{f}(-\bp,t_0),f(\bq,t_0);t-t_0] \; ,
\end{equation}
which entails that $f(\bp,t)$ and $\overline{f}(-\bp,t)$ satisfy
the same evolution equations. Since the evolution equation for
antiparticles (with momentum $-\bp$) is identical to that for
particles (with momentum $\bp$), we now focus solely on the
particle distribution function, the steps can be repeated
straightforwardly for the distribution of antiparticles.

\subsection{DRG resummation of the secular terms}

Combining eq.~(\ref{vlasovpart}) with the collision term obtained
above we find
\begin{equation}
\frac{\partial }{\partial t}f(\bp,t) = -
e\,\boldsymbol{\mathcal{E}}\cdot\!\bnabla_\bp
f(\bp,t_0)+C[f(.,t_0),\overline{f}(.,t_0);t-t_0].\label{boltz1}
\end{equation}
Notice that the right-hand side depends only on time through the
sine functions in the collision term, and the distribution
functions $f(\bp,t_0)$ and $f(-\bp,t_0)$ are fixed at the initial
time [see eq.~(\ref{collgainloss})]. Thus the evolution equation
can be integrated to yield
\begin{equation}\label{secular}
f(\bp,t)=f(\bp,t_0)-e \;
\boldsymbol{\mathcal{E}}\cdot\!\bnabla_\bp f(\bp,t_0) \; (t-t_0)+
\int^t_{t_0}C[f(.,t_0),\overline{f}(.,t_0);t'-t_0] \;  dt' \; .
\end{equation}
The last term  gives upon integration on time,
\begin{equation}
\int^t_{t_0} C[f(.,t_0),\overline{f}(.,t_0);t'-t_0] \;  dt' =
\intpo \; \frac{1-\cos[(p-p_0)(t-t_0)]}{\pi(p-p_0)^2}\;
\mathcal{K}(f(.,t_0),\overline{f}(.,t_0);p_0,p) \;
,\label{intcoll}
\end{equation}
with
\begin{eqnarray}
\mathcal{K}(f(.,t_0),\overline{f}(.,t_0);p_0,p)&=& 2 \left\{
\left[1-f(\bp,t_0)\right] \;
\Sigma_>(p_0,p)-f(\bp,t_0)\Sigma_<(p_0,p)\right\} \; .
\end{eqnarray}
In general the time integral of the collision term will feature
\emph{secular terms}, whose time dependence is determined by the
behavior of the term that multiplies the cosine near the resonance
$p=p_0$. For example, if the coefficient of the cosine term
$\mathcal{K}(f(.,t_0),\overline{f}(.,t_0);p_0,p)$ \emph{is finite
and does not feature any singularity} as $p-p_0\to 0$ then we can
use the long-time asymptotic expression~\cite{scalar}
\begin{equation}\label{asy}
\frac{1-\cos[(p-p_0)(t-t_0)]}{\pi(p-p_0)^2}\to (t-t_0) \;
\delta(p-p_0)
\end{equation}
to extract the secular term, which in this case is linear in time
as is the case in the usual Fermi's Golden rule. However, if
$\mathcal{K}(f(.,t_0),\overline{f}(.,t_0);p_0,p)$ either vanishes
or features singularities near the resonance $p=p_0$ then one must
be more careful in extracting the secular term. Such is the case
with the anomalous logarithms that will emerge in the relaxation
time approximation (see below). Whatever the form of these secular
terms, their main feature is that they grow in time and invalidate
the perturbative expansion, therefore they must be resummed. This
resummation scheme in real time is the goal of the dynamical
renormalization group program~\cite{boyanqed,scalar}.

Just as the usual renormalization group absorbs the divergences
into a renormalization of couplings at an arbitrary momentum
scale, the dynamical renormalization group  absorbs the secular
terms into a \emph{finite renormalization} of the distribution
function at an arbitrary time scale $\tau$. Let us define
\begin{equation}
f(\bp, t_0)=f(\bp,\tau) \;
\mathcal{Z}(\bp,\tau,t_0),\label{rendrg}
\end{equation}
where the renormalization constant $\mathcal{Z}$ will be chosen in
an expansion in $\mathcal{E}$ and the coupling that enters in the
collision term to cancel the secular terms in eq.~(\ref{secular})
at the time $t=\tau$. Choosing to lowest order
\begin{equation}\label{counterdrg} \mathcal{Z}(\bp,\tau,t_0)= 1 +
\frac{e\,\boldsymbol{\mathcal{E}}\cdot\!\bnabla_\bp
f(\bp,\tau)}{f(\bp,\tau)}(\tau-t_0)-
\int^\tau_{t_0}\frac{C_s[f(.,\tau),
\overline{f}(.,\tau);t'-t_0]}{f(\bp,\tau)}\; dt'+
\mathcal{O}(\mathcal{E}^2,C^2_s,\mathcal{E} \; C_s ) \; ,
\end{equation}
where $C_s$ denotes the part of the collision term that leads to
secular terms. Then to lowest order in the external background
field and in the strength of the collision term we find
\begin{equation}
f(\bp,t)=f(\bp,\tau)-e\,\boldsymbol{\mathcal{E}}
\cdot\!\bnabla_\bp f(\bp,\tau)(t-\tau)- \int^\tau_{t}
C_s[f(.,\tau),\overline{f}(.,\tau);t'-t_0] \;  dt'
+\mathcal{O}(\mathcal{E}^2,C^2_s,\mathcal{E} \; C_s ) +
\mbox{non~secular~terms} \; ,\label{renof}
\end{equation}
Choosing the arbitrary scale $\tau$ near $t$, the perturbative
expansion is now valid in a neighborhood of $t=\tau$ by
renormalizing the distribution function.

However, the distribution function $f(\bp,t)$ \emph{cannot} depend
on the arbitrary time scale $\tau$, namely $d f(\bp,t)/d\tau =0$
leads to the dynamical renormalization group equation to this
order (after setting $\tau=t$)
\begin{equation}
\frac{\partial }{\partial t}f(\bp,t)
+e\,\boldsymbol{\mathcal{E}}\cdot\!\bnabla_\bp
f(\bp,t)=C_s[f(.,t),\overline{f}(.,t);t] \; .\label{DRGBE}
\end{equation}
This dynamical renormalization group equation is recognized as the
Boltzmann equation, where the collision term is the part of the
original one that leads to secular terms. (See Ref.~\cite{scalar}
for the scalar theory). Thus in order to write down the Boltzmann
equation we must first work out $C_s$. As we shall see below in
sec. V C, when $\mathcal{K}(f(.,t_0),\overline{f}(.,t_0);p_0,p)$
is regular at $p_0 = p$, we simply find
$$
C_s[f(.,t),\overline{f}(.,t);t]=\mathcal{K}(f(.,t),\overline{f}(.,t);p,p)
\; .
$$

\subsection{Relaxation time approximation}\label{sec:reltimeappx}

In the relaxation time approximation \emph{only} the particles
with momentum $\bp$ and the antiparticles with momentum $-\bp$ are
slightly distorted out of equilibrium, while particles and
antiparticles with all other momenta are assumed to be in
equilibrium. Hence, we write the particle distribution as
\begin{equation}
f(\bp,t_0)= n_F(p)+\delta f(\bp,t_0) \; , \quad
f(\bp+\bq,t_0)=n_F(|\bp+\bq|)\;\;\mathrm{for}\;\;\bq\neq 0 \; ,
\label{dptofeq}
\end{equation}
and similarly for antiparticle distributions. With the above
distributions the collision term given by eq.~(\ref{collgainloss})
simplifies to
\begin{equation}
C[f(.,t_0),\overline{f}(.,t_0);t-t_0] =-2  \;\delta
f(\bp,t_0)\intpo \;
\Sigma_+(p_0,p)\;\frac{\sin\left[(p-p_0)(t-t_0)\right]}{\pi(p-p_0)}
\; , \label{collgainlossrel}
\end{equation}
where we have used the fact that when evaluated with the
equilibrium distribution ($n_F(p)$),
$\mathcal{K}(n_F(.),n_F(.);p_0,p)$ vanishes by virtue of detailed
balance and $\Sigma_>(p_0,p)+\Sigma_<(p_0,p)=\Sigma_+(p_0,p)$ with
$\Sigma_+(p_0,p)$ given by eq.~(\ref{SEplus2}).

As discussed in Sec.~\ref{subsec:anomadamping}, the self-energy
$\Sigma_+(p_0,p)$ near the fermion mass shell is dominated by the
ultrasoft photon exchange that leads to anomalous damping. Using
eq.~(\ref{logadamp}) and the results quoted in
Refs.~\cite{boyanqed,scalar}, the time integral in
eq.~(\ref{secular}) to linear order in the external electric field
and the departure of equilibrium (considered to be of the same
order as the external electric field) is given by
\begin{equation}
\delta f(\bp,t)=\delta f(\bp,t_0)-
e\,\boldsymbol{\mathcal{E}}\cdot\!\bnabla_\bp n_F(p)(t-t_0)-2
\alpha  T \,  (t-t_0)\, \ln\left[(t-t_0)\omega_D\right]\,\delta
f(\bp,t_0) +\mathrm{subleading} \; . \label{seclin}
\end{equation}
This expression clearly displays the anomalous logarithms as
secular terms found previously in the perturbative expression
eq.~(\ref{asyGSE}).

Following the steps described above by introducing the
renormalization constant $\mathcal{Z}(\bp,\tau,t_0)$:
\begin{equation}
\delta f(\bp,t_0)= \mathcal{Z}(\bp,\tau,t_0) \; \delta
f(\bp,\tau)\; ,
\end{equation}
with
\begin{equation}
\mathcal{Z}(\bp,\tau,t_0)=1+e\,\boldsymbol{\mathcal{E}}
\cdot\!\bnabla_\bp n_F(p)(\tau-t_0)+2 \alpha  T\, (\tau-t_0)\,
\ln\left[\omega_D(\tau-t_0)\right]+{\cal O}(\mathcal{E}^2, e^3)\;,
\label{reno}
\end{equation}
we are led to the following DRG equation
\begin{equation}
\frac{\partial}{\partial t}\delta
f(\bp,t)+e\,\boldsymbol{\mathcal{E}}\cdot\!\bnabla_\bp n_F(p)
=-2\; \Gamma(t) \; \delta f(\bp,t) \; ,\label{DRGeqn}
\end{equation}
with $\Gamma(t)$ the \emph{time dependent} relaxation rate, which
in the leading logarithmic approximation is given by (for $t \gg
t_0$)
\begin{equation}\label{gammaoftau}
\Gamma(t)= \alpha \;  T \left[ \ln(\omega_D t)+1 \right] \;.
\end{equation}
The equation for $\delta\overline{f}(-\bp,t)$ is the \emph{same}
as that above for particles. The DRG equation (\ref{DRGeqn}) is
the Boltzmann equation in the relaxation time approximation.

\emph{If} $\Gamma$ were a \emph{constant} the solution to the
above equation would be
\begin{equation} \delta f(\bp,t) = -e\,
\boldsymbol{\mathcal{E}}\cdot\!\bnabla_\bp \; n_F(p) \;
\frac{1-e^{-2\Gamma t}}{2\Gamma} \; .\label{gamaconst}
\end{equation}
Using the relations given by eqs.~(\ref{deflin})-(\ref{idG}) with
this solution we would then find
\begin{equation}\label{Ggammaconst} \mathcal{G}^{ij}(t)= -{4e^2}\int
\frac{d^3p}{(2\pi)^3} \; \frac{dn_F(p)}{dp}\; \hat{p}^i \;
\hat{p}^j \; e^{-2\Gamma t} = \delta_{ij}
 \;   \frac{e^2 \; T^2}{9}  \; e^{-2\Gamma t} \; .
\end{equation}
This is exactly the result that we would find if we considered the
fermion propagator obtained from the Dyson resummation with a
constant width, as discussed in Sec.~\ref{sec:secterms}. Thus we
see that in the case in which the width is constant, the DRG
resummation in the relaxation time approximation is equivalent to
the Dyson resummation of the fermion
propagator~\cite{boyanqed,scalar}. The result (\ref{Ggammaconst})
leads to a steady-state, drift current at long times given by
\begin{equation}
J^i = -\frac{4e^2}{2\Gamma}\intp \; \hat{p}^i  \;
\boldsymbol{\mathcal{E}} \cdot\!\bnabla_\bp n_F(p) = \frac{e^2
T^2}{18\,\Gamma} \; \mathcal{E}^i \; ,\label{currgamaconst}
\end{equation}
and to the electrical conductivity
\begin{equation}
\sigma=\frac{e^2 T^2}{18 \,\Gamma} \; ,\label{condgamaconst}
\end{equation}
which is in complete agreement with that of the analysis of
Sec.~\ref{sec:secterms} for \emph{constant} $\Gamma$.

However, for $\Gamma(t)$ given by eq.~(\ref{gammaoftau}) the
dynamical renormalization group equation (\ref{DRGeqn}) leads to a
very \emph{different} picture. The solution of eq.~(\ref{DRGeqn})
with eq.~(\ref{gammaoftau}) is given by
\begin{equation}
\delta f(\bp,t)=-e\,\boldsymbol{\mathcal{E}}\cdot\!\bnabla_\bp \;
n_F(p)\; e^{-2\alpha T t \ln(\omega_D t)} \int^t_0 \; e^{2\alpha T
t' \ln(\omega_D t')} \;  dt' \; ,\label{trusol}
\end{equation}
where we have assumed that the electric field was switched-on at
$t=0$ and that $\delta f(\bp,t=0)=0$. Taking the time derivative
of this expression, using the identifications
eqs.~(\ref{deflin})-(\ref{idG}) and that the Boltzmann equation
for antiparticles (with momentum $-\bp$) is the same as that for
particles (with momentum $\bp$), we find to $\mathcal{O}(\alpha)$
\begin{equation}
\mathcal{G}^{ij}(t)=4 e^2\int \frac{d^3p}{(2\pi)^3} \frac{d
n_F(p)}{dp} \;  \hat{p}^i \; \hat{p}^j \; \left[1-2 \alpha  T  t
 \ln(\omega_D t)\right] = - \delta_{ij} \;  \frac{e^2 T^2}{9}
\;\left[1-2\alpha T  t  \ln(\omega_D t)\right] \; ,\label{gsecul}
\end{equation}
namely, the solution of the Boltzmann equation in the relaxation
time approximation reproduces the leading secular term to
$\mathcal{O}(\alpha)$ found in the perturbative expansion given by
eq.~(\ref{sum}).

Thus we reach an important result of this study: \emph{the
relaxation time approximation in the Boltzmann equation is
equivalent to \emph{only} considering self-energy corrections to
the internal fermion propagators in the photon polarization}. This
is a reassuring result and establishes a direct link between the
perturbative expansion of self-energy corrections to the photon
polarization and the Boltzmann equation in the relaxation time
approximation.

Of course, these secular terms are an artifact of the
approximation, the full solution given by eq.~(\ref{trusol}) is
bound at all times. Its asymptotic long time behavior can be found
as follows. First it is convenient to split the time integral in
eq.~(\ref{trusol}) as $\int^t_0 dt'= \int^{1/\omega_D}_0
dt'+\int^t_{1/\omega_D}dt'$. In the first integral $t'\ln(\omega_D
t')<0$ hence we can replace $e^{2\alpha T t \ln(\omega_D t)} \sim
1$, leading to an exponentially small contribution at
asymptotically long time. The second integral can be computed by
successive integration by parts which lead to the asymptotic
expansion,
\begin{equation}
\delta f(\bp,t) = -\frac{e\,\boldsymbol{\mathcal{E}}
\cdot\!\bnabla_\bp n_F(p)}{2\alpha T \ln(\omega_D
t)}\left[1+\frac{1}{2\alpha   T  t  \ln^2(\omega_D
t)}+\cdots\right] \; ,\label{asysolu}
\end{equation}
namely the drift current vanishes asymptotically, hence the
logarithmic anomalous damping leads to a \emph{vanishing}
electrical conductivity. The main result of this analysis is that
while a constant damping rate $\Gamma$ would lead to a finite
conductivity given by eq.~(\ref{condgamaconst}), an anomalous
logarithmic time dependence of the damping rate leads to a
\emph{vanishing} electrical conductivity in the asymptotic,
long-time limit.

Again we highlight that the result given by eq.~(\ref{asysolu}) is
a consequence of the relaxation time approximation which is
equivalent to keeping only self-energy corrections. This result
\emph{does not apply} to the hot QED plasma because the vertex
corrections introduce the cancellation analyzed in the previous
section. The correct expression is given below by
eqs.~(\ref{noneqf}) and (\ref{solFK}). We derived here
eq.~(\ref{asysolu}) to show what would be the solution if the
relation time approximation were valid, which is not the case
here.

\subsection{Full and linearized Boltzmann equation}\label{sec:BE}

The above discussion on the relaxation time approximation
highlights that this approximation is equivalent to considering
\emph{only} the self-energy corrections to the photon
polarization. This conclusion is based on the equivalence of the
secular terms between the perturbative result (\ref{sum}) and the
lowest order in $\alpha$ term in the exact solution of the
Boltzmann equation in the relaxation time approximation
(\ref{gsecul}). The anomalous logarithms in time are a consequence
of ultrasoft photon exchange leading to the threshold infrared
divergence.

Just as in the perturbative case we expect that the cancellation
of infrared divergences between self-energy and vertex corrections
will also be manifest in the Boltzmann equation. Thus, we
\emph{assume} for a moment, and will be confirmed below,  that the
coefficient of the cosine term
$\mathcal{K}(f(.,t_0),\overline{f}(.,t_0);p_0,p)$ in
eq.~(\ref{intcoll}) is non-singular and finite at $p_0=p$ and use
the result eq.~(\ref{asy}) to extract the asymptotic long time
limit. In this case the secular term is linear in time, and we
recognize that the collision term that enters in the Boltzmann
equation (\ref{DRGBE}) is given by
\begin{equation}
C_s[f(.,t),\overline{f}(.,t)]=C^\mathrm{sp}_s[f(.,t),\overline{f}(.,t)]+
C^\mathrm{sf}_s[f(.,t),\overline{f}(.,t)] \; ,
\end{equation}
where
\begin{eqnarray}
C^\mathrm{sp}_s[f(.,t),\overline{f}(.,t)]&=&2\pi e^2\intq\intqo
\big\{\big[K^+_1(\bp,\bq) \; {}^\ast\!\rho_L(q_0,q)+
K^+_2(\bp,\bq) \; {}^\ast\!\rho_T(q_0,q)\big]
 \; \delta(p-q_0-|\bp+\bq|) \nn\\
&& \times\big[[1-f(\bp,t)] \; f(\bp+\bq,t) \; n_B(q_0)- f(\bp,t)
[1-f(\bp+\bq,t)] \; [1+n_B(q_0)]\big]\nn\\
&&+\big[K^-_1(\bp,\bq) \; {}^\ast\!\rho_L(q_0,q)+ K^-_2(\bp,\bq)
\; {}^\ast\!\rho_T(q_0,q)\big]
\; \delta(p-q_0+|\bp+\bq|)\nn\\
&& \times\big[[1-f(\bp,t)] \;
[1-\overline{f}(-\bp-\bq,t)]n_B(q_0)-f(\bp,t)
\overline{f}(-\bp-\bq,t)[1+n_B(q_0)]\big]\big\} \; ,\label{truker1}\\
C^\mathrm{sf}_s[f(.,t),\overline{f}(.,t)]&=&\pi
e^2\intq\frac{1}{|\bp-\bq|}\intqo \big\{\big[K^+_3(\bp,\bq) \;
{}^\ast\!\rho_+(q_0,q)+ K^+_3(\bp,\bq) \;
{}^\ast\!\rho_-(q_0,q)\big] \;
\delta(p-q_0-|\bp-\bq|)\nn\\
&&\times\big[[1-f(\bp,t)] \; n_F(q_0) \; n_B(|\bp-\bq|)-
f(\bp,t) \; [1-n_F(q_0)] \; [1+n_B(|\bp-\bq|)]\big] + \nn\\
&&+\big[K^-_3(\bp,\bq) \; {}^\ast\!\rho_+(q_0,q)+ K^-_3(\bp,\bq)
\; {}^\ast\!\rho_-(q_0,q)\big] \;
\delta(p+q_0+|\bp-\bq|)\nn\\
&&\times\big[[1-f(\bp,t)] \; [1-n_F(q_0)] \; [1+n_B(|\bp-\bq|)]-
f(\bp,t) \; n_F(q_0) \; n_B(|\bp-\bq|)\big]\big\} \; .
\label{truker2}
\end{eqnarray}
The reason that in the soft-fermion contribution $C^\mathrm{sf}_s$
there appear the free fermion distribution functions $n_F(q_0)$
and not $f$ is a consequence of the fact that the HTL correction
to the vertex does not lead to secular terms to lowest order as
discussed above and analyzed in detail in Ref.~\cite{aarts}. This
will become clear below when we identify self-energy and vertex
contributions in the collision kernel of the Boltzmann equation
and we compare the perturbative solution of the Boltzmann equation
with the perturbative expansion of the function $\mathcal{G}(t)$
studied in the previous sections.

We notice that the delta functions $\delta(p-q_0+|\bp+\bq|)$ in
eq.~(\ref{truker1}), which multiplies terms that mix particles
($f$) with antiparticles ($\overline{f}$), and
$\delta(p+q_0+|\bp+\bq|)$ in eq.~(\ref{truker2}) can only be
satisfied if $q_0 \sim T$ since $p\sim T$ and $q\sim eT$. However,
in obtaining the kinetic current we have neglected precisely these
type of processes which are responsible for very fast oscillations
on time scales $\sim \mathcal{O}( 1/T)$ [see discussion after
eqs.~(\ref{conveccurr}) and (\ref{spinccurr})].

To be consistent with the kinetic description, we \emph{must}
consider only processes in which the transfer of energy and
momentum is $q_0$, $q \ll T$ and hence neglect the second terms in
eqs.~(\ref{truker1}) and (\ref{truker2}). Therefore, the correct
Boltzmann equation to this order is given by
\begin{eqnarray}
\frac{\partial}{\partial t} f(\bp,t)+
e\,\boldsymbol{\mathcal{E}}\cdot\!\bnabla_\bp f(\bp,t)&=&2\pi e^2
\int\frac{d^3q}{(2\pi)^3}\intqo\bigg\{ \big[K^+_1(\bp,\bq) \;
{}^\ast\!\rho_L(q_0,q)+
K^+_2(\bp,\bq) \; {}^\ast\!\rho_T(q_0,q)\big]\nn\\
&&\times\,\big[[1-f(\bp,t)] \; f(\bp+\bq,t) \; n_B(q_0)
-f(\bp,t) \; [1-f(\bp+\bq,t)] \; [1+n_B(q_0)]\big]\nn\\
&&\times\, \; \delta(p-q_0-|\bp+\bq|)+\frac{1}{2|\bp-\bq|}
\big[K^+_3(\bp,\bq) \; {}^\ast\!\rho_+(q_0,q)+
K^-_3(\bp,\bq) \; {}^\ast\!\rho_-(q_0,q)\big]\nn\\
&&\times\big[[1-f(\bp,t)] \; n_F(q_0) \; n_B(|\bp-\bq|)-
f(\bp,t) \; [1-n_F(q_0)] \; [1+n_B(|\bp-\bq|)]\big]\nn\\
&&\times\, \; \delta(p-q_0-|\bp-\bq|)\bigg\} \; . \label{trueBE}
\end{eqnarray}
The Boltzmann equation for antiparticles (with momentum $-\bp$) is
exactly the same as eq.~(\ref{trueBE}) but with the replacement
$f(\bp,t)\to \overline{f}(-\bp,t)$ on both sides of the equation.

\begin{figure}[t]
\includegraphics[width=2.0in,keepaspectratio=true]{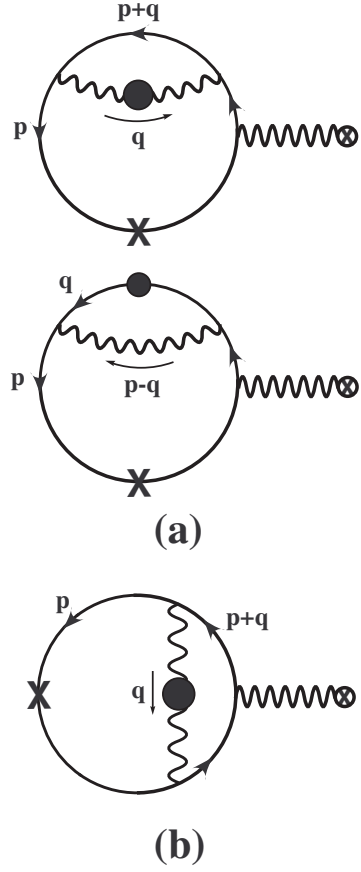}
\caption{Diagrams contributing to the linearized Boltzmann
equation. There are two diagrams of type (a) corresponding to
insertions to each side of the self-energy. The solid line denotes
the free fermion propagator without the background gauge field.
The wiggly line with a circled cross attached to the end denotes
to an insertion of the background gauge field. Otherwise, the
notation is the same as that in Fig.~\ref{fig:kinecurr}.}
\label{fig:linearboltz}
\end{figure}

The linearized approximation of the Boltzmann equation
(\ref{trueBE}) can be obtained by setting
\begin{equation}
f(\bp,t)=n_F(p)+\delta f(\bp,t) \;, \quad
\overline{f}(-\bp,t)=n_F(p)+\delta \overline{f}(-\bp,t) \; ,
\label{linear2}
\end{equation}
where $\delta f$ and $\delta\overline{f}$ are the departure from
equilibrium, and keeping only terms linear in $\delta f$ and
$\delta\overline{f}$ in the collision term. Since the collision
term vanishes for the equilibrium distribution ($n_F$), we find
the final form of the linearized Boltzmann equation
\begin{eqnarray}
\frac{\partial}{\partial t}\delta
f(\bp,t)+e\,\boldsymbol{\mathcal{E}} \cdot\!\bnabla_\bp
n_F(p)&=&-2\pi e^2 \intq\intqo\bigg\{ \big[K^+_1(\bp,\bq) \;
{}^\ast\!\rho_L(q_0,q)+
K^+_2(\bp,\bq) \; {}^\ast\!\rho_T(q_0,q)\big]\nn\\
&& \times\big[\delta f(\bp,t) \; [1+n_B(q_0)-n_F(|\bp+\bq|)]-
\delta f(\bp+\bq,t) \; [n_B(q_0)+n_F(p)]\big]\nn \\
&&\times\,\delta(p-q_0-|\bp+\bq|)+\frac{1}{2|\bp-\bq|}
\big[K^+_3(\bp,\bq) \; {}^\ast\!\rho_+(q_0,q)+
K^-_3(\bp,\bq) \; {}^\ast\!\rho_-(q_0,q)\big]\nn\\
&&\times\,\delta f(\bp,t) \; [1+n_B(|\bp-\bq|)-n_F(q_0)]
 \; \delta(p-q_0-|\bp-\bq|)\bigg\} \; .\label{linBE}
\end{eqnarray}
The departure from equilibrium for antiparticles
$\delta\overline{f}(-\bp,t)$ satisfies a similar equation but with
the replacement $\delta f(\bp,t)\to\delta \overline{f}(-\bp,t)$.
We note that the linearized approximation entails that $\delta f$,
$\delta\overline{f}\propto \mathcal{E}$, hence the replacement
$e\,\boldsymbol{\mathcal{E}}\cdot\!\bnabla_\bp f(\bp,t)\to
e\,\boldsymbol{\mathcal{E}}\cdot\!\bnabla_\bp n_F(p)$ on the
left-hand side of the Boltzmann equation is justified because the
neglected term is of second order in $\mathcal{E}$.

\subsection{Connection with perturbation theory: the emergence of
secular terms}\label{sec:emsecular}

The linearization of the collision term in the Boltzmann equation
has a simple and natural interpretation in terms of Feynman
diagrams. We recall that the collision term to lowest order is
given in Fig.~\ref{fig:kinecurr}, where the fermion propagators
are those in the external background field. The linearization
corresponds to expanding the propagators to linear order in the
background. There are two types of insertions: (i) The external
background field is inserted in the fermion lines \emph{outside}
the self-energy (two diagrams), this corresponds to the relaxation
time approximation which is manifest by the first term in
eq.~(\ref{linBE}) [proportional to $\delta f(\bp,t)$] and is
depicted in Fig.~\ref{fig:linearboltz}(a). (ii) The insertion is
in the fermion propagator \emph{inside} the self-energy, which
corresponds to the second term [proportional to $\delta
f(\bp+\bq,t)$]  and is depicted in Fig.~\ref{fig:linearboltz}(b).
Obviously this last term corresponds to the \emph{vertex}
correction. In fact these diagrams must be compared to those of
the self-energy and vertex corrections in the perturbative
expansion given in Figs.~\ref{fig:pse} and \ref{fig:pv},
respectively, with the external wiggle line in
Fig.~\ref{fig:linearboltz} corresponding to the external photon
line in Figs.~\ref{fig:pse} and \ref{fig:pv}.

In order to see this relation more clearly, we now show that the
perturbative expansion of the solution of the linearized Boltzmann
equation (\ref{linBE}) reproduces the perturbative results found
in Sec.~\ref{sec:pertsec} above. Let us write the solution of the
linearized Boltzmann equation in the following form
\begin{equation}
\delta f(\bp,t)= e \; \delta f^{(1)}(\bp,t)+e^3 \; \delta
f^{(2)}(\bp,t)+\cdots+e^{2n-1}\;\delta f^{(n)}(\bp,t)+\cdots,
\end{equation}
then eq.~(\ref{linBE}) reduces to
\begin{eqnarray}
\frac{\partial}{\partial t}\delta f^{(1)}(\bp,t)&=&
-\boldsymbol{\mathcal{E}}\cdot\bhp \; n_F'(p)\;,\nn\\
\frac{\partial}{\partial t}\delta f^{(2)}(\bp,t)&=&\delta
C[\delta f^{(1)}(\bp,t)]\;,\nn\\
\vdots&&\vdots \label{perblBE}\\
\frac{\partial}{\partial t}\delta f^{(n)}(\bp,t)&=&\delta
C[\delta f^{(n-1)}(\bp,t)]\;,\nn\\
\vdots&&\vdots\nn
\end{eqnarray}
where $n_F'(p)=dn_F(p)/dp$ and $\delta C$ denotes the collision
term of the linearized Boltzmann equation (\ref{linBE})  divided
by $e^2$. The set of equations (\ref{perblBE}) can be solved by
iteration starting from the lowest order one. Up to order $e^3$ we
obtain
\begin{eqnarray}
\delta f^{(1)}(\bp,t)& = & -t\,\boldsymbol{\mathcal{E}}
\cdot\bhp \; n_F'(p)\;, \nn\\
\delta f^{(2)}({\bp},t)& = & 2\pi\left(\frac{t^2}{2}\right)
\intq\intqo\bigg\{\left[K^+_1(\bp,\bq)\;{}^\ast\!\rho_L(q_0,q)+
K^+_2(\bp,\bq) \; {}^\ast\!\rho_T(q_0,q)\right] \nn\\
&&\times \big\{\boldsymbol{\mathcal{E}}\cdot\bhp \; n_F'(p) \;
[1+n_B(q_0)-n_F(|\bp+\bq|)]-\boldsymbol{\mathcal{E}}
\cdot\widehat{\bp+\bq} \; n_F'(|\bp+\bq|) \; [n_B(q_0)+n_F(p)]
\big\}\nn \\&&\times\, \; \delta(p-q_0-|\bp+\bq|)+
\frac{1}{2|\bp-\bq|} \big[K^+_3(\bp,\bq) \;
{}^\ast\!\rho_+(q_0,q)+
K^-_3(\bp,\bq) \; {}^\ast\!\rho_-(q_0,q)\big]\nn\\
&&\times\,\boldsymbol{\mathcal{E}}\cdot\bhp \; n_F'(p) \;
[1+n_B(|\bp-\bq|)-n_F(q_0)] \; \delta(p-q_0-|\bp-\bq|)\bigg\}\;.
\label{expa}
\end{eqnarray}
>From these expressions we can now proceed to compute the current
\begin{equation}
J^i(t)= 4 e \intp  \, \hat{p}^i  \, \delta
f(\bp,t)\;,\label{curre}
\end{equation}
where the factor four accounts for spins, particle and
antiparticle. It is more convenient to compute $dJ^i(t)/dt$ in
order to compare to the perturbative expression of the function
$\mathcal{G}^{ij}(t)$. Inserting eq.~(\ref{expa}) into
eq.~(\ref{curre}), we find up to order $e^4$ that
\begin{eqnarray}
\frac{d}{dt} J^i(t)&=& -4e^2\intp \; \hat{p}^i \; n_F'(p) \;
\boldsymbol{\mathcal{E}}\cdot\bhp+ 8 \pi e^4 \;  t\intp \;
n_F'(p) \;
\boldsymbol{\mathcal{E}}\!\cdot\!\bhp\intq\nn\\
&&\times\intqo \; \bigg\{(\hat{p}^i-\widehat{p+q}^i)
\left[K^+_1(\bp,\bq) \; {}^\ast\!\rho_L(q_0,q)+
K^+_2(\bp,\bq) \; {}^\ast\!\rho_T(q_0,q)\right]\nn\\
&&\times\,[n_B(q_0)+n_F(|\bp+\bq|)]+
\frac{\hat{p}^i}{2|\bp-\bq|}\big[K^+_3(\bp,\bq) \;
{}^\ast\!\rho_+(q_0,q)+K^-_3(\bp,\bq) \;
{}^\ast\!\rho_-(q_0,q)\big]\nn\\
&&\times\,[1+n_B(|\bp-\bq|)-n_F(q_0)]\bigg\}\;
\delta(q_0-\bhp\cdot\bq) \; .\label{djdt}
\end{eqnarray}
In the above expression we have combined terms of
$\mathcal{O}(e^4)$ that arise from the soft-photon contribution.
This is done by relabelling $q_0 \to -q_0$ and using the
properties $1+n_B(-q_0)=-n_B(q_0)$ and
$\rho_{L,T}(-q_0,q)=-\rho_{L,T}(q_0,q)$ in the first term of
$\mathcal{O}(e^4)$, as well as by relabelling $ \bp \to -\bp-\bq $
in the second term of this order. Furthermore, we have also used
$q\ll p\sim T$ to approximate the argument of the delta functions.

We can now compare the above \emph{perturbative} expression for
$dJ^i(t)/dt$ to that obtained from $\mathcal{G}^{ij}(t)$ in
Sec.~\ref{sec:pertsec}. With the identification given by
eq.~(\ref{timederi}) it is clear that the term of
$\mathcal{O}(e^2)$ in eq.~(\ref{djdt}) exactly reproduces the HTL
result eq.~(\ref{ghtl}) and that the terms of $\mathcal{O}(e^4)$
exactly reproduces the two-loop results for the soft-photon and
soft-fermion contributions given by eq.~(\ref{G2lups}). Thus we
see that the perturbative solution of the linearized Boltzmann
equation reproduces the leading secular terms of the perturbative
expansion resulting from the self-energy and vertex corrections.
The cancellation of the contribution of ultrasoft photon exchange
observed at the level of the perturbative expansion is a direct
manifestation of the transport time scale that enters in the
Boltzmann equation. This exercise not only shows this important
result in detail, but it also highlights the following important
identifications:
\begin{itemize}
\item[(i)]{The term with $\delta f(\bp,t)$ in the collision term
of the linearized Boltzmann equation (\ref{linBE}) is associated
with the self-energy correction, as manifest in the previous study
of the relaxation time approximation.}

\item[(ii)]{The term with
$\delta f(\bp+\bq,t)$ in the collision term is associated with the
vertex correction. The cancellation between the self-energy and
vertex corrections in $q\to 0$ limit, is manifest in the
linearized Boltzmann equation (\ref{linBE}) as the cancellation
between the terms with $\delta f(\bp,t)$ and $\delta f(\bp+\bq,t)$
[upon using the fact that the photon spectral functions
${}^\ast\!\rho_{L,T}(q_0,q)$ are odd functions of $q_0$].}
\end{itemize}
This explicit comparison between the perturbative evaluation of
the photon polarization and the perturbative expansion of the
solution of the linearized Boltzmann equation (in real time)
provides also a justification for ignoring the HTL correction for
the vertex [see Fig.~\ref{fig:pv}(b)] at the level of the
collision term. This diagram does not contribute to the leading
logarithm for linear secular terms in the perturbative expansion,
and hence will not contribute leading logarithms to the collision
term of the linearized Boltzmann equation.

\begin{figure}[t]
\includegraphics[width=2.0in,keepaspectratio=true]{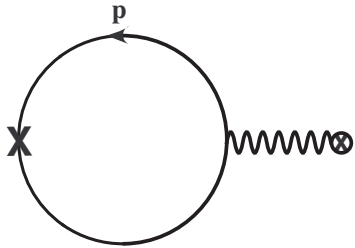}
\caption{Diagram corresponding to the lowest order perturbative
solution $\delta f^{(1)}$ of the linearized Boltzmann equation.}
\label{fig:fp1lup}
\end{figure}

We now argue that the equivalence between the perturbative
solution of the linearized Boltzmann equation and the perturbative
evaluation of the photon polarization in quantum field theory in
real time can be generalized to all orders in perturbation theory,
providing a direct identification of the infinite diagrams in the
photon polarization that are resummed by the linearized Boltzmann
equation. This is achieved by the following observation:
\begin{itemize}
\item[(i)]{The lowest order solution $\delta f^{(1)}$ features
linear secular term and corresponds diagrammatically to the
HTL-type diagram with hard fermion loop, as depicted in
Fig.~\ref{fig:fp1lup}.}

\item[(ii)]{The second order solution $\delta f^{(2)}$, which is
determined by $\delta C[\delta f^{(1)}]$, features quadratic
secular term and corresponds to diagrams obtained from the
HTL-type diagram in Fig.~\ref{fig:fp1lup} by inserting self-energy
and vertex corrections, respectively, into the \emph{hard} fermion
lines and the vertex attached to the \emph{external} wiggly line.}

\item[(iii)] { As discussed  in detail in the previous sections,
the secular terms with powers $t^{n}$ arise from singular
denominators of the form $(\bk \cdot {\bf p}-\omega)^{-n}$ for
$n=1,2...$ in the limit $\bk ; \omega \rightarrow 0$. In turn
these denominators arise from the pinch of two propagators on
opposite side of the loop. Thus the iteration above which features
powers $t,t^2,...$ arises from the insertion of uncrossed ladder
type diagrams with self-energy corrected  side rails. This is the
manifestation in real time of the argument provided in
references\cite{jeon,basagoiti2} that the Boltzmann equation to
leading order sums (uncrossed) ladder type diagrams with
self-energy corrected side rails. }
\end{itemize}
>From this observation and the iterative nature of the perturbative
solution [see eq.~(\ref{perblBE})], we conclude by induction that
the $n$-th order solution $\delta f^{(n)}$ features secular term
of the form $t^n$ and the corresponding diagrams is obtained from
those of the $(n-1)$-th order solution $\delta f^{(n-1)}$ by
performing the same insertion described above. Since the insertion
of the vertex correction involves only the vertex attached to the
external wiggly line, the only set of diagrams that will be
generated are \emph{uncrossed} ladder diagrams with rainbow type
self-energy insertion in the side rails. Consequently, this
analysis reveals that the linearized Boltzmann equation
(\ref{linBE}) resums the uncrossed ladder plus the insertion of
rainbow type self-energy diagrams in the photon polarization in
accord with the results of reference~\cite{basagoiti2,aarts}.

There is an important corollary of the equivalence between the
perturbative solution of the linearized Boltzmann equation and the
perturbative evaluation of the photon polarization in quantum
field theory in real time. Namely, by identifying the
contributions at different orders to the \emph{linear} secular
terms in the perturbative expansion of the function
$\mathcal{G}(t)$, one can understand precisely what are the
diagrams that contribute to the collision term in the linearized
Boltzmann equation. In particular, this equivalence allows a
direct identification of the contribution from self-energy and
vertex corrections to the polarization to the different terms in
the linearized Boltzmann equation. Thus, obtaining the
\emph{linear} secular terms in $\mathcal{G}(t)$ yields direct
information on the collision term of the linearized Boltzmann
equation \emph{to all orders in perturbation theory}.

Furthermore, this discussion clearly highlights that having proved
that the contributions from self-energy and vertex insertions that
leading to secular terms fulfill the Ward identity in section
(\ref{sec:WardId}) , the subsequent series resummed by the
dynamical renormalization group or Boltzmann equation is fully
gauge invariant.

\begin{figure}[t]
\includegraphics[width=2.0in,keepaspectratio=true]{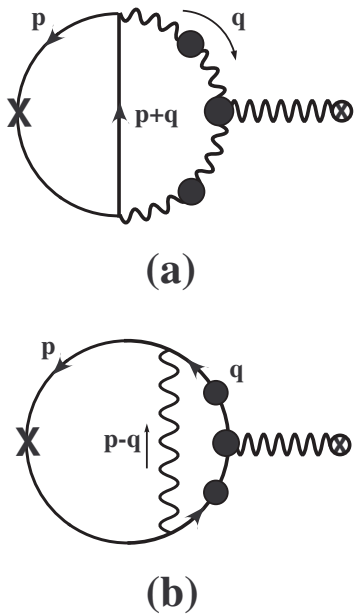}
\caption{Possible contributions to the linearized Boltzmann
equation resulting from a change in the fermion distribution
functions that enter the HTL-resummed propagators in (a) the
soft-photon contribution and (b) the soft-fermion contribution
(see Figs.~\ref{fig:kinecurr} and \ref{fig:linearboltz}).}
\label{fig:3fotonvertex}
\end{figure}

Another important issue that we want to highlight at this point is
that the HTL-resummed spectral functions ${}^\ast\!\rho_{T,L}$ and
${}^\ast\!\rho_\pm$ are obtained from the Dyson resummation of the
hard thermal loop result, namely these spectral densities are
function(als) of the \emph{hard} fermion distribution functions
$f(\bp,t)$ and $\overline{f}(\bp,t)$. Thus the question arises:
Why did we not account for a change in the fermion distribution
functions that enter in the HTL-resummed spectral functions? The
answer to this important question is the following: consider such
a variation in the \emph{hard} fermion distribution functions that
enter in ${}^\ast\!\rho_{L,T}$ and ${}^\ast\!\rho_\pm$, the
resulting contributions are depicted in
Fig.~\ref{fig:3fotonvertex}. For ${}^\ast\!\rho_{L,T}$ such
contribution would correspond to the vertex correction with a
HTL-resummed three-photon vertex [see
Fig.~\ref{fig:3fotonvertex}(a)]. However, such vertex is
\emph{forbidden} by Furry's theorem, a consequence of CPT
invariance in a charge neutral medium (i.e., vanishing fermionic
chemical potential) in the absence of a magnetic field. This
conclusion is obviously supported by the perturbative expansion,
where the absence of such three-photon vertex is manifest. For
${}^\ast\!\rho_\pm$ such contribution would correspond to the HTL
correction for the vertex [see Fig. \ref{fig:3fotonvertex}(b)]
which is subleading at the leading logarithmic order as we have
argued in Secs.~\ref{vcontribution} and \ref{sec:pertsec}. Thus to
leading logarithmic order the departure from equilibrium of the
fermion distribution functions are only associated with the
\emph{hard} fermion propagators that generate the kinetic current,
and \emph{not} with the internal fermion propagators that enter in
the HTL corrections to the \emph{soft} photon and fermion
propagators.

The Boltzmann equation provides a resummation of these secular
divergences. As argued above, by carrying out the perturbative
expansion of the Boltzmann equation to higher order and
identifying secular terms of the form $t^n$ with inverse powers of
frequency $\omega^{n+1}$ in the Fourier transform, one can
identify what type of the diagrams in the polarization and which
parts of the diagrams (e.g., leading logarithms, etc.) that are
being resummed. The resummation of diagrams furnished by the
Boltzmann equation can thus be put on a similar footing as the
resummation of diagrams in the usual renormalization group, where
by solving the equation for the running coupling via the
renormalization group beta function and expanding this solution in
a naive perturbative expansion, one can recognize what type of
diagrams and which part of the diagrams are being resummed. Up to
the leading logarithmic order considered, the linearized Boltzmann
equation resums the leading logarithmic contributions from the
uncrossed ladder plus the insertion of rainbow type self-energy
diagrams in the photon polarization~\cite{basagoiti2,aarts}.

The resummation of the \emph{linear} secular term in real time is
also akin to computing the \emph{decay rate} in time dependent
perturbation theory. By using Fermi's golden rule to extracts the
leading linear secular term at long times and
\emph{exponentiating} the secular term, one obtains the
exponential decay of probabilities. This exponentiation is a
resummation of the perturbative series.

\subsection{Fokker-Planck equation}

For soft momentum transfer $q\ll p\sim T$ we can expand the
integrand in the collision term of the linearized Boltzmann
equation (\ref{linBE}) in powers of $q/p$, as we did previously in
Sec.~\ref{sec:pertsec} to obtain the leading logarithms in the
perturbative expansion. This expansion, valid for small momentum
transfer is akin to that used in astrophysics to obtain the
Kompaneets equation from the kinetic equation for the photon
distribution function, which is widely used to study inverse
Compton scattering by hot electrons~\cite{lightman}. The result is
a partial differential equation for the distribution function, a
Fokker-Planck equation in \emph{momentum space}.

\begin{figure}[t]
\includegraphics[width=2.0in,keepaspectratio=true]{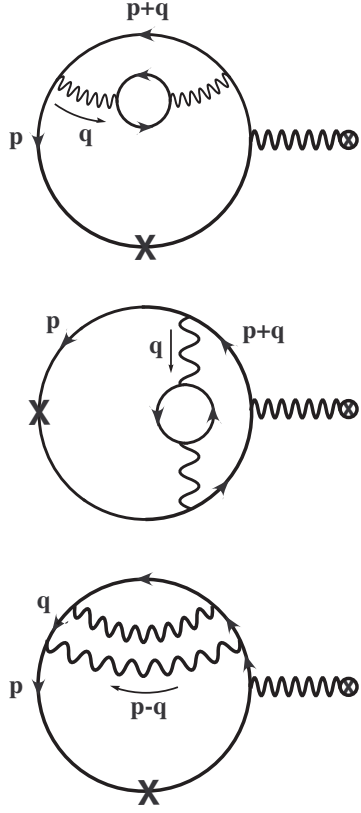}
\caption{Diagrams that yield the leading logarithmic contribution
to the Boltzmann equation.}\label{fig:BEthreelup}
\end{figure}

The first step in the derivation consists in writing the departure
from equilibrium as in eq.~(\ref{deflin}) in the following form
\begin{equation}
\delta f(\bp,t) = n_F'(p) \; \boldsymbol{\mathcal{E}}\cdot\bhp \;
\Delta(p,t) \; . \label{noneqf}
\end{equation}
The main reason for this definition is that the inhomogeneous term
in the Boltzmann equation is proportional to $ n_F'(p)\;
\boldsymbol{\mathcal{E}}\cdot\bhp $. Our analysis of the leading
logarithmic contributions in Sec.~\ref{sec:pertsec} clearly showed
that in order to extract the leading logarithms, the transverse
and longitudinal photon spectral functions can be replaced by
eq.~(\ref{leadlogapx}) and the momentum integrals must be carried
out with an upper cutoff of order $T$ and a lower cutoff of order
$eT$. These spectral functions are the perturbative ones,
corresponding to just one hard fermion loop in the soft photon and
fermion propagators, \emph{not} the full Dyson series of bubbles.
The diagrams that yield the leading logarithmic approximation are
depicted in Fig.~\ref{fig:BEthreelup}. These diagrams are
identified with those of Fig.~\ref{fig:3loop1} in the perturbative
expansion with the external wiggly line in
Fig.~\ref{fig:BEthreelup} corresponding to the external photon
line in Fig.~\ref{fig:3loop1}.

For $q\ll p$ the argument of the delta functions in the collision
term in eq.~(\ref{linBE}) are approximated by
\begin{equation}
|\bp\pm\bq|-p = \left\{ \pm\,q\,\bhp\cdot\bhq +
\frac{p}{2}\left(\frac{q}{p}\right)^2
\left[1-(\bhp\cdot\bhq)^2\right]\right\} \left[1 + {\cal
O}\left(\frac{q^3}{p^3}\right) \right]  \; , \label{expaeps}
\end{equation}
which simplifies the integral over $q_0$ in the collision term in
eq.~(\ref{linBE}). The resulting integrand in the collision term
can be expanded in powers of $q/p$. For the soft-photon
contribution the leading logarithm arises from terms linear in
$q/p$, which leads to a momentum integral of the form $\int^T_{eT}
dq /q = \ln(1/e)$. Terms of $\mathcal{O}(1)$ in $q/p$ are odd in
$\bhp\cdot\bhq$ thus vanish after angular integration, while terms
of higher order in $q/p$ lead to higher powers of $\alpha$ but not
logarithms. For the soft-fermion contribution the leading
logarithm arises from terms of $\mathcal{O}(1)$ in $q/p$, which
leads to the same momentum integral as for the soft-photon case.
Likewise, terms of higher order in $q/p$ lead to higher powers of
$\alpha$ but not logarithms.

After some lengthy but straightforward algebra we find the
linearized collision term in eq.~(\ref{linBE}) to be given by
\begin{equation}
C[\Delta] = \frac{e^4 T^3}{24\pi}\ln\left(\frac{1}{e}\right)
\boldsymbol{\mathcal{E}}\cdot\bhp\,n_F'(p)
\bigg[\Delta''(p,t)+\left(\frac{2}{p}-\tanh\frac{p}{2T}\right)
\Delta'(p,t)-\left(\frac{3}{4pT^2}\coth\frac{p}{2T}
+\frac{2}{p^2}\right)\Delta(p,t)\bigg] \; , \label{kerleadlog}
\end{equation}
where the primes stand for derivatives with respect to $p$. The
Boltzmann equation in this approximation becomes the Fokker-Planck
equation
\begin{equation}
\frac{1}{e}\frac{\partial}{\partial t} \Delta(p,t)
-\frac{e^3T^3}{24\pi}\ln\left(\frac{1}{e}\right)
\bigg[\Delta''(p,t)+\left(
\frac{2}{p}-\tanh\frac{p}{2T}\right)\Delta'(p,t)-
\left(\frac{3}{4pT^2}\coth\frac{p}{2T}
+\frac{2}{p^2}\right)\Delta(p,t)\bigg]=-1 \; . \label{FP}
\end{equation}
It proves convenient to introduce the transport time scale
\begin{equation}
t_\mathrm{tr}=\frac{24\pi}{e^4 T \ln(1/e)} \; , \label{taurel}
\end{equation}
and the dimensionless variables,
\begin{equation}\label{dimless}
\Phi(x,\tau) \equiv  \frac{e^3 T}{24\pi}\ln\left(
\frac{1}{e}\right)\Delta(p,t)\;,\quad
\tau\equiv\frac{t}{t_\mathrm{tr}}\;,\quad x\equiv\frac{p}{T} \; ,
\end{equation}
in terms of which the Fokker-Planck equation becomes
\begin{equation}
\dot{\Phi}(x,\tau)-\Phi''(x,\tau)-\left(\frac{2}{x}-
\tanh\frac{x}{2}\right)\Phi'(x,\tau)+\bigg(
\frac{3}{4x}\coth\frac{x}{2} +\frac{2}{x^2}\bigg)\Phi(x,\tau)=-1
\; ,\label{dimlessFP}
\end{equation}
where the dot and prime stand for derivatives with respect to
$\tau$ and $x$, respectively. This equation clearly reveals that
the time scale for transport is determined by $t_\mathrm{tr}$
defined in eq.~(\ref{taurel}). The real-time approach leads
directly to this Fokker-Planck equation in the leading logarithmic
approximation, which reveals at once the transport time scale and
includes the time dependence of the departure from equilibrium.
Thus this approach has a distinct advantage over the ladder
resummation program advocated in
Refs.~\cite{basagoiti1,basagoiti2,aarts} which only captures the
asymptotic, steady-state solution.

\subsection{Solution of the Fokker-Planck equation: asymptotics}

The solution of the Fokker-Planck equation (\ref{dimlessFP}) can
be written as
\begin{equation}\label{inter}
\Phi(x,\tau) = \phi(x,\tau) + \Phi_{\infty}(x) \; ,
\end{equation}
where $\phi(x,\tau)$ is a solution of the homogeneous
time-dependent Fokker-Planck equation
\begin{equation}\label{FPdept}
\dot{\phi}(x,\tau)-\phi''(x,\tau)-\left(\frac{2}{x}-
\tanh\frac{x}{2}\right)\phi'(x,\tau)+\bigg(
\frac{3}{4x}\coth\frac{x}{2} +\frac{2}{x^2}\bigg)\phi(x,\tau)= 0
\; ,
\end{equation}
while, the steady-state solution $\Phi_{\infty}(x)$ obeys the
time-independent inhomogeneous equation,
\begin{equation}
\Phi''_{\infty}(x)+\left(\frac{2}{x}-\tanh\frac{x}{2}\right)
\Phi'_{\infty}(x)- \bigg(\frac{3}{4x}\coth\frac{x}{2}
+\frac{2}{x^2}\bigg)\Phi_{\infty}(x)=1 \; . \label{steadystate}
\end{equation}
Eq.(\ref{FPdept}) is a parabolic equation that can be solved by
expanding ${\phi}(x,\tau)$ in terms of the eigenfunctions
$\psi_n(x)$ satisfying
\begin{equation}\label{autof}
\left[-\frac{d^2}{dx^2} -
\left(\frac{2}{x}-\tanh\frac{x}{2}\right)\frac{d}{dx}
+\frac{3}{4x}\coth\frac{x}{2} +\frac{2}{x^2}\right]\psi_n(x) =
\lambda_n \, \psi_n(x)
\end{equation}
in the interval $ 0 \leq x \leq \infty $ with $\psi_n(0) =
\psi_n(\infty) = 0$. Setting
\begin{equation}
\psi_n(x) = \frac{\xi_n(x)}{x}\cosh\frac{x}{2}\; ,
\end{equation}
we can rewrite eq.~(\ref{autof}) in a Schr\"odinger-like form
\begin{equation}
\left[-\frac{d^2}{dx^2} + V(x) \right]\xi_n(x)= \lambda_n \,
\xi_n(x)\; ,
\end{equation}
where
\begin{equation}
\label{pot} V(x) \equiv \frac14 - \frac{1}{2 \,
\cosh^2\frac{x}{2}} - \frac{1}{x} \tanh\frac{x}{2} +
\frac{3}{4x}\coth\frac{x}{2} +\frac{2}{x^2}  \; .
\end{equation}
This potential has a repulsive singularity for $ x \to 0 $,
\begin{equation}
V(x)\buildrel{x\to 0}\over= \frac{7}{2 \; x^2}
\end{equation}
and for $ x \gg 1$ approaches its asymptotic value with an
attractive Coulomb tail
\begin{equation}
V(x)\buildrel{x \gg 1 }\over=\frac14 - \frac1{4\; x } +
\mathcal{O}(e^{-x})\;.
\end{equation}
$V(x)$ has its  minimum at $ x = x_\mathrm{min} = 15.997\ldots $
with
\begin{equation}
V(x_\mathrm{min}) = 0.2421873\ldots\;,\quad V''(x_\mathrm{min}) =
0.0000609\ldots \; .\label{mini}
\end{equation}
In addition, $V(x) \geq V(x_\mathrm{min})$ and $V''(x) > 0$ for $0
\leq x \leq \infty $.

The Coulomb-like tail of the potential suggests that the  spectrum
of eq.~(\ref{autof}) is formed by an infinite number of discrete
eigenvalues with an accumulation point at $\lambda=1/4-0$ and a
continuum that starts at $\lambda=1/4$. Therefore the discrete
eigenvalues $\lambda_k$ (with $k=1,2,3, \ldots$) are in the
interval $ V(x_\mathrm{min}) < \lambda_k < \frac14 $. Thus the
solution of eq.~(\ref{FPdept}) can be written as,
\begin{equation}
\phi(x,\tau)= \sum_k c_k \; e^{- \lambda_k \, \tau } \; \psi_k(x)+
\int_{\frac14}^{\infty}  d\lambda \; c(\lambda) \; e^{- \lambda \;
\tau } \; \psi_{\lambda}(x) \;,
\end{equation}
where the Fourier coefficients $c_k$ and $c(\lambda)$ are obtained
from the initial data $ \phi(x,0) $ by scalar product with the
eigenfunctions. The late times asymptotics is dominated by the
ground state,
\begin{equation}
\phi(x,\tau)\buildrel{\tau\to\infty }\over= c_1 \; e^{- \lambda_1
\,\tau } \; \psi_1(x)  \; .
\end{equation}

A simple estimate from eq.~(\ref{mini}), namely, $\lambda_1
\approx V(x_\mathrm{min})+\sqrt{V''(x_\mathrm{min})/2}$ suggests
that $\lambda_1 \approx 0.247664 $. Therefore, $\Delta(p,t)$
approaches the steady-state solution as
\begin{equation}
\Delta(p,t) \buildrel{t\to\infty }\over=\Delta(p,\infty) +
\frac{24\,\pi\,c_1}{e^3 T \ln(1/e)}\; e^{- t/(4.038\ldots \,
t_\mathrm{tr})}\; \psi_1(p)  \; .\label{solFK}
\end{equation}
The steady-state drift current at asymptotically large time $t\gg
t_\mathrm{tr}$ is given by
\begin{equation}
J^i =4 \; e \intp \; \hat{p}^i \; \delta f(\bp) = 4e \intp \;
\hat{p}^i \; \boldsymbol{\mathcal{E}}\cdot\bhp \; n'_F(p) \;
\Delta(p) = \frac23 \; e \; \mathcal{E}^i \int_0^{\infty} dp\; p^2
\; n'_F(p) \;  \Delta(p) \; ,
\end{equation}
where the factor $4$ accounts for spin and the fact that the
departure from equilibrium for the particle and antiparticle
distribution functions are the same. In terms of the steady-state
solution $\Phi_{\infty}(x)$ and the dimensionless quantities
introduced above in eq.~(\ref{dimless}), the conductivity is
finally given by
\begin{equation}
\sigma = \frac{16 \,  T}{\pi e^2 \ln(1/e)} \int_0^{\infty} dx\;x^2
\; n_F'(x)  \; \Phi_{\infty}(x) \; . \label{fincondu}
\end{equation}
>From the differential equation (\ref{steadystate}) we can write
the following identity
\begin{equation}
\Phi_{\infty}(x)=2 \bigg\{\Phi_{\infty}(x)-
\frac{\Phi_{\infty}(x)}{2} \bigg[
\Phi''_{\infty}(x)+\Phi'_{\infty}(x)\left(\frac{2}{x}-
\tanh\frac{x}{2}\right)- \bigg(\frac{3}{4x}\coth\frac{x}{2}
+\frac{2}{x^2}\bigg)\Phi_{\infty}(x)\bigg]\bigg\} \; ,
\label{iden}
\end{equation}
which is now inserted in the expression for the conductivity
eq.~(\ref{fincondu}). After integration by parts we finally find
that
\begin{equation}\label{sigAMY}
\sigma[\Phi_\infty] = \frac{32 \;  T}{\pi e^2
\ln(1/e)}\int_0^{\infty} dx \; n_F'(x)\bigg[\frac{x^2}{2} \;
\left[\Phi'_{\infty}(x)\right]^2
+\bigg(1+\frac{3x}{8}\coth\frac{x}{2}\bigg) \;
\Phi^2_{\infty}(x)+x^2 \; \Phi_{\infty}(x)\bigg] \; .
\end{equation}
It is straightforward to confirm that this expression is exactly
$2Q[\chi]/3$ given by (4.3) in Ref.~\cite{arnold} after the
rescaling of the function $\chi(p)$ in that reference to our
dimensionless function $\Phi_\infty(x)$. Considering the
conductivity as a \emph{functional} of $\Phi_\infty(x)$ given by
eq.~(\ref{sigAMY}), we find that the variational condition
\begin{equation}
\frac{\delta\,\sigma[\Phi_\infty]}{\delta \Phi_\infty}=0
\end{equation}
leads to the steady-state Fokker-Planck equation
(\ref{steadystate}).

Clearly an analytic solution of the steady-state Fokker-Planck
equation (\ref{steadystate}) is not available. However, before we
embark on a numerical study it is important to understand the
small and large $x$ behavior of the solution. For this purpose we
use the small and large $x$ expansion of the hyperbolic functions.
After some algebra we find that the solution regular at the origin
has the power expansion
\begin{eqnarray}
\Phi^<_{\infty}(x)  & = &  a\, H_<(x) + I_<(x) \; ,\nn \\
H_<(x) & = & x^\nu\left[1 + a_1 \; x^2 + a_2 \;  x^4 +
  {\cal O}(x^6) \right] \; ,\nn\\
I_<(x) & = & x^2\left[\frac{2}{5} + \frac{3}{110} \; x^2 +
  \frac{157}{254100} \; x^4 + {\cal O}(x^6) \right] \; ,\label{smallx}
\end{eqnarray}
where \be \nu=\frac{\sqrt{15}-1}{2} = 1.43649\ldots \quad , \quad
a_1 = \frac{32 - 5 \, \sqrt{15}}{176} = 0.0717902\ldots\ \quad,
\quad a_2 = \frac{6099 -1526 \, \sqrt{15}}{84480} =
0.00223517\ldots \, . \ee For $x\gg 1$ the solution that does not
grow exponentially is given by
\begin{equation}
\Phi^>_{\infty}(x) =  b\, H_>(x) + I_>(x)\;,
\end{equation}
where $H_>(x)$ has an asymptotic expansion in positive powers of
$1/x$ and $e^{-x}$. Neglecting the exponentially small
corrections, we find
\begin{eqnarray}
H_>(x) &\buildrel{x\gg 1 }\over=& x \;
\Psi\left(\frac{7}{4},4;x\right) \left[ 1 +{\cal O}(e^{-x})\right]\nn\\
&=& x^{-\frac34}\bigg[ 1 + \sum_{k\geq 1} \frac{(-1)^k}{k! \,
x^k}\left(\frac74\right)_k
\left(-\frac{5}{4}\right)_k \bigg] \left[ 1 +{\cal O}(e^{-x})\right]\nn\\
&=&  x^{-\frac34}\left[1 +\frac{35}{16 \,x} + \frac{385}{512 \,
x^2} + {\cal O}(x^{-3}) \right]\left[ 1 +{\cal O}(e^{-x})
\right].\label{Hasi}
\end{eqnarray}
Here, $\Psi(a,b;z)$ stands for a confluent hypergeometric
function~\cite{gr} and $(\mu)_k \equiv \mu (\mu +1) \cdots
(\mu+k-1)$. The function $I_>(x)$ has the asymptotic behavior
\begin{equation}
I_>(x)= -\frac{4}{7} \; x +\frac{1}{7} \;
e^{-x}+\cdots.\label{Iasi}
\end{equation}
The functions $H_\lessgtr(x)$ are the corresponding solutions of
the homogeneous equation, while $I_\lessgtr(x)$ are the particular
solutions of the inhomogeneous equation that are finite at the
origin and do not grow exponentially at infinity. The asymptotic
linear growth with $x$ of $I_>(x)$ is a direct consequence of the
inhomogeneity. In addition, the homogeneous equation has a
linearly independent (irregular) solution that grows as
$x^{-\frac{\sqrt{15}+1}{2}}$ for $x \to 0$.

\begin{figure}[t]
\includegraphics[width=4.5in,keepaspectratio=true]{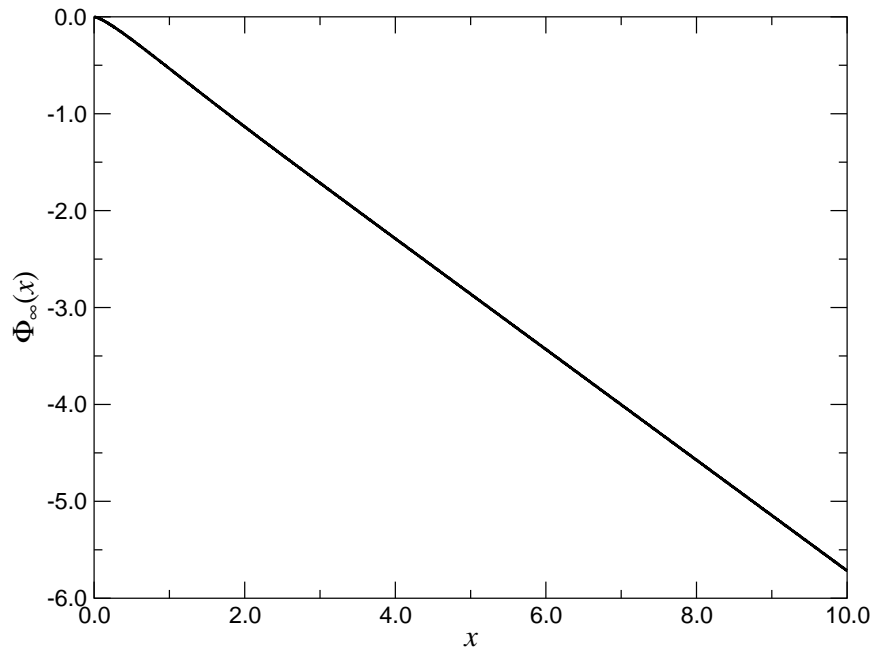}
\caption{Numerical solution of the steady-state Fokker-Planck
equation (\ref{steadystate}).}\label{fig:fiofx}
\end{figure}

The coefficients $a$ and $b$ are found by integrating both the
homogeneous and inhomogeneous equation with the initial conditions
gleaned from eq.~(\ref{smallx}) on the function and its derivative
at the origin for $H_<(x)$ and $I_<(x)$ and matching the function
and derivative to the asymptotic form
eqs.~(\ref{Hasi})-(\ref{Iasi}) at a value of $x\gg 1$. Such
procedure is numerically straightforward, we find that the
asymptotic form is achieved for relatively small values of
$x\simeq 3$ and the coefficients are found to be given by $a =
-0.896\ldots$, $b= -0.01\ldots$. The numerical solution is shown
in Fig.~\ref{fig:fiofx}. With this solution we finally find the
following value of the conductivity
\begin{equation}
\sigma= \frac{15.698\,T}{e^2 \ln(1/e)} \; ,\label{sigmaval}
\end{equation}
which is consistent with the numerical value quoted in
Ref.~\cite{arnold}, $\sigma= \frac{15.6964\,T}{e^2 \ln(1/e)}$.
However we point out that the asymptotic forms for the solution
for small or large $x$ are very different from the variational
basis chosen in Ref.~\cite{arnold}, in particular near the origin
the solution is not an analytic function of $x$.

\subsection{Beyond leading logarithms and the Boltzmann equation}

In Sec.~\ref{sec:pertsec} we have discussed the diagrams that will
contribute to next to leading order, which are necessary to fix
the constant inside the logarithm for the transport relaxation
time and the conductivity. These diagrams will yield new
contributions in the collision term. To next to leading order the
nonequilibrium fermion distribution function $f$ will replace the
equilibrium one $n_F$ in the three-loop diagram for the
polarization in the soft-fermion contribution $C_s^\mathrm{sf}$
[see eq.~(\ref{truker2})], the collision kernel must also include
the first diagram in Fig.~\ref{fig:nolead3loops}, the second
diagram will arise from linearizing in the departure from
equilibrium as discussed above.

The discussion after eq.~(\ref{spinccurr}) indicates that the
validity of the Boltzmann equation for the evolution of the
particle and antiparticle distribution functions is reliable
provided processes in which particles and antiparticles mix can be
neglected, namely, processes involve momentum exchange $q\ll T$.
However, \emph{if} there are processes that mix particles and
antiparticles that lead to secular terms in the perturbative
expansion, then the contribution from the spin current in
eq.~(\ref{spinccurr}) must be accounted for in the evolution
equation. In this case one would have to find the equation of
motion for the bilinears $\langle b^{\dagger} d^{\dagger}\rangle$,
these equations will likely involve the correlators $\langle
b^\dagger b\rangle$ leading to a coupled system of equations. The
new terms are associated with spin precession and in general must
be included for a consistent nonequilibrium
description~\cite{heinz}. The perturbative analysis and the direct
correspondence of the secular terms from the perturbative solution
of the Boltzmann equation indicate that such new contributions do
not appear at leading logarithmic order. Thus the simpler
Boltzmann equation for the particle and antiparticle distribution
functions in this leading logarithmic approximation can be
interpreted as a low-energy limit of the full set of evolution
equations, in the sense that processes that mix particles and
antiparticles are neglected.

\section{Conclusions}\label{conclusions}

In this article we began a program to study transport phenomena in
ultrarelativistic plasmas via the dynamical renormalization group
resummation. We focused on the DC electrical conductivity in a
ultrarelativistic QED plasma and extracted the conductivity from
the long time behavior or a kernel directly related to the
retarded photon polarization.

We began by studying the long-time behavior of this kernel in
perturbation theory in real time. Pinch singularities in the limit
of long wavelength and low frequency that are ubiquitous in the
usual perturbative expansion of the imaginary part in the
polarization in frequency-momentum space, are manifest as
\emph{secular} terms in real time, namely terms that diverge in
the long-time limit. At any finite time the perturbative
contributions to the kernel are finite but the long time behavior
of the perturbative expansion is divergent reflecting these pinch
singularities. We first extracted the leading secular terms at
lowest order in the perturbative expansion which includes both
self-energy and vertex corrections for the fulfillment of the Ward
identity. While the self-energy correction features a logarithmic
singularity on the fermion mass shell arising from the exchange of
ultrasoft photon which leads to the anomalous damping rate of the
hard fermions in the loop, the addition of the vertex correction,
as required to fulfill the Ward identity cancels the contributions
from the anomalous fermion damping. This cancellation is at the
heart of the difference between the quasiparticle and transport
relaxation time. We then extracted the leading logarithmic
contributions to the leading secular terms and identified the
diagrams that yield the leading logarithms and those that yield
subleading contributions without logarithms of the coupling.

After the study of the secular divergences in the perturbative
expansion, we introduce a resummation program via the dynamical
renormalization group applied to the equations of motion of the
gauge invariant single-particle distribution functions of hard
fermions. The resulting dynamical renormalization group equation
in real time \emph{is} the Boltzmann equation. By expanding the
solution of the linearized Boltzmann equation in perturbation
theory, we established a direct link between the perturbative
expansion of the polarization and the linearized Boltzmann
equation by explicitly showing how the Boltzmann equation
reproduces the secular terms in a perturbative expansion. This
comparison allowed to identify the different terms in the
linearized Boltzmann equation with the self-energy and vertex
corrections to the polarization. This detailed study leads to a
deeper understanding of the resummation of the perturbative series
in quantum field theory via the Boltzmann equation. Furthermore,
it allows to identify unambiguously the diagrams that contribute
to the collision term of the linearized Boltzmann equation. Just
as the renormalization group equation in deep inelastic scattering
or critical phenomena provides a resummation of parts (such as
leading logarithms) of select type of diagrams, the Boltzmann
equation also resums parts (leading logarithms) of select diagrams
(\emph{uncrossed} ladder plus rainbow type self-energy
corrections).

Thus the dynamical renormalization group provides directly in real
time a link between the quantum field theoretical and the
Boltzmann kinetic approaches to transport phenomena. Furthermore,
we have explicitly shown how the relaxation time approximation of
the Boltzmann equation is equivalent to neglecting the vertex
corrections in the linearized approximation and features the
anomalous logarithmic time dependence manifest in the hard fermion
damping rate. The conductivity \emph{vanishes} in the relaxation
time approximation as a consequence of the anomalous logarithms in
time. As discussed in Secs.~\ref{sec:secterms} and
\ref{sec:reltimeappx} such an approximation is not valid for the
hot QED plasma.

Recognizing that the leading logarithmic contribution to the
collision term arises from the region of momentum exchange
$eT\lesssim q\lesssim p\sim T$, we expanded the collision kernel
of the linearized Boltzmann equation in powers of $q/p$ to extract
the leading logarithmic contribution. We then obtained a
Fokker-Planck equation in momentum space for the small departure
from equilibrium. This equation clearly reveals that the transport
time scale is $t_\mathrm{tr}=\frac{24 \,\pi}{e^4 T \ln(1/e)}$. The
time dependent Fokker-Planck equation is solved by expanding in
the eigenfuntions of a positive definite Hamiltonian. For late
times, its solution approaches the steady-state solution as $\sim
e^{- t/(4.038\ldots \, t_\mathrm{tr})}$. We solved analytically
the steady-state Fokker-Planck equation for small and large
momenta. We matched numerically the solutions and found that the
DC conductivity is given by $\sigma=
\frac{15.698\,T}{e^2\ln(1/e)}$ to leading logarithmic order. This
result is within less than $0.1\,\%$ that quoted in
Ref.~\cite{arnold}. We established contact with the variational
formulation of Refs.~\cite{arnold,basagoiti2,aarts} by showing
that the steady-state Fokker-Planck equation can be obtained from
a variational functional which is the DC conductivity. We have
identified the diagrams that contribute beyond the leading
logarithmic order and provided a discussion of the corrections
that are necessary to be included in order to go \emph{beyond} the
Boltzmann equation.

The results of this study highlight that the dynamical
renormalization group is the concept that provides a direct bridge
between quantum field theory and the kinetic description in real
time and puts transport phenomena on a similar footing as critical
phenomena, namely, the Boltzmann equation \emph{is} the
renormalization group equation that provides a nonperturbative
resummation of diagrams. It establishes this formulation as an
alternative to studying transport phenomena in relativistic
plasmas firmly based on the microscopic quantum field theory. In
the next step of our program we will apply these ideas to study
transport phenomena in QCD relevant to the physics of the
quark-gluon plasma.

\begin{acknowledgments}
We thank Emil Mottola for collaboration during the early stages of
this work, illuminating discussions and constant interest. D.B.\
thanks C.\ Gale and A.\ Majumder for correspondence. D.B.\ and H.
J. d. V.\ would like to thank the Theoretical Division of Los
Alamos National Laboratory for its hospitality during the early
stages of this work. The authors thank the hospitality of the
Institute for Theoretical Physics at the University of California
at Santa Barbara during part of this work. The work of D.B.\ was
supported in part by the US National Science Foundation under
grants PHY-9988720 and NSF-INT-9905954. The work of S.-Y.W.\ was
supported by the US Department of Energy under contract
W-7405-ENG-36.
\end{acknowledgments}

\appendix

\section{Imaginary-Time Propagators}\label{app:itf}

In this appendix we review the imaginary-time propagators in their
spectral representations. The fermion propagator is given
by~\cite{book:lebellac}
\begin{equation}
S(i\omega_n,\bp)=\intpo \; \frac{\rho_F(p_0,\bp)}{p_0-i\omega_n}
\; , \label{fermionprop}
\end{equation}
where $\omega_n=(2n+1)\pi T$. The spectral function
$\rho_F(p_0,\bp)$ for \emph{hard} (or \emph{free}) fermions is
given by
\begin{eqnarray}
&&\rho_F(p_0,\bp)=\gamma_+(\bhp) \; \rho_+(p_0,p)+\gamma_-(\bhp)
\;
  \rho_-(p_0,p)   \; ,\nn\\
&&\gamma_\pm(\bhp)=(\gamma^0\mp\bgamma\cdot\bhp)/2 \;,\quad
\rho_\pm(p_0,p)=\delta(p_0\mp p) \; ,\label{ferprop}
\label{rhoplusmin}
\end{eqnarray}
where $p=|\bp|$, $\bhp=\bp/p$ and $\rho_\pm(p_0,p)$ are the free
particle ($+$) and antiparticle ($-$) spectral functions,
respectively. The \emph{soft} (or \emph{HTL-resummed}) fermion
propagator ${}^\ast\!S(i\omega,\bp)$ has the same form as that
given in eq.~(\ref{fermionprop}) but in terms of the HTL-resummed
fermion spectral function,
\begin{eqnarray}
&&{}^\ast\!\rho_F(p_0,\bp)=\gamma_+(\bhp) \;
{}^\ast\!\rho_+(p_0,p)+
\gamma_-(\bhp) \; {}^\ast\!\rho_-(p_0,p) \; ,\nn\\
&&{}^\ast\!\rho_\pm(p_0,p)=Z_\pm(p) \; \delta(p_0-\omega_\pm(p))+
Z_\mp(p) \; \delta(p_0+\omega_\mp(p))+\beta_\pm(p_0,p) \;
\Theta(p^2-p_0^2) \; , \label{htlferprop}
\end{eqnarray}
where $Z_\pm(p)$ are the residues of the timelike fermionic
quasiparticle poles at $\omega_\pm(p)$, respectively, and
\begin{eqnarray}
\beta_\pm(p_0,p)&=&\frac{\frac{m_f^2}{2p}\big(1\mp\frac{p_0}{p}\big)}{\left\{p
\big(1\mp\frac{p_0}{p}\big)\pm\frac{m_f^2}{2p}\left[\big(1\mp\frac{p_0}{p}\big)
\ln\big|\frac{p_0+p}{p_0-p}\big|\pm 2\right]\right\}^2+\frac{\pi^2
m_f^4}{4 p^2}\big(1\mp\frac{p_0}{p}\big)^2} \; ,
\end{eqnarray}
with $m_f^2=e^2 T^2/8$ the thermal fermion mass.

The longitudinal and transverse photon propagators are given by
\begin{eqnarray}
D_L(i\nu_n,q)&=&
-\frac{1}{q^2}+\intqo\;\frac{\rho_L(q_0,q)}{q_0-i\nu_n} \; ,\nn\\
D_T(i\nu_n,q)&=&\intqo \; \frac{\rho_T(q_0,q)}{q_0-i\nu_n} \; ,
\label{photonprop}
\end{eqnarray}
respectively, where $\nu_n=2\pi n T$. The frequency independent
term is $D_L$ arises from the instantaneous Coulomb interaction
and is neglected in our calculation because it does not contribute
to the imaginary part. For \emph{hard} photons we have $\rho_L=0$
and
\begin{equation}
\rho_T(q_o,q)=\frac{1}{2q}[\delta(q_0-q)-\delta(q_0+q)] \; .
\end{equation}
The \emph{soft} photon propagators ${}^\ast\!D_L(i\nu_n,q)$ and
${}^\ast\!D_T(i\nu_n,q)$ have the same form as those given by
eq.~(\ref{photonprop}) but in terms of the HTL-resummed spectral
functions~\cite{book:lebellac}
\begin{equation}
\rho_{L(T)}(q_0,q)=\mathrm{sgn}(q_0) \; Z_{L(T)}(q) \;
\delta[q_0^2-\omega_s^2(q)]+ \beta_{L(T)}(q_0,q) \;
\Theta(q^2-q_0^2) \; ,\label{HTLrhos}
\end{equation}
where $Z_{L(T)}(q)$ are the residues of the timelike quasiparticle
poles at $\omega_{L(T)}(q)$ and
\begin{eqnarray}\label{betas}
\beta_L(q_0,q)&=&-\frac{1}{\pi}
\frac{\Im\Pi^\mathrm{HTL}_L(q_0,q)}
{[q^2+\Re\Pi^\mathrm{HTL}_L(q_0,q)]^2
+[\Im\Pi^\mathrm{HTL}_L(q_0,q)]^2} \; ,\nn\\
\beta_T(q_0,q)&=&-\frac{1}{\pi}
\frac{\Im\Pi^\mathrm{HTL}_T(q_0,q)}
{[q_0^2-q^2-\Re\Pi^\mathrm{HTL}_T(q_0,q)]^2
+[\Im\Pi^\mathrm{HTL}_T(q_0,q)]^2} \; .
\end{eqnarray}
In the above expressions, $\Pi^\mathrm{HTL}_{L(T)}(q_0,q)$ is the
longitudinal (transverse) photon self-energy in the HTL
approximation~\cite{book:lebellac}:
\begin{eqnarray}
\Pi^\mathrm{HTL}_L(q_0,q)&=&\frac{e^2 T^2}{6}
\left[2-\frac{q_0}{q}
\ln\left|\frac{q_0+q}{q_0-q}\right|\right]-i\pi \; \frac{e^2
T^2}{6}
\frac{q_0}{q} \; \Theta(q^2-q_0^2) \; ,\nn\\
\Pi^\mathrm{HTL}_T(q_0,q)&=&\frac{e^2 T^2}{12} \left[\frac{2
q_0^2}{q^2}+\frac{q_0}{q} \left(1-\frac{q_0^2}{q^2}\right)
\ln\left|\frac{q_0+q}{q_0-q}\right|\right] -i\pi \; \frac{e^2
T^2}{12}\frac{q_0}{q} \left(1-\frac{q_0^2}{q^2}\right) \;
\Theta(q^2-q_0^2) \; ,\label{piTHTL}
\end{eqnarray}
The frequency independent part in $D_L(i\nu_n,q)$ and
${}^\ast\!D_L(i\nu_n,q)$ arises from the instantaneous Coulomb
interaction. Furthermore, we will adopt the convention that
bosonic Matsubara frequencies are denoted by $\nu_m=2m \pi T$ and
the fermionic ones by $\omega_m=(2m+1)\pi T$.

The spectral representation of the fermion and photon propagators
facilitate the sum over internal Matsubara frequencies. The
convenient formulas to carry out these sums systematically are
given by~\cite{book:lebellac}
\begin{eqnarray}
T\sum_{\nu_n}F(i\nu_n)&=& -\sum_{z_i}\mathrm{Res}F(z_i)\; n_B(z_i) \; ,\nn\\
T\sum_{\omega_n}F(i\omega_n)&=&\sum_{z_i}\mathrm{Res}F(z_i)\;
n_F(z_i) \; , \label{matsums}
\end{eqnarray}
where $\mathrm{Res}$ refers to the residues at the (simple) poles
$z_i$ of the function $F(z)$, $n_B$ is the Bose distribution and
$n_F$ is the Fermi distribution (for vanishing chemical potential)
\begin{equation}
n_B(\omega)=\frac{1}{e^{\omega/T}-1} \; ,\quad
n_F(\omega)=\frac{1}{e^{\omega/T}+1} \; .
\end{equation}

\section{Real-Time Propagators}\label{app:rtf}

In this appendix we summarize the real-time propagators which are
used in the main text to derive the quantum Boltzmann equation.
The free fermion propagators (with zero chemical potential) are
defined by
\begin{eqnarray}
&&\langle \Psi^{a}(\bx,t) \bar{\Psi}^{b}({\bx}',t')\rangle = i
\intp S_\bp^{ab}(t,t')\, e^{i\bp
\cdot({\bx}-{\bx}')} \; ,\nn\\
&&S_\bp^{++}(t,t')= S_\bp^>(t,t') \; \theta(t-t') +S_\bp^<(t,t')
\; \theta(t'-t), \nn\\ &&S_\bp^{--}(t,t')= S_\bp^>(t,t') \;
\theta(t'-t)
+S_\bp^<(t,t') \; \theta(t-t') \; ,\nn\\
&&S_\bp^{\pm\mp}(t,t')= S_\bp^\lessgtr(t,t') \; ,
\label{fermionprop1}
\end{eqnarray}
where $\langle\cdots\rangle$ denotes expectation value with
respect to the initial density matrix and $a,\;b=\pm$ refer to
fields in the forward ($+$) and backward ($-$) time branches. For
\emph{hard} fermions the Wightman functions read
\begin{eqnarray}
S_\bp^>(t,t')&=& -i\big\{\gamma_+(\bhp) \;  [1-f(\bp,t_0)]\;
e^{-ip(t-t')}+\;
\gamma_-(\bhp)\; \overline{f}(-\bp,t_0)\; e^{ip(t-t')}\big\} \; ,\nn \\
S_\bp^<(t,t')&=&i\big\{\gamma_+(\bhp)\; f(\bp,t_0)\;
e^{-ip(t-t')}+\;\gamma_-(\bhp)  \; [1-\overline{f}(-\bp,t_0)]\;
e^{ip(t-t')}\big\} \; , \label{fermionprop2}
\end{eqnarray}
where $f(\bp,t_0)$ and $\overline{f}(\bp,t_0)$ are the
nonequilibrium distribution functions for the hard fermions and
antifermions at the initial time $t_0$, respectively. For
\emph{soft} fermions the HTL-resummed Wightman functions
${}^\ast\!S_p^\gtrless(t,t')$ are expressed in terms of the
spectral functions as
\begin{eqnarray}
{}^\ast\!S_p^>(t,t')&=& -i\intqo\,
{}^\ast\!\rho_F(p_0,p)\,[1-n_F(p_0)]\,e^{-ip_0(t-t^\prime)},\nn\\
{}^\ast\!S_p^<(t,t')&=& i\intqo\, {}^\ast\!\rho_F(p_0,p)\;
n_F(p_0)\; e^{-ip_0(t-t^\prime)},\label{wightmanS}
\end{eqnarray}
where ${}^\ast\!\rho_F(p_0,p)$ is given by eq.~(\ref{htlferprop}).

The longitudinal photon propagators are given by
\begin{eqnarray}
&&\langle A^{a}_0({\bx},t) A^{b}_0({\bx}',t')\rangle = i
\intq\,D_{L,q}^{ab}(t,t') \;
e^{i\bq\cdot({\bx}-{\bx}')} \; ,\nn\\
&&D_{L,q}^{++}(t,t')= \frac{1}{q^2} \; \delta(t-t')+
D_{L,q}^>(t,t') \; \theta(t-t')
+D_{L,q}^<(t,t') \; \theta(t'-t) \; ,\nn \\
&&D_{L,q}^{--}(t,t')= -\frac{1}{q^2} \; \delta(t-t')+
D_{L,q}^>(t,t') \; \theta(t'-t)
+D_{L,q}^<(t,t') \; \theta(t-t') \; ,\nn \\
&&D_{L,q}^{\pm\mp}(t,t')= D_{L,q}^\lessgtr(t,t').
\end{eqnarray}
The Wightman functions are expressed in terms of the spectral
functions as
\begin{eqnarray}
D_{L,q}^>(t,t')&=& -i\intqo\;
\rho_L(q_0,q)\; [1+n_B(q_0)]\; e^{-iq_0(t-t^\prime)} \; ,\nn\\
D_{L,q}^<(t,t')&=& -i\intqo\; \rho_L(q_0,q)\; n_B(q_0)\;
e^{-iq_0(t-t^\prime)} \; . \label{wightmanL}
\end{eqnarray}
where $\rho_L=0$ for the \emph{hard} photon propagator and
$\rho_L={}^\ast\!\rho_L$ for the \emph{soft} photon propagator.

The transverse photon propagators are given by
\begin{eqnarray}
&&\langle A^{i,a}_T({\bx},t) A^{j,b}_T({\bx}',t')\rangle =-i \int
\frac{d^3q}{(2\pi)^3} \; D_{T,q}^{ab}(t,t')\;
\mathcal{P}_T^{ij}(\bq)\; e^{i\bq\cdot({\bx}-{\bx}')} \; ,\nn\\
&&D_{T,q}^{++}(t,t')= D_{T,q}^>(t,t') \; \theta(t-t')
+D_{T,q}^<(t,t') \; \theta(t'-t) \; ,\nn \\
&&D_{T,q}^{--}(t,t')= D_{T,q}^>(t,t') \; \theta(t'-t)
+D_{T,q}^<(t,t') \; \theta(t-t') \; ,\nn\\
&&D_{T,q}^{\pm\mp}(t,t')= D_{T,q}^\lessgtr(t,t') \; ,
\label{gaugeprop1}
\end{eqnarray}
where $\mathcal{P}_T^{ij}(\bhq)=\delta^{ij}-\hat{q}^i \hat{q}^j$
is the transverse projector. The Wightman functions are expressed
in terms of the spectral functions as
\begin{eqnarray}
D_{T,q}^>(t,t')&=&i\intqo
\,{\rho}_T(q_0,q)\; [1+n_B(q_0)]\; e^{-iq_0(t-t')},\nn\\
D_{T,q}^<(t,t')&=&i\intqo\; \rho_T(q_0,q)\; n_B(q_0)\;
e^{-iq_0(t-t')} \; , \label{wightmanT}
\end{eqnarray}
where $\rho_T=\rho_B$ for the \emph{hard} photon propagator and
$\rho_T={}^\ast\!\rho_T$ for the \emph{soft} photon propagator.

\end{document}